\documentclass[11pt,oneside]{article}
\usepackage{lmodern} 
\usepackage{jheppub} 
\usepackage{amsmath,amsfonts}
\usepackage{amssymb}
\usepackage{charter}
\usepackage{booktabs}
\usepackage[dvipsnames]{xcolor}
\usepackage{setspace}
\usepackage{tensor}
\usepackage{xspace}
\usepackage{slashed}
\usepackage{pdflscape}
\usepackage{hyperref}
\usepackage[capitalize]{cleveref}
\usepackage[normalem]{ulem}
\usepackage{fontawesome}

\newcommand\blfootnote[1]{%
  \begingroup
  \renewcommand\thefootnote{}\footnote{#1}%
  \addtocounter{footnote}{-1}%
  \endgroup
}

\usepackage{lipsum}
\setlipsum{%
  par-before = \begingroup\color{gray},
  par-after = \endgroup
}

\usepackage{bbold, bm}
\usepackage{multicol,multirow}
\newcommand{\gray}[1]{{\color{Gray} #1}}

\newcommand{\cD}{\mathcal{D}}

\newcommand{\cL}{\mathcal{L}}

\newcommand{\cO}{\mathcal{O}}
\newcommand{\cP}{\mathcal{P}}

\newcommand{\mQL}{\ensuremath{m^2_{\tilde Q_3}}}
\newcommand{\mtR}{\ensuremath{m^2_{\tilde{t}_R}}}
\newcommand{\mbR}{\ensuremath{m^2_{\tilde{b}_R}}}
\newcommand{\mQLp}[1]{\ensuremath{m^{#1}_{\tilde Q_3}}}
\newcommand{\mtRp}[1]{\ensuremath{m^{#1}_{\tilde{t}_R}}}
\newcommand{\mbRp}[1]{\ensuremath{m^{#1}_{\tilde{b}_R}}}

\newcommand{\dd}{\mathop{}\!\mathrm{d}}

\newcommand{\hc}{\mathrm{H.c.}}

\newcommand{\Dfbu}{\mathord{\buildrel{\lower3pt\hbox{$\scriptscriptstyle{\leftrightarrow \tiny{ \ \ \ } }$}}\over {D^{\mu}}}} 
\newcommand{\Dfbd}{\mathord{\buildrel{\lower3pt\hbox{$\scriptscriptstyle\leftrightarrow$}}\over {D}_{\mu}}} 

\newcommand{\matchete}{\texttt{Matchete}\xspace}

\makeatletter
\g@addto@macro\bfseries{\boldmath} % use boldmath in titles
\makeatother

\allowdisplaybreaks
\title{SUSY meets SMEFT:\\ Complete one-loop matching of the general MSSM}

\author[a]{Sabine Kraml,}
\author[b]{Andre Lessa,}
\author[c]{Suraj Prakash,}
\author[d]{and Felix Wilsch}

\affiliation[a]{Laboratoire de Physique Subatomique et de Cosmologie, Université Grenoble-Alpes, \\CNRS/IN2P3, 53 Avenue des Martyrs, 38026 Grenoble, France}
\affiliation[b]{Centro de Ciências Naturais e Humanas, Universidade Federal do ABC, Santo André, \\09210-580 SP, Brazil}
\affiliation[c]{Departament de Física Teòrica, IFIC (Universitat de València - CSIC), Parc Científic UV,\\ 
C\,/ Catedrático José Beltrán 2, E-46980 Paterna (Valencia), Spain}
\affiliation[d]{Institute for Theoretical Particle Physics and Cosmology, RWTH Aachen University, \\Sommerfeldstr.~16, D-52074 Aachen, Germany}

\emailAdd{sabine.kraml@lpsc.in2p3.fr}
\emailAdd{andre.lessa@ufabc.edu.br}
\emailAdd{suraj.prakash@ific.uv.es}
\emailAdd{felix.wilsch@physik.rwth-aachen.de}

\preprint{%
TTK-25-14 
\\
\hspace*{\fill} P3H-25-036
}

\abstract{%
We present the complete one-loop matching of the Minimal Supersymmetric Standard Model~(MSSM) onto the Standard Model Effective Field Theory~(SMEFT), considering the most general case for the MSSM with conserved $R$-parity, which has 124~free parameters. 
The matching is performed with the \matchete package, which integrates out all superpartners at once with non-degenerate masses, while also retaining the most general flavor structure.
Our results include all correlations among the different SMEFT Wilson coefficients that are governed by supersymmetry and thus provide a basis for future systematic and global studies of the MSSM parameter space employing EFT methods.
A~detailed discussion is provided on the treatment of the Higgs sector and electroweak symmetry breaking, along with the reduction of redundant operators in the EFT Lagrangian to the Warsaw basis.
Furthermore, we validate against existing results in the literature and present a minimal phenomenological example. 
As an alternative low-energy scenario, we also provide the complete one-loop matching of the MSSM onto the two-Higgs-doublet-model EFT, where the second Higgs doublet is retained in the infrared spectrum.
Extensive auxiliary material, including the code utilized for the matching, is available on GitHub~\href{https://github.com/BSM-EFT/MSSM-to-SMEFT}{\faicon{github}}.
\blfootnote{\href{https://github.com/BSM-EFT/MSSM-to-SMEFT}{\hspace*{-0.64cm} \scriptsize \faicon{github} https://github.com/BSM-EFT/MSSM-to-SMEFT}}
}

\begin{document}
\maketitle
\setstretch{1.1}

% Introduction
\section{Introduction}

The quest for new physics~(NP) beyond the Standard Model~(SM) of electroweak and strong interactions is one of the principal challenges for high-energy particle physics and the driving motivation for an extensive experimental program at the Large Hadron Collider~(LHC) and possible future colliders. 
The non-observation, so far, of any direct signals of NP at current experiments indicates a mass gap between the electroweak scale and the NP mass scale. 
This motivates the use of Effective Field Theory~(EFT) techniques to study indirect deviations from SM predictions in a model-independent manner, using the framework of the Standard Model Effective Field Theory~(SMEFT)~\cite{Buchmuller:1985jz}, where the renormalizable part of the SM Lagrangian is supplemented by higher-dimensional operators that capture the indirect effects of the unknown heavy BSM states.
A general EFT-based analysis strategy for measurements at the LHC was outlined in~\cite{Englert:2014cva}. For recent reviews on the SMEFT,  see~\cite{Brivio:2017vri,Isidori:2023pyp}. 

In principle, the SMEFT facilitates placing model-independent constraints on NP by deriving limits on the Wilson coefficients of the higher-dimensional operators from experimental measurements.
Apart from the model independence, a significant advantage of the EFT framework is its ability to systematically combine diverse measurements from various different energies and experiments in a consistent manner, while adequately resumming large logarithmic corrections that appear due to the presence of scale hierarchies.
Several such global SMEFT fits have been performed in recent years, e.g., in~\cite{Ellis:2018gqa,daSilvaAlmeida:2018iqo,Ellis:2020unq,Brivio:2021alv,Ethier:2021bye,Bruggisser:2022rhb,Grunwald:2023nli,Allwicher:2023shc,Bartocci:2023nvp,Bartocci:2024fmm}.

The large number of free parameters in the SMEFT alone ---~ 2499~in the baryon-~($B$) and lepton-number~($L$) conserving sector of the SMEFT at dimension six~\cite{Alonso:2013hga}~--- poses a significant challenge to global fits, however.
Therefore, these fits always employ some simplifying assumptions, such as only considering a subset of operators, or specific flavor assumptions, in order to reduce the number of free parameters.
This comes at the cost of losing model independence and possibly important correlations between different sectors, making the interpretation of global fits in the context of concrete NP scenarios ambiguous.
These shortcomings can be overcome by considering specific BSM theories as ultraviolet~(UV) completions of the SMEFT.
Once the UV theory is matched onto the SMEFT, it imposes specific correlations among the different Wilson coefficients and, by that, significantly reduces the number of parameters.

Several tools that automate this matching process at the one-loop level, for theories with heavy scalars and fermions,\footnote{Integrating out vector bosons is more challenging since they originate either from composite dynamics or from the spontaneous breaking of an extended symmetry group, complicating a fully automatic implementation~\cite{Thomsen:2024abg}.} 
have recently become available:\footnote{See also~\cite{Cohen:2020qvb, Fuentes-Martin:2020udw} for partially automatic one-loop implementations and \cite{Criado:2017khh} for a tree-level matching tool.}  
\texttt{Matchete}~\cite{Fuentes-Martin:2022jrf} which employs functional methods~\cite{Gaillard:1985uh, Chan:1986jq, Cheyette:1987qz, Chan:1985ny, Fraser:1984zb, Aitchison:1984ys, Aitchison:1985pp, Aitchison:1985hu, Cheyette:1985ue, Dittmaier:1995cr, Dittmaier:1995ee, Henning:2014wua, delAguila:2016zcb, Henning:2016lyp, Fuentes-Martin:2016uol, Zhang:2016pja, Cohen:2020fcu, Boggia:2016asg, Dittmaier:2021fls} to determine the matching conditions, 
\texttt{Matchmakereft}~\cite{Carmona:2021xtq} which relies on Feynman diagrams, and 
\texttt{CoDEx}~\cite{DasBakshi:2018vni} based on the Universal One-Loop Effective Action~(UOLEA)~\cite{Drozd:2015rsp,Kramer:2019fwz}, which at its core also uses functional methods.
Up to this point, however, these methods and tools have mainly been applied to relatively basic models, wherein only one or at most a handful of fields are integrated out from the UV theory~\cite{Fuentes-Martin:2016uol, Dawson:2017vgm, Kramer:2019fwz, Haisch:2020ahr, Gherardi:2020det, Ellis:2020ivx, Angelescu:2020yzf, Summ:2020dda, Zhang:2021jdf, Brivio:2021alv, Dedes:2021abc, Du:2022vso, Li:2022ipc, Crivellin:2022fdf, Liao:2022cwh, Banerjee:2023iiv, Chakrabortty:2023yke, terHoeve:2023pvs, Li:2023ohq, DasBakshi:2024krs, Cepedello:2024ogz}.\footnote{See also~\cite{Fuentes-Martin:2019ign,Fuentes-Martin:2020luw,Fuentes-Martin:2020hvc} for the (non-automated) matching of a BSM theory with heavy vector fields, and \cite{Dekens:2019ept} for the one-loop matching of the SMEFT onto its low-energy description given by the Low Energy Effective Theory~(LEFT).}

In this article, we go a significant step further by applying \texttt{Matchete} to a highly complex UV theory. 
Concretely, we derive the complete one-loop matching of the general $R$-parity conserving {Minimal Supersymmetric Standard Model}~(MSSM), see e.g.~\cite{Martin:1997ns}, with its entire set of 124~free parameters, onto the $B$- and $L$-conserving dimension-six SMEFT Lagrangian~\cite{Grzadkowski:2010es}. 

Besides their complexity, supersymmetric models are interesting for their theoretical beauty and relevant phenomenological implications. Supersymmetry is a symmetry between fermions and bosons, thus providing a unified description of matter, gauge, and Higgs fields. It is in fact the natural and unique extension of relativistic space--time symmetries~\cite{Haag:1974qh}. 
Supersymmetry implies a partner particle, a so-called \emph{superpartner}, for every SM field. In its local gauge theory version (supergravity)~\cite{Freedman:1976xh,Deser:1976eh}, 
it also includes spin-2 and spin-3/2 states, 
the graviton and its superpartner, the gravitino, and is hence potentially capable of connecting gravity with the other interactions.
On the phenomenological side, even in the form of split~\cite{Arkani-Hamed:2004ymt,Giudice:2004tc} or unnatural~\cite{Bagnaschi:2014rsa} supersymmetry, it allows for gauge coupling unification and viable dark-matter candidates, and it motivates a light SM-like Higgs state, as observed at 125~GeV~\cite{ATLAS:2012yve,CMS:2012qbp}.  

Concentrating on the MSSM, in this work, all superpartners of the SM~fields, i.e., five types of sfermions, each coming in three generations, three types of gauginos, and the Higgsinos, as well as a second Higgs doublet, are integrated out. 
The matching is performed up to dimension six at a single scale~$\bar\mu$, but the masses of all heavy particles are kept generic and non-degenerate. 
The matching scale is furthermore taken above the electroweak scale~$\bar\mu \gg v$, corresponding to the limit where all NP~degrees of freedom decouple. 
In this case, one of the two Higgs doublets of the MSSM can be identified as the SM-like doublet, leading to the SMEFT describing the correct low-energy limit of the theory. 
The corresponding code, as well as the full matching conditions in the Warsaw basis~\cite{Grzadkowski:2010es}, are provided  
on GitHub~\cite{mssm-to-smeft:github}.
Our results include all correlations among the different SMEFT Wilson coefficients that are governed by supersymmetry. They can thus provide the basis for future systematic and global studies of the MSSM parameter space employing EFT methods. 

It is important to note here that we have used a modified, performance-enhanced version of the \matchete package for this project. For instance, we benefit from optimized routines for redundant operator reduction, resulting in considerable gains in speed. 
More specifically for the MSSM matching, the way in which operator coefficients are simplified internally in \matchete was significantly improved; 
this was particularly crucial for the MSSM, given the immense expressions it entails. 
In addition, the memory consumption had to be improved due to the large number of interaction terms that contribute at one loop. 
Finally, another necessary enhancement of \matchete for this project was the incorporation of a consistent treatment of heavy flavored particles, in our case the sfermions. These features are now part of \texttt{v0.4.0}~\cite{matchete:v03} of \matchete. The analysis code for this project, provided as supplementary material, is not compatible with earlier release versions of the package.

The matching of (parts of) the MSSM as the UV theory onto the SMEFT was previously studied in the literature. In particular, the one-loop matching conditions between the SM and the MSSM were derived in~\cite{Bagnaschi:2014rsa}, see also~\cite{Crivellin:2010er,Crivellin:2011jt,Giudice:2011cg,Bagnaschi:2017xid,Wells:2017vla}, and used to derive constraints from Higgs mass measurements. 
The resulting Wilson coefficients from integrating out heavy stops and/or sbottoms from the MSSM were discussed more widely in~\cite{Henning:2014gca,Henning:2014wua,Drozd:2015kva,Drozd:2015rsp,Huo:2015nka,Kramer:2019fwz}. The effect of neutralinos/charginos was discussed in~\cite{Han:2017cfr}. 
The effects of a scalar top partner and a fermionic dark-matter candidate (i.e., stop- and neutralino-like new states) on $t\bar t$~differential distributions were studied in~\cite{Lessa:2023tqc} in a simplified setting, treating the coupling of the new states with SM top quarks as a free parameter.\footnote{See also~\cite{Battaglia:2021nys} for a study of the effects of a scalar top partner on the Higgs $p_T$~spectrum.}
We validate our results against the analytic expressions existing in the literature. Moreover, we use the scenario from \cite{Lessa:2023tqc} for a minimal phenomenological example and numerical comparison.  

All previous results in the literature applied approximations in order to obtain the MSSM matching conditions. 
For example, only the leading contributions to specific operators were taken into account, or only a subset of the full superpartner spectrum was considered contributing to the matching. 
To our knowledge, no complete matching onto the SMEFT, taking into account all subleading contributions by all superpartners and including all correlations, has been performed thus far. 
This gap is filled by the work presented here. Moreover, the provided implementation in \texttt{Matchete} can be the basis for future studies and global fits, e.g., in the spirit of \cite{terHoeve:2023pvs}.

%%%%%%%%%%%%%%%%%%%%%%%%%%

The remainder of this paper is structured as follows:
In Sec.~\ref{sec:MSSM}, we briefly review the general structure of supersymmetric Lagrangians before introducing the MSSM, for which we will adopt a notation similar to the one commonly employed in the SMEFT literature.
The Higgs sector of the MSSM is discussed in Sec.~\ref{sec:Higgs}, paying special attention to differences in the treatment of electroweak symmetry breaking in the MSSM and the SMEFT. 
A summary of all the steps involved in the matching calculation and the reduction of redundant operators is presented in Sec.~\ref{sec:matching}.  
(Readers familiar with the MSSM and its subtleties may want to skip Sec.~\ref{sec:MSSM}--\ref{sec:Higgs}, and directly jump to Sec.~\ref{sec:matching}. Similarly, readers with experience in EFT matching may want to skip Sec.~\ref{sec:matching}.)
Here, special attention is paid to the peculiarities of the MSSM matching.
The results of the matching are presented in Sec.~\ref{sec:matching-results}, where we compare specific operators to known results from the literature. Moreover, we present a minimal phenomenological example including a numerical evaluation.
We conclude in Sec.~\ref{sec:conclusion}. 

A couple of appendices complete the paper. 
Our conventions, crucial for using our results, are summarized in Appendix~\ref{app:conventions}.
Details on the mapping of the MSSM Lagrangian from the common notation in terms of Weyl spinors to a notation based on Dirac spinors, which are typically used in the SMEFT literature, are provided in Appendix~\ref{app:Weyl-Dirac}. 
Appendix~\ref{app:MSSM-Lagrangian}  provides the full MSSM Lagrangian, expressed in terms of four-component
Dirac and Majorana spinors, and written in the soft-SUSY mass basis, while 
Appendix~\ref{app:Higgs-basis_alignment-limit} focuses on the Higgs basis,  discussing some subtleties of the alignment limit.
An instructive example for how the results of this work can be extended is given in Appendix~\ref{app:2HDM-EFT}, where we consider the case that the second Higgs doublet of the MSSM is light enough to be kept dynamic in the spectrum of the EFT. 
The matching is thus performed onto a two-Higgs-doublet-model EFT (2HDM-EFT)~\cite{Dermisek:2024ohe}. 
This also constitutes, to our knowledge, the first complete one-loop matching of a BSM theory onto an EFT beyond the SMEFT.
Finally, Appendix~\ref{app:reading-results} provides details on how to read and interpret our results. 

Due to their length, the full one-loop matching conditions in the Warsaw basis (and in the 2HDM-EFT basis) are delegated to the auxiliary material provided on GitHub~\cite{mssm-to-smeft:github}, where they can be found in human readable~(PDF files) and electronic (Wolfram language files with \matchete syntax) form.

% Minimal Supersymmetric Standard Model
\section{Minimal Supersymmetric Standard Model}
\label{sec:MSSM}

We begin by briefly reviewing the general structure of supersymmetric Lagrangians, before discussing the MSSM as a special case. 
We focus on those features of the MSSM that are particularly relevant for its matching onto the SMEFT. 
Hence, we translate the common formulation of the MSSM Lagrangian in terms of supermultiplets and Weyl spinors to a notation in terms of four-component Dirac spinors, which are more frequently used in the SMEFT literature. 
The details of this translation can be found in  Appendix~\ref{app:Weyl-Dirac}.
The conventions used in this and the following sections are collected in Appendix~\ref{app:conventions}.
For more details on the construction of the MSSM Lagrangian see, e.g.,~\cite{Martin:1997ns}. 

\subsection{Supersymmetry for generic gauge theories}

We consider a general supersymmetric gauge theory with a chiral supermultiplet $\Phi^i \!=\! (\psi^i, \phi^i)$, where $\psi^i$ is a left-handed Weyl fermion and $\phi^i$ is a complex scalar, transforming under some representation of an $N$-dimensional simple Lie algebra, with generators~$T^a$, where $a=1,\dots,N$ is an index of the adjoint representation. The corresponding coupling constant is labeled~$g$. 
The generic index~$i$ runs over all gauge and flavor degrees of freedom.
The fermionic and scalar components of the corresponding gauge superfield are denoted by $(\lambda^a, A_\mu^a)$, where $\lambda^a$ is a Weyl fermion. 
The generalization to semi-simple algebras and multiple chiral superfields will be straightforward. 
The most general renormalizable supersymmetric Lagrangian density can be decomposed into a chiral and a gauge sector:
\begin{equation}
    \mathcal{L}_\mathrm{SUSY} =  \mathcal{L}_{\mathrm{chiral}} +
    \mathcal{L}_{\mathrm{gauge}} \,.
    \label{eq:generic-SUSY-Lagrangian}
\end{equation}
The corresponding parts of the Lagrangian are given by
\begin{align}
\begin{split}
    \mathcal{L}_{\mathrm{chiral}} 
    &=
    (D_\mu \phi_i)^\dagger (D^\mu \phi_i) + i \psi^{\dagger i} \overline{\sigma}^\mu D_\mu \psi_i + F^{\dagger i} F_i + \left(W^i F_i - \frac{1}{2} W^{ij} \psi_i \cdot \psi_j + \hc \right)
    \,,
    \label{eq:chiral-Lagrangian}
\end{split}
    \\[0.25cm]
    \mathcal{L}_{\mathrm{gauge}} 
    &= 
    -\frac{1}{4 \, g^2} F_{\mu\nu}^a F^{\mu\nu a} + i\lambda^{\dagger a} \overline{\sigma}^\mu D_\mu \lambda^a - \sqrt{2}\, g \big[ \big(\phi^\dagger T^a \psi\big) \cdot \lambda^a + \hc \big] \nonumber \\
    & \hspace{8cm} + g \big(\phi^\dagger T^a \phi\big) \cD^a + \frac{1}{2} \cD^a \cD
    ^a .
    \label{eq:gauge-Lagrangian}
\end{align}
Here we have introduced two static auxiliary fields of mass dimension two: a~real scalar~$\cD^a$ transforming in the adjoint representation, and a complex scalar~$F_i$ transforming in the same representation as $\phi_i$ and~$\psi_i$.
The auxiliary fields satisfy their classical equations of motion
\begin{align}
    \cD^a &= -g \big(\phi^\dagger T^a \phi\big) \,,
    &
    F_i &= -W_i^\dagger \,,
    &
    F^{\dagger i} &= -W^i 
\end{align}
and can thus be eliminated from the Lagrangian.
The advantage of working with the auxiliary fields~$\cD^a$ and~$F_i$ is that the Lagrangian in Eq.~\eqref{eq:generic-SUSY-Lagrangian} is manifestly supersymmetric, both on-shell and off-shell.\footnote{%
On-shell, there are two degrees of freedom, for the scalar as well as fermion components of each supermultiplet. 
Off-shell, however, the Weyl fermions $\psi$ and~$\lambda$ are complex two-component spinors and hence contain four real degrees of freedom, while $\phi$ and $A_\mu$ have two and three degrees of freedom respectively. 
The auxiliary fields~$F_i$ and~$\cD^a$ are therefore introduced to match the number of degrees of freedom off-shell for each supermultiplet.
}

In Eq.~\eqref{eq:chiral-Lagrangian}, $W^i$~and $W^{ij}$ respectively represent the first and second scalar-field derivatives of the holomorphic superpotential~$W$,
\begin{gather}
    W(\phi) = \frac{1}{2} m^{ij} \phi_i \phi_j + \frac{1}{6} y^{ijk} \phi_i \phi_j \phi_k \,,
\\
    W^i = \frac{\partial W}{\partial \phi_i} = m^{ij} \phi_j + \frac{1}{2} y^{ijk} \phi_j \phi_k \,, 
    \qquad \quad 
    W^{ij} = \frac{\partial^2 W}{\partial \phi_i \partial \phi_j} = m^{ij} + y^{ijk} \phi_k \,.
\end{gather}
We see that $m^{ij}$~must be symmetric and provides the same mass matrix for both the fermion and the scalar fields. 
On the other hand, $y^{ijk}$~constitutes a Yukawa coupling between two fermions and a scalar, as well as a quartic scalar coupling.

In order to explicitly break SUSY without evoking new quadratic divergencies, we can introduce super-renormalizable (also called \textit{soft}) breaking terms in the Lagrangian: 
\begin{align}
    \cL_\mathrm{soft}
    &=
    - \left[ \frac{1}{2} \hat{m} \, \lambda^a \cdot \lambda^a 
    + \frac{1}{6} a^{ijk} \, \phi_i \phi_j \phi_k 
    + \frac{1}{2} b^{ij} \, \phi_i \phi_j + \hc \right] - (\hat{\hat{m}}^2)_j^i \, \phi^{\ast j} \phi_i \,,
\end{align}
where the indices on the couplings~$a$ and~$b$ can only be flavor indices. 
Obviously, only gauge-invariant combinations of the operator form above are allowed. 
The breaking terms in~$\cL_\mathrm{soft}$ introduce masses for all gauginos and for all scalars. 
Thus, applied to the SM, only the Higgs boson acquires a mass through soft SUSY breaking, but all other SM fields remain massless.

\subsection{Supersymmetry for the Standard Model}
\label{sec:MSSM-Lagrangians}

\subsubsection{Field content and structure of the MSSM Lagrangian}
To extend the Lagrangian of a generic supersymmetric theory presented in Eqs.~\eqref{eq:generic-SUSY-Lagrangian}--\eqref{eq:gauge-Lagrangian} to the SM field content, we embed all SM fermions ($q,u,d,\ell,e$) into five chiral super-multiplets.
This introduces five scalar sfermion superpartners~($\tilde{q},\tilde{u},\tilde{d},\tilde{\ell},\tilde{e}$) to the SM~fields, each appearing in three generations. 
The Higgs field is also embedded in a chiral supermultiplet, introducing a fermionic Higgsino superpartner.
Moreover, a second Higgs doublet and its corresponding Higgsino have to be introduced to generate both the up- and down-type Yukawa couplings with a holomorphic superpotential, as well as to avoid gauge anomalies.
We label the two Higgs doublets by $H_u$ and~$H_d$, and their partner Higgsinos by $\tilde{H}_u$ and~$\tilde{H}_d$.
In total, we introduce the seven chiral supermultiplets described in Table~\ref{tab:chiral-multiplets}.

\begin{table}[tbp]
	\centering
    \renewcommand{\arraystretch}{1.2}
	\resizebox{\textwidth}{!}{%
	\begin{tabular}{c | c | l | l  c  l | c }
		names & $\hat\Phi$ & \multicolumn{1}{c |}{spin~0} & \multicolumn{3}{c |}{spin~1/2 (Weyl~$\to$~Dirac)} & $\big(\mathrm{SU}(3)_c,\,\mathrm{SU}(2)_L\big)_{\mathrm{U}(1)_Y}$
		\\[0.15cm]\hline\hline
		\multirow{3}{*}{(s)quarks} & $\hat{q}$ & $\tilde{q} \gray{\,=(\tilde{u}_L,\tilde{d}_L)}$ & $\mathsf{q}\gray{\,=(\mathsf{u}_L,\mathsf{d}_L)}$ & $\to$ & $q\gray{\,=(u_L,d_L)}$ & $\left(\mathbf{3},\mathbf{2}\right)_{+\frac{1}{6}}$
		\\
		& $\hat{\overline{u}}$ & $\tilde{u}^{\;\!\dagger} $ & $\mathsf{u}^c$ & $\to$ & $u^c$ & $\left(\mathbf{\bar{3}},\mathbf{1}\right)_{-\frac{2}{3}}$
		\\
		& $\hat{\overline{d}}$ & ${\tilde{d}}^{\;\!\dagger} $ & $\mathsf{d}^c$ & $\to$ & $d^c$ & $\left(\mathbf{\bar{3}},\mathbf{1}\right)_{+\frac{1}{3}}$
		\\[0.15cm]\hline
		\multirow{2}{*}{(s)leptons} & $\hat{\ell}$ & $\tilde{\ell}\gray{\,=(\tilde{\nu}_L,\tilde{e}_L)}$ & $\mathsf{l}\gray{\,=(\mathsf{\nu}_L,\mathsf{e}_L)}$ & $\to$ & $\ell\gray{\,=(\nu_L,e_L)}$ & $\left(\mathbf{1},\mathbf{2}\right)_{-\frac{1}{2}}$
		\\
		& $\hat{\overline{e}}$ & $\tilde{e}^{\;\!\dagger} $ & $\mathsf{e}^c$ & $\to$ & $e^c$ & $\left(\mathbf{1},\mathbf{1}\right)_{+1}$
		\\[0.1cm]\hline
        \multirow{2}{*}{Higgs(inos)} & $\hat{H}_u$ & $H_u\gray{\,=(H_u^+, H_u^0)}$ & $\tilde{\mathsf{H}}_u\gray{\,=(\tilde{\mathsf{H}}_u^+, \tilde{\mathsf{H}}_u^0)}$ & $\to$ & $\tilde{H}_u\gray{\,=(\tilde{H}_u^+,\tilde{H}_u^0)}$ & $\left(\mathbf{1},\mathbf{2}\right)_{+\frac{1}{2}}$
		\\
		& $\hat{H}_d$ & $H_d\gray{\,=(H_d^0, H_d^-)}$ & $\tilde{\mathsf{H}}_d\gray{\,=(\tilde{\mathsf{H}}_d^0, \tilde{\mathsf{H}}_d^-)}$ & $\to$ & $\tilde{H}_d\gray{\,=(\tilde{H}_d^0,\tilde{H}_d^-)}$ & $\left(\mathbf{1},\mathbf{2}\right)_{-\frac{1}{2}}$
	\end{tabular}}
	\caption{%
    Chiral supermultiplets~$\hat\Phi$ of the MSSM. The third and fourth column show the scalar and fermion components, respectively. 
    For the latter, we show both their representation in terms of left-handed two-component Weyl spinors (sans serif font) as well as four-component Dirac spinors (italic font).
	We take $q,\ell,H_{u,d}$ as left-handed four-component spinors, whereas $u,d,e$ are right-handed four-component spinors.
    The charge conjugate of a spinor~$\psi$ is denoted by~$\psi^c$, which has opposite chirality to~$\psi$. 
    Thus, all fermions listed in this table are left handed.
    The last column shows the gauge quantum numbers of the supermultipltes.
    For the weak doublets we also show their individual $\mathrm{SU}(2)_L$~components in gray.
    }
	\label{tab:chiral-multiplets}
\end{table}

Since the SM is based on a product group of three gauge groups, we also have to introduce three gauge supermultiplets, which are listed in Table~\ref{tab:vector-multiplets}.
Each of these contains one of the SM gauge bosons ($G,W,B$) and the corresponding fermionic superpartners~($\tilde{G},\tilde{W},\tilde{B}$) called gauginos.

\begin{table}[tbp]
	\centering
    \renewcommand{\arraystretch}{1.2}
	\begin{tabular}{c | c  c  c | c | c }
		names & \multicolumn{3}{c |}{spin~1/2 (Weyl~$\to$~Majorana)} & spin~1 & $\big(\mathrm{SU}(3)_c,\,\mathrm{SU}(2)_L\big)_{\mathrm{U}(1)_Y}$
		\\[0.1cm]\hline\hline
		gluinos, gluon & $~~~~~~\tilde{\mathsf{G}}$ & $~~~~~\to$ & $\tilde{G}$ & $G$ & $\left(\mathbf{8},\mathbf{1}\right)_{0}$
		\\[0.1cm]\hline
		winos, W~boson & $~~~~~~\tilde{\mathsf{W}}$ & $~~~~~\to$ & $\tilde{W}$ & $W$ & $\left(\mathbf{1},\mathbf{3}\right)_{0}$
		\\[0.1cm]\hline
		bino, B~boson & $~~~~~~\tilde{\mathsf{B}}$ & $~~~~~\to$ & $\tilde{B}$ & $B$ & $\left(\mathbf{1},\mathbf{1}\right)_{0}$
	\end{tabular}
	\caption{Vector/gauge supermultiplets of the MSSM. The SM gauge bosons are $G,W,B$, and we take the gaugeinos $\tilde{G},\tilde{W},\tilde{B}$ as four-component Majorana spinors, whereas $\tilde{\mathsf{G}},\,\tilde{\mathsf{W}},\,\tilde{\mathsf{B}}$ label the corresponding two-component Weyl spinors.}
	\label{tab:vector-multiplets}
\end{table}

We can now construct a minimal supersymmetric version of the SM by including one copy of the Lagrangian terms in Eq.~\eqref{eq:chiral-Lagrangian} for each of the seven chiral supermultiplets, and one copy of the Lagrangian terms in Eq.~\eqref{eq:gauge-Lagrangian} for each of the three gauge supermultiplets.
In addition, we have to specify the superpotential which can be written in terms of the chiral supermultiplets as (see~\cite{Martin:1997ns} for details) 
\begin{align}
	\mathcal{W}_\mathrm{MSSM} &= \int \dd^2\theta \left[ 
    \tilde{\mu} \hat{H}_u \varepsilon \hat{H}_d
    + (\hat{\overline{u}} y_u \hat{q}) \varepsilon \hat{H}_u 
    - (\hat{\overline{d}} y_d \hat{q}) \varepsilon \hat{H}_d 
    - (\hat{\overline{e}} y_e \hat{\ell}) \varepsilon \hat{H}_d  
    \right] \,,
\end{align}
where we suppress all indices for a compact notation.
Here, $\tilde{\mu}$~acts as a mass mixing between $\hat{H}_u$ and~$\hat{H}_d$, whereas $y_{u,d,e}$~denote the $3 \times 3$~Yukawa matrices, and $\varepsilon$ is the full anti-symmetric tensor of~$\mathrm{SU}(2)$.
Finally, in order to obtain a realistic mass spectrum, the following soft SUSY-breaking terms are introduced for the MSSM
\begin{align}
\begin{split}
	\cL_\mathrm{soft}^\mathrm{Weyl}
	= &-\frac{1}{2} \left[ 
    m_3 \, \tilde{\mathsf{G}}^A \cdot \tilde{\mathsf{G}}^A 
    + m_2 \, \tilde{\mathsf{W}}^I \cdot \tilde{\mathsf{W}}^I 
    + m_1 \, \tilde{\mathsf{B}} \cdot \tilde{\mathsf{B}} + \mathrm{H.c.} \right]
	\\
	&-\left[ \big(\tilde{q}^{\;\!\dagger} a_u \tilde{u}\big) \varepsilon H_u^\ast 
    - \big(\tilde{q}^{\;\!\dagger} a_d \tilde{d} \,\big) \varepsilon H_d^\ast 
    - \big(\tilde{\ell}^{\;\!\dagger} a_e \tilde{e}\big) \varepsilon H_d^\ast 
    + \mathrm{H.c.} \right]
	\\
	&- \left[ 
    \tilde{q}^{\;\!\dagger} m_{\tilde{q}}^2 \tilde{q} 
    + \tilde{u}^{\;\!\dagger} m_{\tilde{u}}^2 \tilde{u}
    + \tilde{d}^{\;\!\dagger} m_{\tilde{d}}^2 \tilde{d}
    + \tilde{\ell}^{\;\!\dagger} m_{\tilde{\ell}}^2 \tilde{\ell} 
    + \tilde{e}^{\;\!\dagger} m_{\tilde{e}}^2 \tilde{e} 
    \right]
	\\
	&- m_{H_u}^2 H_u^\dagger H_u 
    - m_{H_d}^2 H_d^\dagger H_d 
    - \left[ b H_u^\intercal \varepsilon H_d 
    + \mathrm{H.c.} \right] \,,
\end{split}
\end{align}
where $m_{1,2,3}$ furnish mass terms for the bino, wino, and~gluino; $a_{u,d,e}$~are complex $3 \times 3$~matrices in flavor space of mass-dimension one that we take of~$\cO(\Lambda_\text{UV})$ for the power counting, with $\Lambda_\text{UV}$ being the heavy UV~scale of soft SUSY breaking; and $m_{\tilde{q},\tilde{u},\tilde{d},\tilde{\ell},\tilde{e}}$ are Hermitian $3 \times 3$ mass matrices for the squarks and sleptons, which we take to be diagonal without loss of generality, by fixing the sfermion flavor basis. 
Finally, $m_{H_{u,d}}$~and~$b$ are mass terms for the Higgs doublet system.

\subsubsection{MSSM Lagrangian with Dirac spinors}
\label{sec:MSSM-Lagrangian-Dirac}
While the notation in terms of supermultiplets is very compact, and the formulation in terms of Weyl spinors is commonly used in the SUSY literature, it is less practical for matching onto the SMEFT.
First of all, the \matchete code~\cite{Fuentes-Martin:2022jrf} only allows for four-component spinors as input.
Second, the Warsaw basis~\cite{Grzadkowski:2010es} is usually expressed in terms of Dirac spinors, and most SMEFT analyses also employ this notation.
Hence, to be compatible with the common SMEFT notation, we present the MSSM Lagrangian in terms of Dirac and Majorana four-component spinors.
The mapping between the Weyl spinor notation presented in the previous section and the Dirac spinor notation used in the following can be found in Appendix~\ref{app:Weyl-Dirac}.

While the SM fermion fields are given by Dirac fermions with their usual chiralities, the Higgsinos can be defined as left-handed Dirac fermions, whereas the gauginos are taken as Majorana fermions.
With that, the gauge sector of the MSSM Lagrangian assumes the form
\begin{align}
\begin{split}
	\cL_\mathrm{gauge}^\mathrm{Majorana}
	&= \sum_{F\in\{G,W,B\}} -\frac{1}{4 \, g_F^2} F_{\mu\nu}^A F^{\mu\nu\,A} + \sum_{\lambda\in\{\tilde{G},\tilde{W},\tilde{B}\}} \frac{i}{2} \, \overline{\lambda}^{A} \slashed{D} \lambda^A 
	\\
	&\qquad - \sqrt{2} \, \left[ g_3 \Omega_3^A \tilde{G}^A + g_2 \Omega_2^I \tilde{W}^I + g_1 \Omega_1 \tilde{B} + \mathrm{H.c.} \right] 
	\\
	&\qquad - \frac{1}{2} \left[ g_3^2 \Lambda_3^A \Lambda_3^A + g_2^2 \Lambda_2^I \Lambda_2^I + g_1^2 \Lambda_1 \Lambda_1 \right]
	\,,
\end{split}
\label{eq:L-gauge}
\end{align}
where $g_F$ denotes the gauge coupling belonging to~$F_{\mu\nu}$ and  $\lambda^A$ are now four-component Majorana fermions. 
Furthermore, we have defined
\begin{subequations}
\begin{align}
	\Omega_3^A &= \big[(\overline{q} T^A \tilde{q}) P_R - (\overline{u} T^A \tilde{u}) P_L - (\overline{d} T^A \tilde{d}) P_L \big] \,,
	\\
	\Omega_2^I &= \big[(\overline{q} T^I \tilde{q}) + (\overline{\ell} T^I \tilde{\ell}) + (\overline{\tilde{H}}_u T^I H_u) + (\overline{\tilde{H}}_d T^I H_d) \big] P_R \,,
	\\
	\Omega_1 &= \big[ \mathsf{y}_q (\overline{q} \tilde{q}) P_R - \mathsf{y}_{u} (\overline{u} \tilde{u}) P_L - \mathsf{y}_{d} (\overline{d} \tilde{d}) P_L +\mathsf{y}_\ell (\overline{\ell} \tilde{\ell}) P_R - \mathsf{y}_{e} (\overline{e} \tilde{e}) P_L 
    \nonumber\\
	&\qquad
    +\mathsf{y}_{H_u} (\overline{\tilde{H}}_u H_u) P_R +\mathsf{y}_{H_d} (\overline{\tilde{H}}_d H_d) P_R \big] \,,
	\\[0.3cm]
	\Lambda_3^A &= (\tilde{q}^\dagger T^A \tilde{q}) - (\tilde{u}^\dagger T^A \tilde{u}) - (\tilde{d}^\dagger T^A \tilde{d}) \,,
	\\
	\Lambda_2^I &= (\tilde{q}^\dagger T^I \tilde{q}) + (\tilde{\ell}^\dagger T^I \tilde{\ell}) + (H_u^\dagger T^I H_u) + (H_d^\dagger T^I H_d) \,,
	\\
	\Lambda_1 &= \mathsf{y}_q (\tilde{q}^\dagger \tilde{q}) - \mathsf{y}_{u} (\tilde{u}^\dagger \tilde{u}) - \mathsf{y}_{d} (\tilde{d}^\dagger \tilde{d}) + \mathsf{y}_\ell (\tilde{\ell}^\dagger \tilde{\ell}) - \mathsf{y}_{e} (\tilde{e}^\dagger \tilde{e}) 
	\nonumber\\
	&\qquad + \mathsf{y}_{H_u} (H_u^\dagger H_u) + \mathsf{y}_{H_d} (H_d^\dagger H_d) \,,
\end{align}
\end{subequations} 
where $\mathsf{y}_\phi$ denotes the Hypercharge of the field~$\phi$ and $T^A$, $T^I$ are the generators of $\mathrm{SU}(3)_c$ and $\mathrm{SU}(2)_L$ respectively.

The chiral sector of the Lagrangian takes the form
\begin{align}
	&\cL_\mathrm{chiral}^\mathrm{Dirac}
	=  
	\nonumber\\
	&\sum_{\phi\in\{\tilde{q},\tilde{u},\tilde{d},\tilde{\ell},\tilde{e},H_u,H_d\}} (D_\mu \phi)^\dagger (D^\mu \phi)
    + \sum_{\psi\in\{q,\ell,\tilde{H}_u,\tilde{H}_d\}} i \overline{\psi} \slashed{D} P_L \psi 
    + \sum_{\psi\in\{u,d,e\}} i \overline{\psi} \slashed{D} P_R \psi 
	\nonumber\\
	&- \tilde{\mu}^\ast \tilde{\mu} \left[ H_u^\dagger H_u + H_d^\dagger H_d \right]
		\nonumber\\
		&+ \left[ 
            \tilde{\mu} \big(\tilde{q}^{\;\!\dagger} y_u \tilde{u}\big) H_d 
            + \tilde{\mu} \big(\tilde{q}^{\;\!\dagger} y_d \tilde{d} \, \big) H_u 
            + \tilde{\mu} \big(\tilde{\ell}^{\;\!\dagger} y_e \tilde{e}\big) H_u
            + \mathrm{H.c.} 
        \right]
		\nonumber\\
		&- \left[ 
            \big(H_u^\dagger H_u\big) \big(\tilde{u}^{\;\!\dagger} y_u^\dagger y_u \tilde{u}\big) 
            + \big(H_d^\dagger H_d\big) \big(\tilde{d}^{\;\!\dagger} y_d^\dagger y_d \tilde{d}\,\big) 
            + \big(H_d^\dagger H_d \big) \big(\tilde{e}^{\;\!\dagger} y_e^\dagger y_e \tilde{e}\big) 
            - \left\{ \!\big(H_d^\dagger H_u\big) \big(\tilde{u}^{\;\!\dagger} y_u^\dagger y_d \tilde{d} \, \big) \!+ \!\mathrm{H.c.} \!\right\} \!
        \right]
		\nonumber\\
		&+ \left[ 
            \big(\tilde{q}^{\;\!\dagger} \varepsilon H_u^\ast \big) y_u y_u^\dagger \big(H_u^\intercal \varepsilon \tilde{q}\big) 
            + \big(\tilde{q}^{\;\!\dagger} \varepsilon H_d^\ast\big) y_d y_d^\dagger \big(H_d^\intercal \varepsilon \tilde{q}\big) 
            + \big(\tilde{\ell}^{\;\!\dagger} \varepsilon H_d^\ast\big) y_e y_e^\dagger \big(H_d^\intercal \varepsilon \tilde{\ell}\,\big)
        \right]
		\nonumber\\
		&- \left[ 
            \big(\tilde{q}^{\;\!\dagger} y_u \tilde{u}\big) \big(\tilde{u}^{\;\!\dagger} y_u^\dagger \tilde{q}\big) 
            + \big(\tilde{q}^{\;\!\dagger} y_d \tilde{d}\,\big) \big(\tilde{d}^{\;\!\dagger} y_d^\dagger \tilde{q}\big)
            + \big(\tilde{\ell}^{\;\!\dagger} y_e \tilde{e}\big) \big(\tilde{e}^{\;\!\dagger} y_e^\dagger \tilde{\ell}\,\big)
            + \left\{ \big(\tilde{q}^{\;\!\dagger} y_d \tilde{d}\,\big) \big(\tilde{e}^{\;\!\dagger} y_e^\dagger \tilde{\ell}\,\big) + \mathrm{H.c.} \right\} 
        \right]
		\nonumber\\
		&- \bigg\{ \tilde{\mu} \overline{\tilde{H}_u^c} \varepsilon \tilde{H}_d 
            + \big(\overline{q} y_u u\big) \varepsilon H_u^\ast 
            + \big(\tilde{u}^{\;\!\dagger} y_u^\dagger \overline{q}^c\big) \varepsilon \tilde{H}_u 
            + \big(\overline{u} y_u^\dagger \tilde{q}\big) \varepsilon \tilde{H}_u
            - \big(\overline{q} y_d d\big) \varepsilon H_d^\ast
            - \big(\tilde{d}^{\;\!\dagger} y_d^\dagger \overline{q}^c\big) \varepsilon \tilde{H}_d 
		\nonumber\\
		&\qquad 
            - \big(\overline{d} y_d^\dagger \tilde{q}\big) \varepsilon \tilde{H}_d 
            - \big(\overline{\ell} y_e e\big) \varepsilon H_d^\ast 
            - \big(\tilde{e}^{\;\!\dagger} y_e^\dagger \overline{\ell}^c\big) \varepsilon \tilde{H}_d 
            - \big(\overline{e} y_e^\dagger \tilde{\ell}\,\big) \varepsilon \tilde{H}_d 
        + \mathrm{H.c.} \bigg \} \,,
\label{eq:L-chiral_W}
\end{align}
where brackets~``$(\cdots)$'' indicate the sums over flavor indices and, as before, the anti-symmetric contraction of $\mathrm{SU}(2)_L$-gauge indices is indicated by~$\varepsilon$.\footnote{Notice that for two scalars we thus have $\phi\varepsilon\varphi=-\varphi\varepsilon\phi$.} 

Lastly, the soft SUSY-breaking sector becomes 
\begin{align}
\begin{split}
	\cL_\mathrm{soft}
	= &-\frac{1}{2} \left[ 
    m_3 \overline{\tilde{G}}{}^A\tilde{G}^A 
    + m_2 \overline{\tilde{W}}{}^I\tilde{W}^I 
    + m_1 \overline{\tilde{B}}\tilde{B} \right]
	\\
	&-\left[ 
    \big( \tilde{q}^{\;\!\dagger} a_u \tilde{u} \big) \varepsilon H_u^\ast
    - \big( \tilde{q}^{\;\!\dagger} a_d \tilde{d} \,\big) \varepsilon H_d^\ast
    - \big( \tilde{\ell}^{\;\!\dagger} a_e \tilde{e} \big) \varepsilon H_d^\ast
    + \mathrm{H.c.} \right]
	\\
	&-  \left[ 
    \tilde{q}^{\;\!\dagger} m_{\tilde{q}}^2 \tilde{q} 
    + \tilde{u}^{\;\!\dagger} m_{\tilde{u}}^2 \tilde{u}
    + \tilde{d}^{\;\!\dagger} m_{\tilde{d}}^2 \tilde{d}
    + \tilde{\ell}^{\;\!\dagger} m_{\tilde{\ell}}^2 \tilde{\ell} 
    + \tilde{e}^{\;\!\dagger} m_{\tilde{e}}^2 \tilde{e} 
    \right]
	\\
	&
    - m_{H_u}^2 H_u^\dagger H_u 
    - m_{H_d}^2 H_d^\dagger H_d 
    - \big[ b H_u^\intercal \varepsilon H_d + \mathrm{H.c.} \big]
\end{split}
\label{eq:L-break}
\end{align}
in this notation.

For completeness, the fully expanded MSSM Lagrangian 
\begin{align}
    \cL_\mathrm{MSSM} = \cL_\mathrm{gauge}^\mathrm{Majorana} + \cL_\mathrm{chiral}^\mathrm{Dirac} + \cL_\mathrm{soft}
    \label{eq:L-MSSM-full}
\end{align}
in the form used for the matching is also presented in Appendix~\ref{app:MSSM-Lagrangian} in Eqs.~\eqref{eq:full-MSSM-Lag}--\eqref{eq:full_non-Hermitian_MSSM_Lagrangian}.
At the loop level, the MSSM Lagrangian above is only supersymmetric (with a soft breaking) if its couplings are given in a SUSY-preserving regularization scheme such as dimensional reduction~\cite{Siegel:1979wq}. 
We henceforth assume that the high-scale input parameters for the MSSM Lagrangian are given in the $\overline{\mathrm{DR}}$~scheme.
The conversion to dimensional regularization~\cite{tHooft:1972tcz} and the $\overline{\mathrm{MS}}$~scheme~\cite{Martin:1993yx}, which is used for the matching and subsequently in the SMEFT, is discussed at the end of Sec.~\ref{sec:one-loop matching}.

The MSSM Lagrangian given in Eq.~\eqref{eq:L-MSSM-full} is in a generic basis.
For matching it onto its low-energy EFT, we have to first rotate it into the mass basis. 
Since we want to match onto the SMEFT, which is defined in the unbroken electroweak phase, we have to use the mass basis before electroweak symmetry breaking.
That is, we only take into account mass terms from the superpotential and the soft SUSY breaking sector, but no masses generated by the Higgs fields acquiring vacuum expectation values.

While the gauginos are trivially given in the mass basis, we employ unitary rotations to arrive at the mass basis for the squarks and sleptons.
In order to preserve the flavor universality of the gaugino-sfermion-fermion interactions, we rotate the entire chiral supermultiplet (fermions and sfermions) by the same matrix:
\begin{align}
    \psi^\prime &= U_{\tilde{\psi}} \psi \,, 
    &
    \tilde{\psi}^\prime &= U_{\tilde{\psi}} \tilde{\psi} \,,
    \label{eq:flavor-rotation}
\end{align}
for $\psi \in \{{q},{u},{d},{\ell},{e}\}$, where primed fields are in a generic flavor basis, while their non-primed counterparts are in the sfermion mass basis. Notice that this entirely fixes the fermion basis after EWSB. 
Therefore, from here on, we assume that $m_{\tilde{\psi}}^2$ is diagonal for $\psi \in \{q,u,d,\ell,e\}$ and all the other parameters are given in this basis.
In particular, the Yukawa matrices ($y_{u,d,e}$) will be assumed to be defined in this basis and, in general, will not be diagonal.

It is important to point out that even if the Wilson coefficients obtained below may appear flavor diagonal, a further rotation still is necessary to diagonalize the SM fermion mass matrices in the SMEFT. 
This rotation may induce flavor changing coefficients. 
For example, we find:
\begin{equation}
    \left[C_{ud}^{(1)}\right]_{prst} (\bar{u}_p \gamma_\mu u_r) (\bar{d}_s \gamma_\mu d_t)
    = \delta_{pr}\delta_{st} g_3^4 L_{ps} (\bar{u}_p \gamma_\mu u_p) (\bar{d}_s \gamma_\mu d_s) + \ldots
\end{equation}
where $L_{ps} = F(m_3,m_{\tilde u}^p,m_{\tilde d}^s)$ is a loop function which depends on the squark and gluino masses.
After rotating to the quark mass basis ($\psi \to U_{\psi}^{\dagger} \psi$), we have
\begin{align}
    \left[C_{ud}^{(1)}\right]_{prst} (\bar{u}_p \gamma_\mu u_r) (\bar{d}_s \gamma_\mu d_t) & \to
    g_3^4 L_{ps} [U_u]_{p^\prime p} [U_u^\dagger]_{p r^\prime} [U_d]_{s^\prime s} [U_d^\dagger]_{s t^\prime} (\bar{u}_{p^\prime} \gamma_\mu u_{r^\prime}) (\bar{d}_{s^\prime}\gamma_\mu d_{t^\prime}) + \ldots\nonumber\\
    & \equiv \left[C_{ud}^{\prime (1)}\right]_{p^{\prime}r^{\prime}s^{\prime}t^{\prime}} (\bar{u}_{p^\prime} \gamma_\mu u_{r^\prime}) (\bar{d}_{s^\prime} \gamma_\mu d_{t^\prime}) +\ldots\,,
    \label{eq:generic-fermion-basis-rotation}
\end{align}
and, in general, the presence of flavor non-universal sfermion masses introduces flavor-violating operators.

Notice that the matching conditions presented in Sec.~\ref{sec:matching-results} and provided on \href{https://github.com/BSM-EFT/MSSM-to-SMEFT}{GitHub} are given in the sfermion mass basis.
To use our results in combination with other (s)fermion bases, two observations are important.
First, if the UV inputs, i.e., the high-scale MSSM parameters, are given in a different basis one needs to rotate to the sfermion mass basis before computing the values of the SMEFT Wilson coefficients using the matching conditions we provide, thereby determining also the unitary matrices~$U_{\tilde\psi}$.
Second, the Wilson coefficients computed that way are hence also given in the sfermion mass basis, and not in the basis where the Yukawas are either up or down aligned, which are conventionally employed in the SMEFT.
Therefore, it is required to perform another unitary rotation for the SMEFT fermion fields after the matching to arrive at the up- or down-aligned Yukawa basis.
For example, assuming the UV inputs were already given in the up- or down-aligned Yukawa basis, the required rotations for the fermions are given by the inverse transformation~$\psi \to U_{\tilde\psi}^\dagger \psi$. 

Rotating to the mass basis in the Higgs sector is also non-trivial due to the mixing of the two Higgs(ino) fields present in the MSSM. 
It is hence essential to understand the Higgs sector and the patterns of electroweak symmetry breaking~(EWSB) in the MSSM, which we discuss in the following, before performing the matching in Sec.~\ref{sec:matching}.

% Higgs sector of the MSSM (2HDM)
\section{Higgs sector of the MSSM}
\label{sec:Higgs}

The Higgs sector of the MSSM constitutes a type-II two-Higgs-doublet model~(2HDM), with the two doublets transforming as $H_u \sim \left(\mathbf{1},\mathbf{2}\right)_{1/2}$ and $H_d \sim \left(\mathbf{1},\mathbf{2}\right)_{-1/2}$ under the gauge group. 
As a first step, let us define the charge conjugate of~$H_d$, namely
\begin{align}
    H_d^c &\equiv \varepsilon H_d^\ast 
    \,,
    &
    &\text{transforming as}
    &
    H_d^c &\sim \left(\mathbf{1},\mathbf{2}\right)_{\frac{1}{2}}
    \,,
\end{align}
where $\varepsilon$ is the anti-symmetric tensor of~$\mathrm{SU}(2)_L$.
Thus, $H_d^c$ and $H_u$ transform in the same representation of the gauge group and we can perform a rotation with generic angle~$\vartheta$ to a new basis of doublets
\begin{align}
    \begin{pmatrix}
        H_u \\ H_d^c
    \end{pmatrix}
    &=
    \begin{pmatrix}
        s_\vartheta & c_\vartheta
        \\
        -c_\vartheta & s_\vartheta
    \end{pmatrix}
    \begin{pmatrix}
        H \\ \Phi
    \end{pmatrix}
    \label{eq:rotation-to-matching-basis}
\end{align}
with the standard definitions $s_\vartheta \equiv \sin\vartheta$ and $c_\vartheta \equiv \cos\vartheta$.
In the following, we investigate such a rotation to the mass basis, before discussing electroweak symmetry breaking in this theory and the decoupling of one of the Higgs doublets.

The Higgs potential of the MSSM expressed through $H_{u,d}$ takes the form
\begin{align}
    \begin{split}
        V_{\mathrm{Higgs}}^{\mathrm{MSSM}}[H_u, H_d] ={}& (|\tilde\mu|^2 + m_{H_u}^2) H_u^\dagger H_u + (|\tilde\mu|^2 + m_{H_d}^2) H_d^\dagger H_d + \left[ b H_u^\intercal \varepsilon H_d + \mathrm{H.c.} \right]\\
        &+ \frac{g_1^2 + g_2^2}{8} \left( H_u^\dagger H_u - H_d^\dagger H_d \right)^2 + \frac{g_2^2}{2} (H_u^\dagger H_d) (H_d^\dagger H_u) \,.
    \end{split}
    \label{eq:Higgs-potential}
\end{align}
Applying the rotation given in Eq.~\eqref{eq:rotation-to-matching-basis} with the generic angle~$\vartheta$, we obtain its form in terms of $H$ and~$\Phi$
\begin{align}
\begin{split}
    V_{\mathrm{Higgs}}^{\mathrm{MSSM}}[H, \Phi]
    =
    &+ \underbrace{\Big[ |\tilde\mu|^2 + s_\vartheta^2 m_{H_u}^2 + c_\vartheta^2 m_{H_d}^2 - s_\vartheta c_\vartheta (b+b^\ast) \Big]}_{\equiv m_H^2} (H^\dagger H)
    \\
    &+ \underbrace{\Big[ |\tilde\mu|^2 + c_\vartheta^2 m_{H_u}^2 + s_\vartheta^2 m_{H_d}^2 + s_\vartheta c_\vartheta (b+b^\ast) \Big]}_{\equiv m_\Phi^2} (\Phi^\dagger \Phi)
    \\
    &+ \Big\{\underbrace{\Big[ (m_{H_u}^2 - m_{H_d}^2) s_\vartheta c_\vartheta - b c_\vartheta^2 + b^\ast s_\vartheta^2 \Big]}_{\equiv \Delta} (H^\dagger \Phi) + \mathrm{H.c.} \Big\} 
    + \ldots \,,
\end{split}
\label{eq:MSSM_Lagrangian-matching-basis}
\end{align}
where the ellipsis indicate the quartic terms and we have introduced the mass terms~$m_{H,\Phi}^2$ and the mass mixing $\Delta$ parameter.

% % % % % % % % % % % % % % % % % % % % % % % % 
\subsection{MSSM Higgs mass basis after soft SUSY breaking}
\label{sec:soft-SUSY_Higgs-mass-basis}

By choosing the rotation angle~$\vartheta$ such that the mixing vanishes at tree level ($0=\Delta\rvert_{\vartheta=\gamma}$), and naming this specific choice~$\gamma$, we obtain what we dub the \emph{soft-SUSY mass basis}, where $H$ and~$\Phi$ are the mass eigenstates after soft SUSY breaking, but before electroweak symmetry breaking~(EWSB).
That is, the masses $m_{H}$ and $m_\Phi$ contain only contributions from the superpotential and the soft breaking terms, but not from the Higgs fields acquiring a vacuum expectation value~(VEV).
Since $b$ can always be made real by a phase redefinition, 
the rotation angle to the mass basis is given by
\begin{align}
    \tan 2\gamma &= \frac{2b}{m_{H_u}^2-m_{H_d}^2} 
    \,.
    \label{eq:rotation-angle_mass-basis}
\end{align}

As there is experimental evidence only for a single (SM-like) Higgs boson thus far, we assume that one of the two doublets has a heavy mass and hence decouples.
We then take $H$ to be the \emph{light}, SM-like Higgs doublet that will remain in the spectrum of the EFT, whereas $\Phi$ becomes the \emph{heavy} doublet that we integrate out.
The scenario where both Higgs doublets are treated as light and remain in the EFT spectrum is discussed in Appendix~\ref{app:2HDM-EFT}.

Note that different choices of~$\vartheta$ can be used for the matching.
The only assumption we make is that the Higgs-doublet mass matrix has one light and one heavy eigenvalue.
In that case, any value of~$\vartheta$ is acceptable as long as it leads to a state~$\Phi$ that is sufficiently aligned with the ``heavy eigenvector'' of the Higgs-doublet mass matrix.
This follows directly from the decoupling theorem~\cite{Appelquist:1974tg}, which guarantees that any interpolating fields can be used for the matching.
Different choices yield equivalent results up to higher-order terms in the EFT power counting, following from the commutation of field redefinitions and matching~\cite{Criado:2018sdb}, see also~\cite{Banta:2023prj}.
Thus, different choices for~$\vartheta$ can be exploited to asses the impact of missing higher-dimensional operators.

An example of an alternative choice for~$\vartheta$ can be found in the MSSM literature, where EWSB is often discussed within the two-Higgs-doublet context.
It is then conventional to define another rotation angle~($\beta$), which relates the contribution of each Higgs doublet VEV to the electroweak scale ($v$). This angle is given by $\tan\beta = v_u/v_d$, with $\langle H_u^0 \rangle \equiv v_u/\sqrt{2}$, $\langle H_d^0 \rangle \equiv v_d/\sqrt{2}$ and $v^2 = v_u^2 + v_d^2$.
Therefore, when comparing the results obtained here to the ones in the literature it will be useful to relate the $\beta$ and $\gamma$ angles. This relation
(see Sec.~\ref{sec:EWSB-MSSM}) is given by:
\begin{align}
    \tan(2\gamma) 
    &=  \tan(2\beta) \left(1 - \frac{g_1^2+g_2^2}{4} \frac{v^2}{m_\Phi^2}  \right) + \mathcal{O}(v^4/m_\Phi^4) \label{eq:gamma-beta}
\end{align}
Note that both angles differ by $\mathcal{O}(v^2/m_{\Phi}^2)$ and in most cases can be taken as equal at leading order in the EFT expansion.

In principle, one could now perform the matching onto the SMEFT, which contains~$H$ as the only scalar degree of freedom, and we will do so in Sec.~\ref{sec:matching}.
The reader interested only in the matching and its results can proceed directly there, whereas the remainder of this section concerns a more detailed discussion of electroweak symmetry breaking in the SMEFT and the MSSM, and a comparison of the two.
This discussion is intended as an illustration and therefore constrained to the tree level. 
As mentioned before, the conclusions apply, however, to all orders due to the decoupling theorem and since the difference between different bases is suppressed in the EFT power counting, cf. Eq.~\eqref{eq:gamma-beta}.
The same applies also to Appendix~\ref{app:Higgs-basis_alignment-limit}, where the matching in the ``Higgs basis''~($\vartheta=\beta$) is discussed.

\subsection{Electroweak symmetry breaking in the MSSM and its EFT}
In this section, we compare the EWSB patterns in the MSSM and the SMEFT. 
To this end, we work at the tree level and include only the leading correction in the EFT power counting, that is, we only consider dimension-six operators.

\subsubsection{EWSB in the SMEFT}
It is instructive to first investigate the patterns of EWSB in the SMEFT, after the heavy doublet~$\Phi$ has been integrated out.
At this stage, the most general SMEFT Higgs potential, including up to dimension-six operators, is given by
\begin{align}\label{eq:higgs-pot-smeft}
    V_{\mathrm{Higgs}}^{\mathrm{SMEFT}}
    &= 
    M_H^2 |H|^2 + \frac{\lambda}{2} |H|^4 - C_H |H|^6
\end{align}
with the potential parameters determined by the matching conditions, which will be discussed in Sec.~\ref{sec:matching-results}, and are given at tree level by:
\begin{align}
    M_H^2 &= m_H^2\,,
    &
    \lambda &=   \cos^2(2\gamma) \frac{ g_1^2 + g_2^2 }{4} \,,
    &
    &\mbox{ and } 
    &
    C_H &=    \frac{\sin^2(4\gamma)}{m_\Phi^2} \frac{\left( g_1^2 + g_2^2 \right)^2}{64} \,.
    \label{eq:soft-SUSY-mass-basis_matching-conditions}
\end{align}
Since the potential must have a minimum at the electroweak scale, i.e., 
 $\langle |H|^2 \rangle \equiv v^2/2$ with $v  = 246$~GeV, we can impose this condition to replace $M_H = m_H$ by $v$:
\begin{align}
    M_{H}^2 = m_{H}^2 = - \lambda \frac{v^2}{2} + \frac{3 v^4}{4} C_H \, , \label{eq:mHdef}
\end{align}
with the mass parameter being tachyonic in our conventions ($M_H^2<0$).
It is useful to employ the above result and re-write the potential as:
\begin{align}
    V_{\mathrm{Higgs}}^{\mathrm{SMEFT}}
    &= 
    \frac{\lambda}{2} \left(|H|^2 - \frac{v^2}{2}\right)^2 + \frac{C_H}{4} |H|^2 \left( 3 v^4 - 4 |H|^4 \right) ,
\end{align}
where we have neglected a constant term.
Finally, assuming the unitary gauge, 
\begin{align}
    H \to \begin{pmatrix}
        0 
        \\
        \frac{1}{\sqrt{2}}\left(v + h^0\right)
    \end{pmatrix}
    \,,
\end{align}
we find:
\begin{align}
\begin{split}
    V_{\mathrm{Higgs}}^{\mathrm{SMEFT}}
    &= 
    \frac{\lambda v^2}{2} \left(1 -\frac{3 v^2}{\lambda}C_H\right) (h^0)^2 + \frac{\lambda v}{2} \left(1 -\frac{5 v^2}{\lambda}C_H\right) (h^0)^3 \\
    &+ \frac{\lambda}{8}\left(1-\frac{15 v^2}{\lambda}C_H \right) (h^0)^4 - \frac{3}{4} v\, C_H (h^0)^5 - \frac{1}{8} C_H (h^0)^6 ,
\end{split}
\label{eq:physical-Higgs-potential_SMEFT}%
\end{align}
where once again a constant term was neglected and
\begin{equation*}
    \frac{C_H}{\lambda} = \frac{\sin^2(2\gamma)}{m_\Phi^2} \frac{\left( g_1^2 + g_2^2 \right)}{4}\; .
\end{equation*}
Note that in the limit $C_H \to 0$ we recover the usual SM Higgs potential.

Using Eqs.~\eqref{eq:physical-Higgs-potential_SMEFT} and~\eqref{eq:soft-SUSY-mass-basis_matching-conditions}, we can readily compute the Higgs mass:\footnote{Note that all the results discussed in this section are at tree level and receive significant loop corrections from the other MSSM fields, such as the stops.}
\begin{align}
\begin{split}
    \textrm{m}_{h^0}^2
    & = \lambda v^2 \left(1 -\frac{3 v^2}{\lambda}C_H\right) \\
    & = \frac{g_1^2 + g_2^2}{4} \cos^2(2\gamma) v^2 - \frac{3}{16}\left(g_1^2 + g_2^2\right)^2 \sin^2(2\gamma) \cos^2(2\gamma) \frac{v^4}{m_\Phi^2}\\
    & = \cos^2(2\gamma) \textrm{m}_Z^2\left[1 - 3 \sin^2(2\gamma) \frac{\textrm{m}_Z^2}{m_\Phi^2}\right] \; ,
\end{split}
\label{eq:physical-Higgs-mass}
\end{align}
where we have used the tree level definition of the $Z$~mass,
\begin{align}\label{eq:Z-mass-tree}
\textrm{m}_Z^2 \equiv \frac{g_1^2 + g_2^2}{4} v^2.    
\end{align}
Equation~\eqref{eq:physical-Higgs-mass} displays the well-known result that the Higgs mass is smaller than the $Z$-boson mass at tree level in the MSSM. In addition, the tree-level corrections from the heavy Higgs doublet can only reduce the Higgs mass.

\subsubsection{EWSB in the MSSM}
\label{sec:EWSB-MSSM}
While in the approach detailed above, we first integrated out the heavy scalar doublet~($\Phi$) and then broke the electroweak symmetry within the SMEFT, in the literature, EWSB is often considered within the full MSSM.
It is therefore instructive to compare this scenario with the one presented in the previous section.
While in the SMEFT only a single Higgs doublet~($H$) is present and acquires a VEV, the MSSM contains two doublets that, in general, can both acquire VEVs.
We can express the two MSSM Higgs doublets in terms of their $\mathrm{SU}(2)_L$~components
\begin{align}
    H_u &= \begin{pmatrix}
        H_u^+ \\ H_u^0
    \end{pmatrix}
    \,,
    & 
    H_d &= \begin{pmatrix}
        H_d^0 \\ H_d^-
    \end{pmatrix} 
\end{align}
with $H_u^- \equiv {H_u^+}^\ast$ and $H_d^+ \equiv {H_d^-}^\ast$\,.
Since the charged components do not acquire VEVs, we can restrict our discussion to the scalar potential for the neutral components. 
Furthermore, since both doublets have opposite hypercharges, we can perform a $\mathrm{U}(1)_Y$~gauge transformation to set the VEVs of both neutral scalar components to be real and positive, i.e., $\langle H_{u,d}^0\rangle \geq 0$. 
After these simplifications, the relevant scalar potential can be written as~\cite{Martin:1997ns}:
\begin{align}
    V_{\mathrm{Higgs}} ={}& \left(|\tilde\mu|^2 + m_{H_u}^2\right) \left(H_u^0\right)^2 + \left(|\tilde\mu|^2 + m_{H_d}^2\right) \left(H_d^0\right)^2 \nonumber\\
    &- 2 b\big(H_u^0 H_d^0 \big) + \frac{g_1^2 + g_2^2}{8} \Big[\big(H_u^0\big)^2 - \big(H_d^0\big)^2 \Big]^2  \,.
    \label{eq:EWSB-Higgs-potential}
\end{align}
The quartic terms in Eq.~\eqref{eq:EWSB-Higgs-potential} stabilize the Higgs potential apart from the flat direction $|H_u^0|=|H_d^0|$.

Once again, the condition that the potential must have a minimum which breaks the electroweak symmetry at the electroweak scale imposes stringent relations between the parameters.
In order for the potential to have a stable minimum, the condition 
\begin{align}
    b < \frac{2|\tilde\mu|^2 + m_{H_u}^2 + m_{H_d}^2}{2}
    \label{eq:condition-stabel-minimum}
\end{align}
must be satisfied.
The mass matrix for the two real doublet components~$H_{u,d}^0$ then reads
\begin{align}
    m^2 &= \begin{pmatrix}
        |\tilde\mu|^2 + m_{H_u}^2 & -b \\
        -b & |\tilde\mu|^2 + m_{H_d}^2
    \end{pmatrix} 
\end{align}
and must have at least one negative eigenvalue for EWSB to occur. 
Consequently, $m^2$~must not be positive semi-definite, meaning that at least one of the principal minors of~$m^2$ must be negative.
Taking $|\tilde\mu|^2 + m_{H_{u,d}}^2 > 0$, we must have~$\det(m^2)<0$, which leads to
\begin{align}
    b^2 > \left( |\tilde\mu|^2 + m_{H_{u}}^2 \right) \left( |\tilde\mu|^2 + m_{H_{d}}^2 \right) \,.
    \label{eq:condition-EWSB}
\end{align}
EWSB can therefore only take place in the MSSM if Eqs.~\eqref{eq:condition-stabel-minimum} and~\eqref{eq:condition-EWSB} are satisfied.

As before, we define the VEVs of the neutral Higgs field components as $\smash{\langle H^0_{u,d} \rangle \equiv v_{u,d}\big/\sqrt{2}}$, with $\smash{v^2 \equiv v_u^2 + v_d^2} = \left(246 \textrm{ GeV}\right)^2$.
The VEV ratio is given by $\tan \beta = {v_u}\big/{v_d}$,
which allows us to write $v_u = v \sin\beta$ and $v_d = v \cos\beta$ with $0<\beta<\frac{\pi}{2}$ and $v_{u,d} \geq 0$.
From the minimization conditions we find
\begin{subequations}
\begin{align}
    0 &= |\tilde\mu|^2 + m_{H_u}^2 - b \cot(\beta) - \frac{g_1^2 + g_2^2}{8} v^2 \cos{(2\beta)} \,,
    \\
    0 &= |\tilde\mu|^2 + m_{H_d}^2 - b \tan(\beta) + \frac{g_1^2 + g_2^2}{8} v^2 \cos{(2\beta)} \,.
\end{align}
\end{subequations}
Summing or multiplying these two equations, we can show that the conditions for EWSB given in Eqs.~\eqref{eq:condition-stabel-minimum} and~\eqref{eq:condition-EWSB} are indeed satisfied.
Furthermore, we find the tree-level expressions for the rotation angle~$\beta$ and the VEV~$v$ in terms of the SUSY parameters:
\begin{align}
    \sin (2\beta) &= \frac{2b}{2|\tilde\mu|^2 + m_{H_u}^2 + m_{H_d}^2} \,,
    \label{eq:beta-definition}
    \\
    \frac{g_1^2 + g_2^2}{4} v^2 
    &=
    \frac{m_{H_u}^2 - m_{H_d}^2}{\cos (2\beta)} - \left( 2|\tilde\mu|^2 + m_{H_u}^2 + m_{H_d}^2 \right) \,.
    \label{eq:Z-mass_Higgs-basis}
\end{align}

\paragraph{Electroweak mass eigenstates of the Higgs sector.}
Next, we can determine the masses of the physical scalar degrees of freedom in the MSSM Higgs sector.
For this, we rotate the gauge eigenstates into the mass eigenstates by
\begin{align}
    \begin{pmatrix}
        H_u^0 \\ H_d^0
    \end{pmatrix}
    &= \frac{1}{\sqrt{2}}\left[ 
    R_\beta \begin{pmatrix}
        v \\ 0
    \end{pmatrix} + R_\alpha \begin{pmatrix}
        h^0 \\ H^0
    \end{pmatrix} + i\,R_{\beta_0} \begin{pmatrix}
        G^0 \\ A^0
    \end{pmatrix}\right]
    \,,
    & 
    \begin{pmatrix}
        H_u^+ \\ H_d^+
    \end{pmatrix} 
    &= R_{\beta_\pm} \begin{pmatrix}
        G^+ \\ H^+
    \end{pmatrix} \,,
    \label{eq:rotation-EWSB}
\end{align}
where the generic rotation matrices take the form
\begin{align}
    R_{\beta} &= \begin{pmatrix}
        s_{\beta} & -c_{\beta} \\ c_{\beta} & s_{\beta}
    \end{pmatrix},
    &
    R_\alpha &= \begin{pmatrix}
        c_\alpha & s_\alpha \\ -s_\alpha & c_\alpha
    \end{pmatrix},
    &
    R_{\beta_0} &= \begin{pmatrix}
        s_{\beta_0} & c_{\beta_0} \\ -c_{\beta_0} & s_{\beta_0}
    \end{pmatrix},
    &
    R_{\beta_\pm} &= \begin{pmatrix}
        s_{\beta_\pm} & c_{\beta_\pm} \\ -c_{\beta_\pm} & s_{\beta_\pm}
    \end{pmatrix}.
    \label{eq:rotation-matrices_Higgs-basis}
\end{align}
Inserting the above relations into the Higgs potential in Eq.~\eqref{eq:EWSB-Higgs-potential} and requiring it to be at its minimum allows us to derive the following relations
\begin{subequations}
\begin{align}
    \beta &= \beta_0 = \beta_\pm \,,
    \label{eq:beta-MSSM}
    \\[0.1cm]
    \tan (2\alpha) &= \tan (2\beta) \, \frac{\textrm{m}_{A^0}^2 + \textrm{m}_{Z}^2}{\textrm{m}_{A^0}^2 - \textrm{m}_{Z}^2} 
    = \tan (2\beta) \left[ 1 + 2 \frac{\textrm{m}_Z^2}{\textrm{m}_{A^0}^2} + \cO\!\left(\frac{\textrm{m}_Z^4}{\textrm{m}_{A^0}^4}\right) \right]
    \,,
    \label{eq:alpha-MSSM}
    \\[0.1cm]
    \textrm{m}_{G^0}^2 &= \textrm{m}_{G^\pm}^2 = 0 \,,
    \\[0.1cm]
    \textrm{m}_{A^0}^2 &= 2|\tilde\mu|^2 + m_{H_u}^2 + m_{H_d}^2 \,,
    \label{eq:mA0}
    \\[0.1cm]
    \textrm{m}_{H^\pm}^2 &= \textrm{m}_{A^0}^2 + \frac{v^2 g_2^2}{4} = \textrm{m}_{A^0}^2 + \textrm{m}_W^2 \,,
    \\[0.1cm]
    \begin{split}
        \textrm{m}_{h^0}^2 
        &= \frac{1}{2} \left[ \textrm{m}_{A^0}^2 + \textrm{m}_{Z}^2 - \sqrt{\left(\textrm{m}_{A^0}^2 - \textrm{m}_{Z}^2\right)^2 + 4 \textrm{m}_{A^0}^2 \textrm{m}_{Z}^2 \sin^2 (2\beta)} \right] 
        \\
        &= \cos^2(2\beta) \textrm{m}_Z^2 \left[ 1 - \sin^2(2\beta) \frac{\textrm{m}_Z^2}{\textrm{m}_{A^0}^2} + \cO\!\left(\frac{\textrm{m}_Z^4}{\textrm{m}_{A^0}^4}\right) \right]
        \,,
        \label{eq:Higgs-basis-Mh0}
    \end{split}
    \\[0.1cm]
    \begin{split}
        \textrm{m}_{H^0}^2 
        &= \frac{1}{2} \left[ \textrm{m}_{A^0}^2 + \textrm{m}_{Z}^2 + \sqrt{\left(\textrm{m}_{A^0}^2 - \textrm{m}_{Z}^2\right)^2 + 4 \textrm{m}_{A^0}^2 \textrm{m}_{Z}^2 \sin^2 (2\beta)} \right] 
        \label{eq:Higgs-basis-MH0} \\
        &= \textrm{m}_{A^0}^2 + \sin^2(2\beta) \textrm{m}_Z^2 \left[ 1 + \cos^2(2\beta) \frac{\textrm{m}_Z^2}{\textrm{m}_{A^0}^2} + \cO\!\left(\frac{\textrm{m}_Z^4}{\textrm{m}_{A^0}^4}\right) \right]
        \,.
    \end{split}
\end{align}
\label{eq:physical-masses}%
\end{subequations}
where we have used the tree-level definition for the $Z$~mass in Eq.~\eqref{eq:Z-mass-tree}.

We now proceed to integrate out the heavy scalars ($H^0$, $H^\pm$ and $A^0$) under the assumption that their masses are well above the electroweak scale~($v$).
For simplicity, we consider again the unitary gauge~($G^{0,\pm} = 0$).
In this case, since the potential preserves CP~symmetry and electric charge, only the heavy CP-even scalar~$H^0$ can contribute to the effective light Higgs potential at tree level.
Hence, after using Eqs.~\eqref{eq:rotation-EWSB} and~\eqref{eq:rotation-matrices_Higgs-basis} to rotate the potential from Eq.~\eqref{eq:Higgs-potential} to the physical basis, we keep only the terms containing~$h^0$ or~$H^0$.

Using Eq.~\eqref{eq:alpha-MSSM} to replace $\alpha$ in favor of~$\beta$, together with Eqs.~\eqref{eq:beta-definition}, \eqref{eq:Z-mass_Higgs-basis}, \eqref{eq:beta-MSSM}, \eqref{eq:mA0} and~\eqref{eq:Higgs-basis-MH0}, allows us to express the resulting potential entirely in terms of the parameters $\beta$, $v^2$, $\textrm{m}_{A^0}^2$, and the gauge couplings. 
We can then integrate out the heavy scalar~$H^0$ at tree level using its equations of motion, which yields the scalar potential for the physical Higgs boson (in unitary gauge):
\begin{align}
\begin{split}
    V_{h^0}^{\mathrm{HEFT}}
    &=
    \left(h^0\right)^2 \cos^2(2\beta) \, \frac{g_1^2+g_2^2}{8} \, v^2 \left[ 1 - \sin^2(2\beta) \, \frac{g_1^2+g_2^2}{4} \, \frac{v^2}{\textrm{m}_{A^0}^2} \right]
    \\
    &+
    \left(h^0\right)^3 \cos^2(2\beta) \, \frac{g_1^2+g_2^2}{8} \, v \left[ 1 - 3 \sin^2(2\beta) \, \frac{g_1^2+g_2^2}{4} \, \frac{v^2}{\textrm{m}_{A^0}^2} \right]
    \\
    &+
    \left(h^0\right)^4 \cos^2(2\beta) \,  \frac{g_1^2+g_2^2}{32} \left[ 1 - 13 \sin^2(2\beta) \, \frac{g_1^2+g_2^2}{4} \, \frac{v^2}{\textrm{m}_{A^0}^2} \right]
    \\
    &-
    \left(h^0\right)^5 3 \sin^2(4\beta) \, \frac{\left(g_1^2 + g_2^2\right)^2}{256} \, \frac{v}{\textrm{m}_{A^0}^2}
    \\
    &-
    \left(h^0\right)^6 \sin^2(4\beta) \, \frac{\left(g_1^2 + g_2^2\right)^2}{512} \, \frac{1}{\textrm{m}_{A^0}^2} 
    +
    \cO\!\left(\textrm{m}_{A^0}^{-4}\right)
    \,,
\end{split}
\label{eq:physical-Higgs-potential_HEFT}%
\end{align}
Before we can compare the Higgs EFT (HEFT)~\cite{Feruglio:1992wf,Alonso:2012px,Buchalla:2013rka} result above to the scalar potential computed with the SMEFT approach in Eq.~\eqref{eq:physical-Higgs-potential_SMEFT}, we need to find the relations between the $m_\Phi$ and $\gamma$ parameters defined in the soft-SUSY mass basis to the $\textrm{m}_{A^0}$ and $\beta$ parameters 
defined within the MSSM.

Using the definitions of $m_\Phi$ and~$m_H$ from Eq.~\eqref{eq:MSSM_Lagrangian-matching-basis} and the definition of $\textrm{m}_{A^0}$ from Eq.~\eqref{eq:mA0}, we have:
\begin{align}
    \textrm{m}_{A^0}^2 = m_\Phi^2 + m_H^2 = m_\Phi^2 \left(1 + \frac{m_H^2}{m_\Phi^2}\right) = m_\Phi^2 \left[1 - \frac{\lambda v^2}{2 m_\Phi^2} + \mathcal{O}\!\left(\frac{v^4}{m_\Phi^4}\right)\right]\; ,
\end{align}
where we have used Eq.~\eqref{eq:mHdef}.
To relate the $\gamma$ and $\beta$ angles we can make use of the minimization conditions, Eqs.~\eqref{eq:beta-definition} and~\eqref{eq:Z-mass_Higgs-basis}, and the definition of $\gamma$ in Eq.~\eqref{eq:rotation-angle_mass-basis}, resulting in
\begin{align}
\begin{split}
    \tan(2\gamma) 
    &= \frac{2b}{m_{H_u}^2-m_{H_d}^2} 
    = \tan(2\beta) \frac{2|\tilde\mu|^2 + m_{H_u}^2 + m_{H_d}^2}{2|\tilde\mu|^2 + m_{H_u}^2 + m_{H_d}^2 + \left(g_1^2 + g_2^2\right) \frac{v^2}{4}} \label{eq:mPhimA0}
    \,.
\end{split}
\end{align}
Finally, using the expressions for $\textrm{m}_{A^0}^2$, Eq.~\eqref{eq:mA0}, and Eq.~\eqref{eq:mPhimA0} we have:
\begin{align}
\begin{split}
     \tan(2\gamma)     
    = \tan(2\beta) \frac{\textrm{m}_{A^0}^2}{\textrm{m}_{A^0}^2 + \left(g_1^2 + g_2^2\right) \frac{v^2}{4}} =  \tan(2\beta) \left(1 - \frac{g_1^2+g_2^2}{4} \frac{v^2}{m_\Phi^2}  \right) + \mathcal{O}\!\left(\frac{v^4}{m_\Phi^4}\right)
    \label{eq:tan2gamma-tan2beta}
\end{split}
\end{align}
From this result we see that $\gamma\sim\beta$ up to power corrections in the EFT. 

Using Eqs.~\eqref{eq:mPhimA0} and~\eqref{eq:tan2gamma-tan2beta}, we can compare the scalar potential in SMEFT~($V_{\mathrm{Higgs}}^{\mathrm{SMEFT}}$) from Eq.~\eqref{eq:physical-Higgs-potential_SMEFT} to the one in HEFT~($V_{h^0}^{\mathrm{HEFT}}$), given in Eq.~\eqref{eq:physical-Higgs-potential_HEFT}. 
In particular, the quadratic term can be written as:
\begin{align}
    V_{\mathrm{Higgs}}^{\mathrm{SMEFT}}
    &= 
    \frac{\lambda v^2}{2} \left(1 -\frac{3 v^2}{\lambda}C_H\right) (h^0)^2 + ...  
    \nonumber\\
    &= \frac{\left( g_1^2 + g_2^2 \right)}{8} \cos^2(2 \gamma) v^2 \left[1 - 3 \sin^2(2\gamma) \frac{\left( g_1^2 + g_2^2 \right)}{4} \frac{v^2}{m_\Phi^2}  \right] (h^0)^2 + ... 
    \label{eq:Vcomp2}\\
    &= \frac{\left( g_1^2 + g_2^2 \right)}{8} \cos^2(2 \beta) v^2 \left[1 - \sin^2(2\beta) \frac{\left( g_1^2 + g_2^2 \right)}{4} \frac{v^2}{\textrm{m}_{A^0}^2}  + \mathcal{O}\!\left(\frac{v^4}{\textrm{m}_{A^0}^4}\right) \right] (h^0)^2 + ... 
    \,, \nonumber
\end{align}
where in the last step we have used:
\begin{align}
    \cos^2(2 \gamma) = \cos^2 (2\beta) \left[1  + 2 \sin^2(2\beta) \frac{\left( g_1^2 + g_2^2 \right)}{4} \frac{v^2}{\textrm{m}_{A^0}^2} \right] + \mathcal{O}\!\left(\frac{v^4}{\textrm{m}_{A^0}^4}\right) \,,
\end{align}
following from Eq.~\eqref{eq:tan2gamma-tan2beta}.
Comparing Eq.~\eqref{eq:Vcomp2} to the first term in Eq.~\eqref{eq:physical-Higgs-potential_HEFT}, we see that both agree to the leading order in the EFT expansion.
In a similar fashion, it can be shown that all the other terms also agree.
This validates that the SMEFT description obtained in our approach by integrating out the entire doublet~$\Phi$ {\it before} EWSB, indeed results in the correct low-energy Higgs sector of the full model if the EFT series expansion holds.

The agreement between both approaches is not necessarily trivial, since there are two independent rotation angles for translating between the gauge and the mass eigenbases. 
In particular, the rotation angle~$\alpha$ of the neutral CP-even sector differs from the rotation angle~$\beta$ for the neutral CP-odd and charged Higgs sectors.
Therefore, the real physical Higgs boson~$h^0$ and the three SM Goldstone bosons~$G^{0,\pm}$ cannot generally be embedded into a single \textit{SM-like} $\mathrm{SU}(2)_L$~Higgs doublet.
However, if we require the BSM masses to be well above the electroweak scale, i.e., $\textrm{m}_{A^0},\textrm{m}_{H^0},\textrm{m}_{H^\pm} \gg \textrm{m}_{h^0},\textrm{m}_{Z}$, Eq.~\eqref{eq:alpha-MSSM} imposes the condition:
\begin{subequations}
\begin{align}
    \tan (2\alpha) &= \tan (2\beta) \left[ 1 + 2 \frac{\textrm{m}_Z^2}{\textrm{m}_{A^0}^2} + \cO\!\left(\frac{\textrm{m}_Z^4}{\textrm{m}_{A^0}^4}\right) \right] 
    \\
    \Rightarrow \delta & \equiv \alpha - \beta + \frac{\pi}{2} = \frac{\textrm{m}_Z^2}{\textrm{m}_{A^0}^2} \frac{\sin (4\beta)}{2} + \cO\!\left( \frac{\textrm{m}_Z^4}{\textrm{m}_{A^0}^4} \right)
    \,.
    \label{eq:misalignment}
\end{align}
\end{subequations}
The case $\delta = 0$ is known as the alignment limit and leads to a SM-like Higgs doublet~\cite{Haber:1989xc,Gunion:2002zf,Haber:2006ue,Bernon:2015qea}. 
Since the current data favor an SM-like Higgs, the alignment limit seems to be at least partially satisfied, thus motivating us to match directly to SMEFT and avoid all complications of the HEFT.

Further details on the alignment limit as well as an alternative to matching in the soft-SUSY mass basis are discussed in Appendix~\ref{app:Higgs-basis_alignment-limit}.
A~more thorough analysis of the correct low-energy limits of 2HDM theories and their matching on SMEFT or HEFT can be found in Ref.~\cite{Dawson:2023ebe}. 
Furthermore, Ref.~\cite{Dawson:2022cmu} demonstrates that the SMEFT provides an excellent approximation for the type-II 2HDMs, such as present in the MSSM.

% MSSM-to-SMEFT matching
\section{MSSM-to-SMEFT matching}
\label{sec:matching}

In the previous section, we discussed how to rotate the two Higgs doublets to the mass basis after soft-SUSY breaking (but before EWSB).
This allowed us to properly identify and distinguish the heavy and light degrees of freedom in the Higgs sector, and to decouple the former.
Once all heavy degrees of freedom of the MSSM have been identified, we can proceed to integrate them out, which is the topic of this section.

\subsection{Higgsino sector}
\label{sec:Higgsinos}
Before doing so, we provide some technical details on the treatment of the Higgsino sector, which has a mixed mass term of the form $\tilde\mu \overline{\tilde{H}_u^c} \, \varepsilon \, \tilde{H}_d$.
All other fields appearing in the MSSM Lagrangian, presented in Sec.~\ref{sec:MSSM-Lagrangian-Dirac}, are already in the mass basis.
The Higgsino sector is simpler to handle than its $R$-even counter part (that is, the Higgs sector discussed above) since both Higgsinos are heavy and integrated out of the theory.
We start from the free part of the Higgsino Lagrangian
\begin{align}
    \cL_{\tilde{H}} 
    &=
    \overline{\tilde{H}}_u \slashed{D} P_L \tilde{H}_u
    + \overline{\tilde{H}}_d \slashed{D} P_L \tilde{H}_d
    + \tilde\mu \overline{\tilde{H}_d^c} \, \varepsilon \, \tilde{H}_u 
    - \tilde\mu^\ast \overline{\tilde{H}}_u \, \varepsilon \, \tilde{H}_d^c 
    \,.
    \label{eq:Higgsino-Lagrangian}
\end{align}
Since the Higgsinos are chiral non-singlet fermions, only a mixed mass term is allowed.\footnote{Therefore, the two Higgsinos can only be decoupled simultaneously.}
In principle, we could simply set $\tilde{H}_{u,d}$ equal to their equations of motion to integrate them out (at tree level).
Continuing this to the loop level is however less obvious, and automatic matching codes, such as \texttt{Matchete}~\cite{Fuentes-Martin:2022jrf} employed here, require the input Lagrangian to be in its mass basis without any mass mixing terms proportional to UV scales such as~$\tilde\mu$.

In a manner similar to the down-type Higgs doublet~$H_d$, we can redefine the down-type Higgsino by $\tilde{H}_d^\prime = \varepsilon \tilde{H}_d^c$, such that $\tilde{H}_u$ and $\tilde{H}_d^\prime$ both transform as $(\mathbf{1},\mathbf{2})_{1/2}$.
Since $\tilde{H}_u$ is a left-handed fermion while $\tilde{H}_d^\prime$ is right-handed, we can define a vector-like fermion~$\Sigma$ by
\begin{align}
    \Sigma &= P_L \tilde{H}_u + P_R \tilde{H}_d^\prime \,,
    \label{eq:vectorlike-Higgsino}
\end{align}
such that the two Higgsino fields are embedded as the two chiralities of~$\Sigma$.
Inverting the equation above yields
\begin{align}
    \tilde{H}_u = P_L \Sigma \,, \qquad
    \tilde{H}_d = - \varepsilon P_L \Sigma^c \,.
    \label{eq:Higgsino-replacements}
\end{align}
Inserting these relations in Eq.~\eqref{eq:Higgsino-Lagrangian} and assuming for simplicity $\tilde\mu\in\mathbb{R}$,\footnote{The case of a complex $\tilde\mu$, can be accounted for by absorbing the corresponding phase through including a factor of $\exp(-i \arg(\tilde\mu))$ in the right-handed component of $\Sigma$ in Eq.~\eqref{eq:vectorlike-Higgsino}.} we find
\begin{align}
    \cL_{\tilde{H}} 
    &=
    i \overline{\Sigma} \slashed{D} \Sigma - \tilde\mu \overline{\Sigma} \Sigma \,.
\end{align}
This form of the Lagrangian is more convenient, because $\Sigma$ is already in its mass basis contrary to the Higgsino fields~$\tilde{H}_{u,d}$ in Eq.~\eqref{eq:Higgsino-Lagrangian}. 
Therefore, the vector-like fermion~$\Sigma$ can be integrated out with standard methods.
The Lagrangian obtained by inserting Eq.~\eqref{eq:Higgsino-replacements} into the full MSSM Lagrangian, Eq.~\eqref{eq:L-MSSM-full}, is then in a basis suitable for the matching, since all heavy degrees of freedom are in their mass basis.

\subsection{Functional one-loop matching of the MSSM with Matchete}
\label{sec:one-loop matching}
To enable the complete one-loop matching of the MSSM Lagrangian with \matchete~\cite{Fuentes-Martin:2022jrf}, different modifications of the code were required. As already mentioned, several performance and memory optimizations were introduced. 
In addition, an efficient method for integrating out flavored particles (squarks and sleptons in this specific case) is crucial and has been included in \matchete; this has not been available before in \matchete and similar codes to our knowledge. 

The one-loop matching in \matchete relies on functional methods~\cite{Dittmaier:1995ee,Henning:2014wua,delAguila:2016zcb,Henning:2016lyp,Fuentes-Martin:2016uol,Zhang:2016pja,Cohen:2020fcu,Fuentes-Martin:2020udw}.
These allow to express the tree-level SMEFT Lagrangian as
\begin{align}
    S^{(0)}_{\scriptscriptstyle\mathrm{SMEFT}}[\hat\eta_L] &= S^{(0)}_{\scriptscriptstyle\mathrm{MSSM}}\big[\hat\eta_L,\hat\eta_H[\hat\eta_L]\big] \,,
\end{align}
where the hat denotes classical fields~$\hat \eta_{H,L}$ and the subscript separates the fields into light and heavy states with respect to the EFT power counting.
Thus, $\hat\eta_H[\hat\eta_L]$ is the solution to the classical equations of motion of the heavy fields~$\hat \eta_H$, expanded as a series in the EFT power counting and therefore expressed in terms of the light fields~$\hat \eta_L$.
At the one-loop level, the matching master formula is given by~\cite{Cohen:2020fcu,Fuentes-Martin:2020udw}
\begin{align}
    S^{(1)}_{\scriptscriptstyle\mathrm{SMEFT}}[\hat\eta_L]
    &=
    \int \mathrm{d}^D x\,\mathcal{L}_{\scriptscriptstyle\mathrm{SMEFT}}^{(1)} 
    = 
    S_{\scriptscriptstyle\mathrm{MSSM}}^{(1)}[\hat{\eta}] + \left. \frac{i}{2} \, \mathrm{STr} \, \log \left( \left. \frac{\delta^2 S_{\scriptscriptstyle\mathrm{MSSM}}}{\delta\bar\eta_i \delta\eta_j} \right|_{\eta=\hat\eta} \right) \right\rvert_\mathrm{hard} 
    \,,
    \label{eq:one-loop_EFT-action}
\end{align}
where $\mathrm{STr}$ denotes a functional supertrace.\footnote{%
The supertrace $\mathrm{STr}\,Q[\hat\eta]$ is defined as the trace over all internal degrees of freedom of the operator~$Q[\hat\eta]$, including the loop momentum~$k$
\begin{align*}
    \mathrm{STr} \, Q[\hat\eta] &= \pm \!\int\! \frac{\mathrm{d}^D k}{(2\pi)^D} \langle k \rvert \, \mathrm{tr} \, Q[\hat\eta] \, \lvert k \rangle \,,
\end{align*}
where the sign depends on the spin and $\mathrm{tr}$~represents the remaining operator trace.}
This uses the fact that all local contributions to the one-loop Wilson coefficients of the SMEFT are entirely encoded in the hard region of the loop integrals contained in the supertrace~\cite{Fuentes-Martin:2016uol,Zhang:2016pja}, which can be shown using the \textit{method of regions}~\cite{Beneke:1997zp,Jantzen:2011nz}.
The evaluation of the functional supertraces is performed in a manifestly gauge-covariant manner based on Wilson lines~\cite{Fuentes-Martin:2024agf}, see also~\cite{Avramidi:1990je,Barvinsky:1985an,Kuzenko:2003eb}.

The entire matching procedure is performed fully automatically; the only input required is the definition of the MSSM Lagrangian in the form of an input model file. 
The concrete implementation used here (and provided on \href{https://github.com/BSM-EFT/MSSM-to-SMEFT}{GitHub}~\cite{mssm-to-smeft:github}) is based upon the four-component spinor formulation presented in Sec.~\ref{sec:MSSM-Lagrangian-Dirac} and Appendix~\ref{app:MSSM-Lagrangian}. 
The Higgs sector is implemented in the soft-SUSY mass basis (see Sec.~\ref{sec:soft-SUSY_Higgs-mass-basis}) and the Higgsinos are written in terms of one vector-like fermion (c.f.\ Sec.~\ref{sec:Higgsinos}). 
As already mentioned, all fields are assumed to be in the mass-eigenstate basis after soft-SUSY breaking but before EWSB. 
Furthermore, all masses are taken as non-degenerate, while the matching is performed at a single scale~$\bar\mu$, which, to avoid large logarithmic corrections of the form $\log (\bar\mu^2/M_{\scriptscriptstyle\mathrm{SUSY}}^2)$ in the matching conditions and to improve the convergence of the perturbative series, should be chosen close to the soft SUSY-breaking scale~$M_{\scriptscriptstyle\mathrm{SUSY}}$.

The supertraces in the functional one-loop matching in Eq.~\eqref{eq:one-loop_EFT-action} provide the entire EFT Lagrangian, containing all generated higher-dimensional operators, as well as their coefficients expressed in terms of the underlying MSSM model parameters.
Therefore, an a~priori knowledge of the allowed EFT operators is not required when employing functional methods, contrary to other techniques.
However, the operators obtained from the supertraces are not yet in a minimal basis of the EFT but exhibit many redundancies.
To simplify the result, many of these operators can be related to each other using, for example, integration-by-parts identities or field redefinitions.
Before discussing the operator reduction to the Warsaw basis in the next section, we first provide additional details on the employed regularization and renormalization schemes.

\subsubsection*{Regularization Scheme Change: $\overline{\mathrm{DR}} \leftrightarrow \overline{\mathrm{MS}}$}
The one-loop integrals present in the supertraces are performed using dimensional regularization~\cite{tHooft:1972tcz} in $D=4-2\epsilon$ spacetime dimensions. 
This regulator breaks supersymmetry explicitly by introducing a different number of degrees of freedom for vectors and their fermionic partners in $D \neq 4$ dimensions;
one should therefore use a SUSY-preserving regulator, such as dimensional reduction~\cite{Siegel:1979wq}.
For matching supersymmetric theories onto the SMEFT, one should thus perform the scheme change from dimensional reduction to dimensional regularization before the matching (see, for example, the treatment in \cite{Giudice:2011cg,Bagnaschi:2014rsa}).
That is, we have to perform the scheme change for the MSSM Lagrangian in Eq.~\eqref{eq:L-MSSM-full}, which is assumed to be renormalized in~$\overline{\mathrm{DR}}$, to the $\overline{\mathrm{MS}}$~scheme before matching.

The difference between dimensional regularization and dimensional reduction originates from loop diagrams containing vector bosons.
It is therefore crucial to distinguish the gauge couplings associated to vector bosons~($g$) from the gauge couplings associated to gaugino interactions~($\hat{g}$) and $D$-terms~($\hat{\hat{g}}$).
Similarly, we distinguish the Yukawa couplings appearing in the Yukawa interactions~($y$) from the ones in the $F$-terms~($\hat{\hat{y}}$).
In $\overline{\mathrm{DR}}$, SUSY guarantees $g^{\scriptscriptstyle \overline{\mathrm{DR}}} = \hat{g}^{\scriptscriptstyle \overline{\mathrm{DR}}} = \hat{\hat{g}}^{\scriptscriptstyle \overline{\mathrm{DR}}}$ and $y^{\scriptscriptstyle \overline{\mathrm{DR}}} = \hat{\hat{y}}^{\scriptscriptstyle \overline{\mathrm{DR}}}$, where as in $\overline{\mathrm{MS}}$, we generally have $g^{\scriptscriptstyle \overline{\mathrm{MS}}} \neq \hat{g}^{\scriptscriptstyle \overline{\mathrm{MS}}} \neq \hat{\hat{g}}^{\scriptscriptstyle \overline{\mathrm{MS}}}$ and $y^{\scriptscriptstyle \overline{\mathrm{MS}}} \neq \hat{\hat{y}}^{\scriptscriptstyle \overline{\mathrm{MS}}}$ due to the SUSY breaking by the regulator.
For more details, see the discussion around Eq.~(2.4) and~(2.5) in~\cite{Martin:1993yx}, where the finite counterterms required for this scheme change have been determined.
The one-loop relations between the $\overline{\mathrm{DR}}$ and $\overline{\mathrm{MS}}$ regularized couplings relevant for the present work are
\begin{subequations}
\label{eq:DR-MS_shifts}
\begin{align}
    g_3^{\scriptscriptstyle \overline{\mathrm{MS}}} &= g_3^{\scriptscriptstyle \overline{\mathrm{DR}}} \left( 1 - \frac{g_3^2}{32 \pi^2} \right) \,,
    \qquad
    g_2^{\scriptscriptstyle \overline{\mathrm{MS}}} = g_2^{\scriptscriptstyle \overline{\mathrm{DR}}} \left( 1 - \frac{g_2^2}{48 \pi^2} \right) \,,
    \qquad
    g_1^{\scriptscriptstyle \overline{\mathrm{MS}}} = g_1^{\scriptscriptstyle \overline{\mathrm{DR}}} \,,
    \label{eq:DR-MS_gauge-shift}
    \\
    y_u^{\scriptscriptstyle \overline{\mathrm{MS}}} &= y_u^{\scriptscriptstyle \overline{\mathrm{DR}}} \left[ 1 + \frac{1}{16 \pi^2} \left( \frac{4}{3} g_3^2 - \frac{3}{8} g_2^2 - \frac{1}{72} g_1^2 \right) \right] \,,
    \label{eq:DR-MS_Yu-shift}
    \\
    y_d^{\scriptscriptstyle \overline{\mathrm{MS}}} &= y_d^{\scriptscriptstyle \overline{\mathrm{DR}}} \left[ 1 + \frac{1}{16 \pi^2} \left( \frac{4}{3} g_3^2 - \frac{3}{8} g_2^2 - \frac{13}{72} g_1^2 \right) \right] \,,
    \\
    y_e^{\scriptscriptstyle \overline{\mathrm{MS}}} &= y_e^{\scriptscriptstyle \overline{\mathrm{DR}}} \left[ 1 - \frac{1}{16 \pi^2} \left( \frac{3}{8} g_2^2 - \frac{3}{8} g_1^2 \right) \right] \,,
    \\
    \lambda^{\scriptscriptstyle \overline{\mathrm{MS}}} &=  \lambda^{\scriptscriptstyle \overline{\mathrm{DR}}} - \frac{1}{16 \pi^2} \left( \frac{3}{4} g_2^4 + \frac{1}{4} g_1^4 + \frac{1}{2} g_1^2 g_2^2 \right) \,,
    \label{eq:DR-MS_lambda-shift}
\end{align}
\end{subequations}
with the normalization $\cL \supset -\frac{\lambda}{2}(H^\dagger H)^2$ and where we have 
\begin{align}
    \lambda^{\scriptscriptstyle \overline{\mathrm{DR}}} 
    &= 
    \cos^2(2\gamma) \frac{\left(\hat{\hat{g}}_1^{\scriptscriptstyle \overline{\mathrm{DR}}}\right)^2 + \left(\hat{\hat{g}}_2^{\scriptscriptstyle \overline{\mathrm{DR}}}\right)^2}{4}
    \label{eq:lambda-MSSM}
\end{align}
from the $D$-terms in the MSSM Lagrangian.
In addition, we find $\hat{g}_i^{\scriptscriptstyle \overline{\mathrm{MS}}}\!=\!\hat{\hat{g}}_i^{\scriptscriptstyle \overline{\mathrm{MS}}}\!=\!g_i^{\scriptscriptstyle \overline{\mathrm{DR}}}$ 
for $i\in\{1,2,3\}$, 
and $\hat{\hat{y}}_n^{\scriptscriptstyle \overline{\mathrm{MS}}}\!=\!\hat{\hat{y}}_n^{\scriptscriptstyle \overline{\mathrm{DR}}}$ for $n\in\{u,d,e\}$.
Combining these results yields 
\begin{align}
    \hat{g}_3^{\scriptscriptstyle \overline{\mathrm{MS}}} 
    &= 
    \hat{\hat{g}}_3^{\scriptscriptstyle \overline{\mathrm{MS}}}
    =
    g_3^{\scriptscriptstyle \overline{\mathrm{MS}}} \left( 1 + \frac{g_3^2}{32 \pi^2} \right)
    ,
    &
    \hat{g}_2^{\scriptscriptstyle \overline{\mathrm{MS}}} 
    &= 
    \hat{\hat{g}}_2^{\scriptscriptstyle \overline{\mathrm{MS}}}
    =
    g_2^{\scriptscriptstyle \overline{\mathrm{MS}}} \left( 1 + \frac{g_2^2}{48 \pi^2} \right)
    ,
    &
    \hat{g}_1^{\scriptscriptstyle \overline{\mathrm{MS}}} 
    &= 
    \hat{\hat{g}}_1^{\scriptscriptstyle \overline{\mathrm{MS}}}
    =
    g_1^{\scriptscriptstyle \overline{\mathrm{MS}}}
    \,,
    \label{eq:D-term_gaugino_shift}
\end{align}
and
\begin{align}
    \lambda^{\scriptscriptstyle \overline{\mathrm{MS}}} 
    &= 
    \frac{\cos^2(2\gamma)}{4}\left[ \left(g_1^{\scriptscriptstyle \overline{\mathrm{MS}}}\right)^2 + \left(g_2^{\scriptscriptstyle \overline{\mathrm{MS}}}\right)^2 \left( 1 + \frac{g_2^2}{48 \pi^2} \right)^2\right] - \frac{1}{16 \pi^2} \left( \frac{3}{4} g_2^4 + \frac{1}{4} g_1^4 + \frac{1}{2} g_1^2 g_2^2 \right) 
    \label{eq:lambdaShift-MSSM}
\end{align}
which allows us to express our results entirely in terms of the $\overline{\mathrm{MS}}$-renormalized vector-boson gauge couplings~($g^{\scriptscriptstyle \overline{\mathrm{MS}}}$).
The results in Eqs.~\eqref{eq:DR-MS_shifts} have been derived from~\cite{Martin:1993yx} and verified against~\cite{Bagnaschi:2014rsa}.\footnote{Since \cite{Bagnaschi:2014rsa} uses a GUT normalization for~$g_1$, one has to apply the shift $g_1 \!\to\! \sqrt{\frac{5}{3}}\,g_1$ to match our normalization.}

Since in this work we only consider up to one-loop results, the scheme change shifts above only impact the tree-level matching.
The operators $(H^\dagger H)(\Phi^\dagger \Phi)$, $(H^\dagger \Phi)(\Phi^\dagger H)$, and $(\Phi^\dagger \Phi)^2$ also receive finite counterterms relevant to the tree-level matching, but these only start contributing at mass-dimension eight (or higher) and are hence also neglected. 
Finally, we note that the operator $(H^\dagger H)(H^\dagger \Phi)$ (and its Hermitian conjugate) is not directly affected by the scheme change, because the difference between both schemes stems from diagrams with a closed vector-boson loop with two vertices of the type scalar-scalar-vector-vector~\cite{Martin:1993yx} and kinetic mixing between the two Higgs bosons is absent in the MSSM.
However, its coefficient contains $\hat{\hat{g}}_2^{\scriptscriptstyle \overline{\mathrm{MS}}}$ and is thus affected by the shift in Eq.~\eqref{eq:D-term_gaugino_shift} when expressing everything in terms of~$g_2^{\scriptscriptstyle \overline{\mathrm{MS}}}$.
The full list of all scheme-change counterterms relevant for the one-loop matching of the MSSM onto the SMEFT at mass-dimension six are also collected in Eq.~\eqref{eq:MSSM_scheme-ct_Lagrangian} of the appendix. 
From here onward, unless stated otherwise, we will drop the distinction between $g$, $\hat{g}$, and~$\hat{\hat{g}}$ and always express results in terms of the $\overline{\mathrm{MS}}$-renormalized vector-boson gauge couplings~$g_i^{\scriptscriptstyle \overline{\mathrm{MS}}}$.
Notice that this is required when using \matchete, since the code performs all computations in $\overline{\mathrm{MS}}$.

To facilitate the proper scheme change, we implemented a function automating the $\overline{\mathrm{DR}} \to \overline{\mathrm{MS}}$ conversion for all renormalizable operators in a given Lagrangian.
This function is included in the \matchete~\texttt{v0.4.0} under the name \texttt{DR2MS} (see the \matchete documentation for further details), and called within our MSSM model file. 
The function is general enough for handling the scheme change also in scenarios where not all BSM states are integrated out and hence additional finite counterterms are needed at the one-loop level.
This is, for example, the case in the MSSM--to--2HDM-EFT matching presented in Appendix~\ref{app:2HDM-EFT}, where the \texttt{DR2MS} function is applied as well.

\subsection{Mapping onto the Warsaw basis}
For phenomenological studies, it is favorable to eliminate all redundant operators in the EFT Lagrangian to have a minimal description of the theory with the least amount of free parameters. 
The most common choice for an operator basis in the SMEFT is the so-called \textit{Warsaw basis}~\cite{Grzadkowski:2010es}, which is employed in a wide range of phenomenological studies and is used by most phenomenological tools.
The elimination of redundant operators and projection onto the Warsaw basis is handled in a fully automated manner by \matchete.

The operator reduction can be conceptually separated into two pieces: off-shell and on-shell reduction. 
The former involves the application of identities that leave all off-shell Green's functions invariant (e.g. integration-by-parts, Fierz relations).
The latter incorporates field redefinitions to reduce those operator combinations that are proportional to the equations of motion~(EOMs) of the EFT fields. 
In the literature, performing field redefinitions is often also referred to as \textit{applying} the EOMs, which is, however, only equivalent at the leading power in the EFT expansion~\cite{Criado:2018sdb}. 
Such on-shell reduction only leaves physical observables invariant but alters the off-shell Green's functions. 
To map onto a minimal four-dimensional physical basis, such as the Warsaw basis, it is required to apply both types of identities, until all redundant operators have been eliminated. 
To that end, we first use integration-by-parts to recast (when possible) derivative-rich higher-dimensional operators in terms of the kinetic parts of the EOMs of the EFT.
These \textit{EOM-redundant} operators are then eliminated in favor of operators with fewer derivatives with the help of field redefinitions. 
For details on the automatic operator reduction in \matchete, see~\cite{Fuentes-Martin:2022jrf}.

Special care has to be taken in order to retain a consistent EFT truncation for the Higgs boson mass. 
The mass term of the SMEFT Higgs doublet~$H$ receives one-loop matching corrections that are proportional to the masses of the heavy states, which are integrated out.
This distorts the power counting of the EFT, since in the SMEFT the Higgs doublet~$H$ is assumed to be light with a mass of the order of the electroweak scale, while contributions proportional to the high SUSY-breaking scale are generated through the matching.
This is nothing but the little hierarchy problem that is introduced by assuming that the superpartners of the SM fields and the second heavy Higgs doublet~$\Phi$ are heavy enough to decouple and to be integrated out.
This is solved in \matchete by introducing an effective Higgs mass that is defined by the sum of all contributions to the Higgs mass term generated by the matching.
Regarding the EFT power counting, the effective Higgs mass is then defined as an IR scale and thus does not upset the EFT assumption. 
Effectively, this assumes that the little hierarchy problem is solved by the required amount of fine tuning. 
For more details on the application of field redefinitions within the \matchete framework, see again~\cite{Fuentes-Martin:2022jrf}.

In the following, we elaborate on two subtleties in the operator reduction, namely the threshold corrections for gauge couplings, and the proper treatment of evanescent operators.

\subsubsection{Gauge coupling threshold corrections \label{sec:gauge-coupligs}}
In the matching, threshold corrections to the kinetic terms of all fields can be generated.
However, the SMEFT Lagrangian is conventionally defined employing a canonical normalization for all fields. 
Except for the gauge fields in our convention, kinetic terms are free of coupling constants, and we thus have to perform field redefinitions in order to obtain a canonical normalization, which is required to obtain the Warsaw basis Wilson coefficients.
For gauge fields, on the other hand, we can achieve a canonical normalization by shifting the gauge couplings.

Suppose that, after the matching, the kinetic term of a gauge field~$A_\mu$ in the EFT Lagrangian reads
\begin{align}
    \cL &\supset -\frac{1}{4\,g^2} \left( 1 + \frac{\Delta}{16\pi^2} \right) F_{\mu\nu} F^{\mu\nu} \,,
    \label{eq:gauge-kinetic}
\end{align}
where $g$ is the corresponding $\overline{\mathrm{MS}}$ gauge coupling and $\Delta$ is given by a product of BSM couplings.
Furthermore, we have $F_{\mu\nu} = i [D_\mu,D_\nu]$ with the covariant derivative $D_\mu = \partial_\mu - i A_\mu$.
The kinetic term can be brought to a canonical normalization with a redefinition of the gauge coupling according to $g \to (1+\frac{\Delta}{32\pi^2}) g$.
Note that in this convention neither the gauge field nor the field-strength tensor is getting shifted due to this redefinition.\footnote{If one uses the convention where the gauge coupling is not absorbed into the gauge, i.e., where the gauge kinetic term reads $-\frac{1}{4}F_{\mu\nu}F^{\mu\nu}$ and the covariant derivative is $D_\mu = \partial_\mu - i g A_\mu$, one has to use a gauge field redefinition in order to obtain a canonical normalization for the kinetic terms. However, this also necessitates shifting the gauge coupling to retain a manifestly covariant form of the Lagrangian. This highlights the advantages of our chosen convention, where only a gauge coupling shift is sufficient.}

Since the shift of the gauge couplings is necessarily loop suppressed, the gauge coupling threshold correction is only relevant if the gauge couplings appear explicitly in the EFT Lagrangian at tree level, when a consistent truncation at one-loop order is applied.
For gauge couplings appearing in the one-loop matching conditions, the threshold correction can be neglected as a two-loop effect.
Therefore, in most theories where only fermions and scalars are integrated out, these gauge coupling corrections can be entirely neglected in one-loop matching calculations.
In these cases, the gauge couplings are usually not present explicitly in the underlying UV Lagrangian but only contribute through the propagators of gauge fields inside loop diagrams. 
Thus, the gauge couplings do not appear explicitly in the tree-level EFT Lagrangian, and thus any one-loop correction to them can be ignored. 
However, in the case of the MSSM, due to the supersymmetric nature of the theory, gauge couplings are also explicitly present in the UV Lagrangian, cf. Eq.~\eqref{eq:L-gauge}. 
Gauge-coupling contributions to the tree-level matching are hence possible, and we consistently consider the threshold matching corrections for the gauge couplings in the present analysis, finding
\begin{subequations}
\label{eq:gauge-coupling-corrections}
\begin{align}
    g_1 \equiv g_1^{\scriptscriptstyle \mathrm{SMEFT}} 
    &= 
    g_1^{\scriptscriptstyle \overline{\mathrm{MS}}} - \frac{g_1^3}{16\pi^2} \left\{ \sum_p \left[ 
    \frac{1}{36}\log\frac{\bar\mu^2}{{(m_{\tilde{q}}^p)}^2} 
    +\frac{2}{9}\log\frac{\bar\mu^2}{{(m_{\tilde{u}}^p)}^2}
    +\frac{1}{18}\log\frac{\bar\mu^2}{{(m_{\tilde{d}}^p)}^2} 
    \right.\right.
    \\[0.1cm]
    &\qquad\qquad\qquad\qquad\qquad\left.\left.
    +\frac{1}{12}\log\frac{\bar\mu^2}{{(m_{\tilde{\ell}}^p)}^2}
    +\frac{1}{6}\log\frac{\bar\mu^2}{{(m_{\tilde{e}}^p)}^2} \right]
    +\frac{1}{12}\log\frac{\bar\mu^2}{m_\Phi^2}
    +\frac{1}{3}\log\frac{\bar\mu^2}{\tilde\mu^2}
    \right\}
    ,
    \nonumber
    \\[0.2cm]
    g_2 \equiv g_2^{\scriptscriptstyle \mathrm{SMEFT}} 
    &= 
    g_2^{\scriptscriptstyle \overline{\mathrm{MS}}} - \frac{g_2^3}{16\pi^2} \left\{ 
    \sum_p \left[ 
    \frac{1}{4}\log\frac{\bar\mu^2}{{(m_{\tilde{q}}^p)}^2} 
    +\frac{1}{12}\log\frac{\bar\mu^2}{{(m_{\tilde{\ell}}^p)}^2} \right]
    \right.
    \label{eq:g2-shift}
    \\[0.1cm]
    &\qquad\qquad\qquad\qquad\qquad\left.
    +\frac{2}{3}\log\frac{\bar\mu^2}{m_2^2}
    +\frac{1}{12}\log\frac{\bar\mu^2}{m_\Phi^2}
    +\frac{1}{3}\log\frac{\bar\mu^2}{\tilde\mu^2}  
    \right\}
    ,
    \nonumber
    \\[0.2cm]
    g_3 \equiv g_3^{\scriptscriptstyle \mathrm{SMEFT}} 
    &= 
    g_3^{\scriptscriptstyle \overline{\mathrm{MS}}} - \frac{g_3^3}{16\pi^2} \left\{ 
    \sum_p \left[ 
    \frac{1}{6}\log\frac{\bar\mu^2}{{(m_{\tilde{q}}^p)}^2} 
    +\frac{1}{12}\log\frac{\bar\mu^2}{{(m_{\tilde{u}}^p)}^2}
    +\frac{1}{12}\log\frac{\bar\mu^2}{{(m_{\tilde{d}}^p)}^2}\right]
    +\log\frac{\bar\mu^2}{m_3^2}
    \right\}
    ,
\end{align}
\end{subequations}
where the sum over flavor indices~$p$ is made explicit for convenience. 
The~$g_i^\mathrm{SMEFT}$ on the left-hand side of Eqs.~\eqref{eq:gauge-coupling-corrections} are the IR/SMEFT couplings, while the right-hand side are the UV/MSSM couplings, both renormalized in the~$\overline{\rm MS}$ scheme.
The results in Eq.~\eqref{eq:gauge-coupling-corrections} agree with the expressions in \cite{Bagnaschi:2014rsa} once we express the $\overline{\rm MS}$ couplings in terms of the $\overline{\rm DR}$ ones following Eq.~\eqref{eq:DR-MS_gauge-shift}.

Note that the \texttt{DR2MS} function used in our \matchete implementation to perform the $\overline{\mathrm{DR}}$--$\overline{\mathrm{MS}}$ scheme change returns a Lagrangian expressed entirely in terms of the $\overline{\mathrm{MS}}$-renormalized MSSM vector-boson gauge couplings~$g^{\scriptscriptstyle \overline{\mathrm{MS}}}$ (c.f. end of Sec.~\ref{sec:one-loop matching}) and not their $\overline{\mathrm{DR}}$-renormalized versions.
Therefore, the scheme-change counterterms are not included in the matching conditions for the SMEFT gauge couplings provided directly by \matchete, but have to be added manually using Eq.~\eqref{eq:DR-MS_gauge-shift}, as is the case in Eqs.~\eqref{eq:gauge-coupling-corrections}.

In principle, the threshold correction of the strong coupling constant~$g_3$ can be neglected, since~$g_3$ does not appear in the tree-level matching conditions, cf.~Eq.~\eqref{eq:tree-level_matching-conditions}. 
This is because the only particle that contributes to the tree-level matching is the second Higgs doublet~$\Phi$, which is an $\mathrm{SU}(3)_c$~singlet.

\subsubsection{Evanescent operators}
\label{sec:evanescent}
Another subtlety arises in the operator reduction when certain intrinsically four-dimensional identities are applied.
Fierz identities for the Lorentz group are an important example of this kind of identities that are also relevant for the MSSM matching. Fierz identities are valid only in $D=4$ space-time dimensions and do not hold in $D=4-2\epsilon$ dimensions, which we use for the one-loop matching calculation performed in dimensional regularization. The reason is that Fierz identities project onto a basis of Dirac structures, which is finite-dimensional in $D=4$ dimensions but infinite-dimensional in non-integer dimensions, where it is defined only by analytic continuation. Therefore, a projection onto a finite basis of Dirac structures is possible only in four space-time dimensions.

In order to project the $D=4-2\epsilon$ dimensional EFT Lagrangian, obtained by the one-loop matching, onto a physical, finite basis of Dirac structures such as the SMEFT Warsaw basis~\cite{Grzadkowski:2010es}, we need to apply the intrinsically four-dimensional identities. 
The remediation of the error, introduced by applying a four-dimensional identity in a $D$-dimensional computation, necessitates the introduction of the so-called \textit{evanescent operators}. Schematically, we can write
\begin{align}
    \cP R &= Q + E \,,
    \label{eq:projection}
\end{align}
where $R$ represents a redundant operator from the $D$-dimensional EFT Lagrangian after the matching and $Q$ is an operator of the physical Warsaw basis.
The projection onto a $D=4$ basis using four-dimensional identities is denoted by~$\cP$.
The evanescent operator~$E$ is then implicitly defined as the difference of the former two operators $E \equiv \cP R - Q$.
Thus, evanescent operators are formally of rank~$\epsilon$ and hence vanish at the tree level.

Using a projection like the one in Eq.~\eqref{eq:projection}, we can express all redundant operators~$\{R\}$ from the EFT Lagrangian in terms of the Warsaw basis operators~$\{Q\}$ and a set of evanescent operators~$\{E\}$. 
In principle, the set of evanescent operators is infinite, but only a finite number of evanescent structures is present at fixed loop order.
If one computes a one-loop matrix element within the EFT, both the physical~$\{Q\}$ and the evanescent operators~$\{E\}$ have to be taken into account.
The latter can contribute through UV-divergent one-loop diagrams, where an evanescent operator of rank~$\epsilon$ can combine with the $1/\epsilon$~pole to produce a finite and local one-loop contribution.
It is well known that instead of considering the contributions of evanescent operators to one-loop matrix elements, one can make the coefficients of all evanescent operators in the EFT vanish by changing the renormalization scheme.
More precisely, we can introduce additional finite counterterms for the Warsaw basis operators such that we can drop the evanescent operators but all one-loop matrix elements of the EFT remain invariant. 
This was originally discussed for the computation of the anomalous dimension of the weak effective Hamiltonian~\cite{Buras:1989xd,Dugan:1990df,Herrlich:1994kh}.
In this scheme, no evanescent operators have to be considered for the computation of matrix elements, but the renormalization group evolution is affected at the two-loop level. 
We neglect the latter fact in the present analysis, which is restricted to the one-loop level.
More recently, evanescent operators were reconsidered in the context of one-loop basis transformations for the LEFT~\cite{Aebischer:2022tvz,Aebischer:2022aze,Aebischer:2022rxf}, and for the matching of BSM theories onto the SMEFT~\cite{Fuentes-Martin:2022vvu}.

Here, we follow~\cite{Fuentes-Martin:2022vvu}, which discusses a consistent prescription for the treatment of evanescent contributions in one-loop matching computations.
In the case of the MSSM, only three evanescent structures appear. 
These are generated by the application of Fierz identities to  the four-fermion operators produced in the tree-level matching. 
For the four-fermion operators that are generated in the one-loop matching, we can simply apply the four-dimensional Fierz identities, since the evanescent operators generated by them only produce physical effects at the two-loop level and can hence be neglected.
The resulting tree-level SMEFT operators that are not directly in the Warsaw basis are
\begin{align}
    & (\overline{q} P_R u) (\overline{u} P_L q) \,,
    &
    & (\overline{q} P_R d) (\overline{d} P_L q) \,,
    &
    & (\overline{\ell} P_R e) (\overline{e} P_L \ell) \,.
    \label{eq:redundant-operators}
\end{align}
They are all produced when integrating out the heavy Higgs doublet~$\Phi$.
A single Fierz relation of the form
\begin{align}
    (\overline{\psi}_1 P_R \psi_2) (\overline{\psi}_3 P_L \psi_4) 
    &=
    -\frac{1}{2} (\overline{\psi}_1 \gamma_\mu P_L \psi_4) (\overline{\psi}_3 \gamma^\mu P_R \psi_2) 
    \,,
\end{align}
projects  the redundant structures of Eq.~\eqref{eq:redundant-operators} onto the Warsaw basis operators $Q_{qu}^{(1,8)}$, $Q_{qd}^{(1,8)}$, and $Q_{\ell e}$ respectively.\footnote{For the hadronic operators, we also need to apply the $\mathrm{SU}(3)_c$~identity $(T^A)_{ij} (T^A)_{kl} = \frac{1}{2} \left(\delta_{il} \delta_{kj} - \frac{1}{3} \delta_{ij} \delta_{kl}\right)$.}
The finite counterterms required to absorb the effects of the evanescent operators, generated by this procedure, can be found in the ancillary material of~\cite{Fuentes-Martin:2022vvu}. 
\matchete~\texttt{v0.4.0}~\cite{matchete:v03} used for the MSSM matching presented here is also capable of computing the required evanescent shifts fully automatically, which we use for our analysis.\footnote{We have compared the automatically generated evanescent shifts to the results provided in the supplementary material of Ref.~\cite{Fuentes-Martin:2022vvu}, finding full agreement up to an opposite sign for evanescent shifts of the Yukawa terms.}
Our one-loop matching conditions are provided in an evanescent-free version of the $\overline{\text{MS}}$~scheme, where the coefficients of all evanescent operators vanish. 
For more details on this scheme and the consistent treatment of evanescent operators in general, see~\cite{Fuentes-Martin:2022vvu}.

% Matching results
\section{Matching conditions and validation of results}
\label{sec:matching-results}

Following the procedure discussed in preceding sections, we can now compute the SMEFT coefficients generated by integrating out the MSSM fields. We use this to derive results previously obtained in the literature for validation, and to provide numerical results for a minimal phenomenological example.  
Our full results are available on \href{https://github.com/BSM-EFT/MSSM-to-SMEFT}{GitHub}~\cite{mssm-to-smeft:github}, where we also describe the steps necessary to reproduce them.

For clarity of notation, recall that all IR/SMEFT parameters are renormalized in the evanescent-free $\overline{\mathrm{MS}}$~scheme and usually denoted by capital letters (Higgs mass:~$M_H$, Yukawa matrices:~$Y_{u,d,e}$). 
For the remainder of this section $\lambda$~denotes the Higgs quartic coupling of the SMEFT, and $g_{1,2,3}$~the SMEFT gauge couplings ($g_{i}^{\rm SMEFT}$).
The UV/MSSM parameters are all renormalized in the $\overline{\mathrm{DR}}$~scheme (except for the gauge couplings, which are~$\overline{\mathrm{MS}}$) and labeled with lower-case symbols (Higgs doublet masses~$m_{H,\Phi}$, Yukawa couplings~$y_{u,d,e}$).
Notice also that the angle~$\gamma$ is defined in Eq.~\eqref{eq:rotation-angle_mass-basis} to diagonalize the Higgs-doublet mass matrix only at tree level but not at one loop. 
For a summary of all parameters, see also Tab.~\ref{tab:parameters} in appendix~\ref{app:threshold-corrections-details}.

\subsection{Threshold corrections to renormalizable parameters}

The one-loop matching of the MSSM onto the SMEFT generates, as with any BSM theory, threshold corrections to the renormalizable couplings ($d \leq 4$). The corrections to the gauge couplings were discussed in the previous section, see Eqs.~\eqref{eq:gauge-coupling-corrections}.
For the remaining parameters, namely the Higgs mass parameter, the Higgs quartic coupling and the Yukawa matrices, we obtain:
\begin{subequations}\label{eq:threshold-corrections}
\begin{align}
 \begin{split}
     M_H^2(\bar{\mu}) 
     &= 
     m_H^2 + \frac{1}{16\pi^2} \,\Delta_{H^2}(\bar{\mu}) \,, 
 \end{split}
    \\[0.1cm]
\begin{split}
    \lambda(\bar{\mu}) 
    &= 
    \cos^2(2\gamma) \, \frac{g_1^2 + g_2^2}{4} + \frac{1}{16\pi^2} \left[\Delta_{\lambda}(\bar{\mu})+\Delta_{\lambda}^{\mathrm{reg}}\right]
    \\
    &= 
     \cos^2(2\beta) \, \frac{g_1^2 + g_2^2}{4} \left[1  + 2 \sin^2(2\beta) \frac{\left( g_1^2 + g_2^2 \right)}{4} \frac{v^2}{m_\Phi^2} \right] + \frac{1}{16\pi^2} \left[\Delta_{\lambda}(\bar{\mu})+\Delta_{\lambda}^{\mathrm{reg}}\right]
    \,, 
\end{split}
    \\[0.1cm]
\begin{split}
    Y_u^{pr}(\bar{\mu}) 
    &= 
    \sin(\gamma) \, {y_u}^{pr} + \frac{1}{16\pi^2} \left[\Delta_{Y_u}(\bar{\mu}) +{\Delta_{Y_u}^\mathrm{reg}\,}^{pr}\right]
    \\
    &= 
    \sin(\beta) \, {y_u}^{pr} \left[ 1 - \cos^2(\beta) \cos(2\beta)  \frac{\left( g_1^2 + g_2^2 \right)}{4} \frac{v^2}{m_\Phi^2}  \right] + \frac{1}{16\pi^2} \left[\Delta_{Y_u}(\bar{\mu})+{\Delta_{Y_u}^\mathrm{reg}\,}^{pr}\right]
    \,, 
\end{split}
    \\[0.1cm]
\begin{split}
    Y_{d,e}^{pr}(\bar{\mu}) 
    &= 
    \cos(\gamma) \, y_{d,e}^{pr} + \frac{1}{16\pi^2}\left[\Delta_{Y_{d,e}}(\bar{\mu})+{\Delta_{Y_{d,e}}^\mathrm{reg}}^{pr}\right]
    \\
    &=
    \cos(\beta) \, y_{d,e}^{pr} \left[ 1 + \sin^2(\beta) \cos(2\beta) \frac{\left( g_1^2 + g_2^2 \right)}{4} \frac{v^2}{m_\Phi^2}   \right] + \frac{1}{16\pi^2}\left[\Delta_{Y_{d,e}}(\bar{\mu})+{\Delta_{Y_{d,e}}^\mathrm{reg}}^{pr}\right]
    .
\end{split}
\end{align}
\end{subequations}
In the above, we have written out the tree-level matching results but represented the one-loop corrections only schematically, since the full expressions for $\Delta_i$ ($i= \lambda, Y_{u, d, e}$) are notably lengthy. The complete one-loop contributions can be found in the supplementary material. 
A~detailed discussion of the incorporation of these corrections within \matchete is given in Appendix~\ref{app:threshold-corrections-details}.
The finite counterterms~$\Delta_i^\mathrm{reg}$ originate from the $\overline{\mathrm{DR}}$--$\overline{\mathrm{MS}}$ conversion explained at the end of Sec.~\ref{sec:one-loop matching} and are given by:
\begin{subequations}
\label{eq:d4-scheme-change-ct}
\begin{align}
    \Delta_{\lambda}^{\mathrm{reg}} &= -\frac{1}{4} \left[ g_1^4 + 2 g_1^2 g_2^2 + g_2^4 \left( 3 - \frac{2}{3}\cos^2(2\beta) \right) \right]\,,
    \label{eq:quartic-scheme-change-ct}\\
    {\Delta_{Y_{u}}^{\mathrm{reg}}\,}^{pr}
    &= \sin(\beta)\, y_u^{pr} \left( \frac{4}{3} g_3^2 - \frac{3}{8} g_2^2 - \frac{1}{72} g_1^2 \right)\,,
    \\
    {\Delta_{Y_{d}}^{\mathrm{reg}}\,}^{pr}
    &= \cos(\beta)\, y_d^{pr} \left( \frac{4}{3} g_3^2 - \frac{3}{8} g_2^2 - \frac{13}{72} g_1^2 \right)\,,
    \\
    {\Delta_{Y_{e}}^{\mathrm{reg}}\,}^{pr}
    &= \cos(\beta)\, y_e^{pr} \frac{3}{8} \left( g_1^2 - g_2^2 \right)\,.
\end{align}
\end{subequations}
These and all other SMEFT parameters are renormalized in the (evanescent-free) $\overline{\mathrm{MS}}$~scheme, including the SUSY-preserving counterterms~$\Delta_i^\mathrm{reg}$.

Note also that, for practical purposes, it is useful to express the Wilson coefficients in terms of the low energy (SMEFT) gauge and Yukawa couplings, since these
are directly constrained by low energy observables.
Concretely, we take as inputs the SMEFT values for $g_{1,2,3}$ and $Y_{u,d,e}$ (see Sec.~\ref{sec:gauge-coupligs} and Appendix~\ref{app:threshold-corrections-details}) together with the high-scale (MSSM) values for the soft masses, trilinear $a$-terms, and $\tan\beta$.

When comparing Eqs.~\eqref{eq:threshold-corrections} to the results in~\cite{Bagnaschi:2014rsa}, one has to consider that the angle~$\beta$ in that work actually matches our definition of~$\gamma$ instead of the common definition of~$\beta$ in terms of $\tan\beta=v_u/v_d$ that we employ here; see the discussion on the renormalization of $\tan\beta$ in~\cite{Bagnaschi:2014rsa}. Thus, we find agreement at tree level.

The threshold corrections $\Delta_\lambda$ were also computed at one loop in \cite{Giudice:2011cg,Bagnaschi:2014rsa} and at two loops in \cite{Bagnaschi:2017xid}. 
Rewriting the MSSM Yukawa couplings $y_a^{pr}$ in terms of their SMEFT counterparts (using their tree-level relations) and retaining only the leading terms in the EFT power counting that are proportional to~$\left(Y_u^{33}\right)^4 \equiv Y_t^4$ in the full expression for $\Delta_\lambda$, we find
\begin{eqnarray}
\Delta_\lambda(\bar\mu) &=&
 3\,Y_t^4 \log\left(\frac{\mQL\mtR}{\bar{\mu}^4}\right)
 + 6\,Y_t^4 X_t^2 \left[ 
 \frac{1}{\mQL-\mtR} \log\left(\frac{\mQL}{\mtR}\right) \right] \nonumber\\
& & +\, 6\,Y_t^4 X_t^4 \left[ \frac{1}{\left(\mQL-\mtR\right)^2} -\frac{1}{2}\frac{\mtR+\mQL}{\left(\mQL-\mtR\right)^3}\log\left(\frac{\mQL}{\mtR}\right) \right],
\label{eq:lambH1loop}
\end{eqnarray}
where $X_t \equiv a_t \sin\beta/Y_t - \tilde{\mu} \cot\beta$, $\mQLp{} \equiv m_{\tilde{q}}^3$ and  $\mtRp{} \equiv m_{\tilde{u}}^3$.
This matches the results found in Ref.~\cite{Giudice:2011cg} up to lower-order terms proportional to the gauge couplings.
In addition, we have verified that our complete results for $\Delta_\lambda(\bar\mu)$ allow to fully reproduce all one-loop contributions that have been previously computed in~\cite{Bagnaschi:2014rsa} in the appropriate limits. 
A~detailed comparison can be found on GitHub~\cite{mssm-to-smeft:github}.\footnote{Notice that our definition of~$\gamma$ in Eq.~\eqref{eq:rotation-angle_mass-basis} diagonalizes the Higgs doublet masses only at tree level, whereas the definition of the corresponding angle~$\beta$ in~\cite{Bagnaschi:2014rsa} diagonalizes the masses also at one loop, complicating the comparison of the results. Details on this cross check are provided in the \texttt{Mathematica} notebook on GitHub~\cite{mssm-to-smeft:github}.}

The one-loop corrections to $\lambda$ are also relevant for computing the leading corrections to the Higgs mass,~$m_h$. These are relevant for increasing~$m_h$ above its tree-level value and reproducing the experimental value. Keeping only the leading terms, proportional to~$Y_t^4 v^2$ and taking the limit $\mQLp{}\to \mtRp{} = M_S$:
\begin{eqnarray}
    \Delta \textrm{m}_{h^0}^2(\bar{\mu}) &\equiv & \frac{1}{16\pi^2} \Delta_\lambda(\bar{\mu})\,v^2 =  \frac{3\,Y_t^4 v^2}{8 \pi^2}\Bigg[  \log\left(\frac{M_S^2}{\bar{\mu}^2}\right)
 + \frac{X_t^2}{M_S^2}  - \frac{X_t^4}{12 M_S^4} \Bigg]\nonumber
 \\
 & = &  \frac{3\,m_t^4}{2 \pi^2 v^2 }\Bigg[  \log\left(\frac{M_S^2}{\bar{\mu}^2}\right)
 + \frac{X_t^2}{M_S^2}\left(1  - \frac{X_t^2}{12 M_S^2}\right) \Bigg].
\end{eqnarray}
The result above agrees with the well-known one-loop stop corrections to the Higgs mass within the MSSM~\cite{Djouadi:2013jqa,Slavich:2020zjv}.

\subsection{Wilson coefficients in the Warsaw basis}

\subsubsection{Tree-level matching results}
As mentioned previously, the tree-level matching conditions only receive contributions from integrating out the second Higgs doublet~$\Phi$, but are unaffected by all other heavy states.
This comes from the fact that we imposed $R$-parity invariance on the MSSM Lagrangian. 
For the MSSM, $R$-parity acts as a $\mathbb{Z}_2$~symmetry under which all SM fields and both scalar Higgs doublets~($H,\Phi$) are even, whereas all superpartners of these fields are odd.
The latter must therefore appear in even powers in the Lagrangian terms and cannot contribute to the tree-level matching.
The heavy Higgs~$\Phi$ is the only UV state that is even under this $\mathbb{Z}_2$~symmetry and can hence contribute to the tree-level matching conditions, which therefore take a rather simple form:
\begin{align}
    C_H &= \frac{\sin^2(4\beta)}{m_\Phi^2} \frac{\left( g_1^2 + g_2^2 \right)^2}{64} 
    \,, 
    & 
    C_{uH}^{pr} &=
    - \frac{\sin(4\beta) \, \cot\beta}{m_\Phi^2} \frac{g_1^2 + g_2^2}{8} \, {Y_u}^{pr}  
    \,,
    \nonumber\\
    C_{le}^{prst} &=
    -\frac{\tan^2 \beta}{2 m_\Phi^2} \, {Y_e^\ast}^{rs} Y_e^{pt} 
    \,, 
    &
    C_{dH}^{pr} &= \frac{\sin(4\beta) \, \tan \beta}{m_\Phi^2} \frac{g_1^2 + g_2^2}{8} \, {Y_d}^{pr}  
    \,,
    \nonumber\\
    C_{ledq}^{prst} &=  \frac{\tan^2 \beta}{m_\Phi^2} {Y_e}^{pr} \, {Y_d^\ast}^{ts} 
    \,, 
    &
    C_{eH}^{pr} &= \frac{\sin(4\beta) \, \tan \beta}{m_\Phi^2} \frac{g_1^2 + g_2^2}{8} \, {Y_e}^{pr}  
    \,, 
    \label{eq:tree-level_matching-conditions}
    \\
    {C_{lequ}^{(1)}}^{prst} &=
    \frac{1}{m_\Phi^2} \, {Y_e}^{pr} {Y_u}^{st} 
    \,, 
    &
    {C_{qu}^{(1)}}^{prst} &=
    \frac{1}{6} {C_{qu}^{(8)}}^{prst} =
    -\frac{\cot^2 \beta}{6 m_\Phi^2} \, {Y_u^\ast}^{rs} Y_u^{pt} 
    \,,
    \nonumber\\
    {C_{quqd}^{(1)}}^{prst} &=
    - \frac{1}{m_\Phi^2} \, {Y_u}^{pr} {Y_d}^{st} 
    \,,
    &
    {C_{qd}^{(1)}}^{prst} &=
    \frac{1}{6}{C_{qd}^{(8)}}^{prst} =
    -\frac{\tan^2 \beta}{6 m_\Phi^2} \, {Y_d^\ast}^{rs} Y_d^{pt} 
    \,.
    \nonumber
\end{align}
Here, we have utilized the relations in Eqs.~\eqref{eq:gauge-coupling-corrections} and~\eqref{eq:threshold-corrections} to rewrite the expressions on the right-hand side in terms of the SMEFT Yukawa~($Y_{u,d,e}$) and SMEFT gauge couplings~($g_{1,2}$), along with the MSSM parameters ($\beta$ and $m_\Phi$).
The one-loop shifts resulting from this rewriting of the tree-level matching conditions in terms of SMEFT parameters are then properly included in the one-loop contributions to the matching conditions for the Wilson coefficients.
Similarly, the shifts due to the scheme change are also included in the one-loop parts of these matching condition as shown below in Eqs.~\eqref{eq:DR-MS_ct_d-6}. 
The rewriting of the matching conditions in terms of the SMEFT parameters within \matchete is described in more detail in Appendix~\ref{app:threshold-corrections-details}.
Recall that the rotation angle~$\gamma$ to the soft-SUSY mass basis agrees with the rotation angle~$\beta$ to the Higgs basis up to higher orders in the EFT power counting, cf.\ Eq.~\eqref{eq:gamma-beta}.
Therefore, when working at leading power, we consistently replace $\gamma \to \beta$ everywhere in the dimension-six matching conditions of Eq.~\eqref{eq:tree-level_matching-conditions} and below to facilitate comparison with the literature.

Since only $m_\Phi$ appears on the right hand side of Eq.~\eqref{eq:tree-level_matching-conditions}, these results can be validated by comparing them with previously obtained expressions from tree-level matching between the Two Higgs Doublet model and the SMEFT. In particular, we find agreement with the results in \cite{deBlas:2017xtg} for the case with a second Higgs doublet when substituting $y^u_\varphi = -\cot\beta\,Y_u$, $y^{d,e}_\varphi = -\tan\beta\,Y_{d,e}^\dagger$, $\lambda_\varphi = -(1/8)\sin (4\beta) (g_1^2 + g_2^2)$ and $M_\varphi = m_\Phi$. Likewise, identifying $\eta_u = 1$, $\eta_{d,e} = -\tan^2\beta$, and $Z_6 = -(1/8)\sin (4\beta) (g_1^2 + g_2^2)$, our matching conditions for $C_H$, $C_{uH}^{pr}$, $C_{dH}^{pr}$ and $C_{eH}^{pr}$ agree with those computed in~\cite{Dawson:2022cmu}. 

When extending our results to the one-loop level in the following, it is important to include the finite counterterm contributions from the $\overline{\mathrm{DR}}$--$\overline{\mathrm{MS}}$ scheme change (cf.~end of Sec.~\ref{sec:one-loop matching}) to these tree-level generated Wilson coefficients. 
Parameterizing the scheme-change contribution to the Wilson coefficient~$C_X$ as $\Delta_{C_X}^\mathrm{reg}/16\pi^2$ we find:
\begin{subequations}
\label{eq:DR-MS_ct_d-6}
\begin{align}
    \Delta_{C_{H}}^\mathrm{reg} &= \frac{1}{48 \, m_\Phi^2} \sin^2(4\beta) \, g_2^4 \left(g_1^2+g_2^2\right) \,,
    \\
    {\Delta_{C_{uH}}^\mathrm{reg}}^{pr} &= -\frac{Y_u^{pr}}{12 \, m_\Phi^2} \frac{\sin(4\beta)}{\tan(\beta)} g_2^4 \,,
    \\
    {\Delta_{C_{dH}}^\mathrm{reg}}^{pr} &= \frac{Y_d^{pr}}{12 \, m_\Phi^2} \sin(4\beta) \tan(\beta) \, g_2^4 \,,
    \\
    {\Delta_{C_{eH}}^\mathrm{reg}}^{pr} &= \frac{Y_e^{pr}}{12 \, m_\Phi^2} \sin(4\beta) \tan(\beta) \, g_2^4 \,.
\end{align}
\end{subequations}
Notice that no other $d=6$ operators receive corrections from the scheme change, since we express their coefficients in terms of the effective low-energy inputs, particularly in terms of the SMEFT Yukawas~($Y_\psi$) instead of the MSSM Yukawas~($y_\psi$).
By their definition in Eqs.~\eqref{eq:threshold-corrections}--\eqref{eq:d4-scheme-change-ct}, the~$Y_\psi$ absorb all effects of the scheme-change counterterms for the high-scale~$y_\psi$ given in Eqs.~\eqref{eq:DR-MS_shifts}.
Also, the threshold corrections to~$g_i$, Eq.~\eqref{eq:gauge-coupling-corrections}, correspond to a two-loop effect when combined with the $\Delta^\text{reg}$, and thus impact the results in Eq.~\eqref{eq:DR-MS_ct_d-6} only at higher orders.

\subsubsection{One-loop results}
At the one-loop level, matching contributions are present from all BSM states and we find that all operators of the Warsaw basis are generated, except for the ones containing dual-field-strength tensors and baryon- or lepton-number violating operators. 
The latter are obviously absent, since we assume $R$-parity conservation, which forbids any $B$- or $L$-violating  terms in the renormalizable part of the MSSM Lagrangian we start with.
Furthermore, operators with dual-field-strength tensors cannot be generated given the field content and coupling structure of the MSSM. 
 
In the following, we present some examples of matching conditions for specific operators.  
Concretely, we concentrate on the operators which induce modifications of the Higgs couplings to gauge bosons,
\begin{align}
\mathcal{L}_{\rm SMEFT} \supset {} 
    & C_{HG} \, \frac{1}{g_3^2} \, (H^\dagger H) G_{\mu\nu}^{A} G^{A,\mu\nu} 
    + C_{HW} \, \frac{1}{g_2^2} \, (H^\dagger H) W_{\mu\nu}^I W^{I,\mu\nu}
    \nonumber \\ 
    & 
    + C_{HB} \, \frac{1}{g_1^2} \, (H^\dagger H) B_{\mu\nu} B^{\mu\nu}
    + C_{HWB} \, \frac{1}{g_1 g_2} \, (H^\dagger \sigma^I H)  W_{\mu\nu}^I B^{\mu\nu} \,, 
\end{align}
for which we can compare with existing (approximate) results in the literature. 
The full matching condition for~$C_{HG}$ reads
\begin{align}
C_{HG} = {}  
    & \frac{1}{16\pi^2} \frac{g_3^2}{144} \Bigg\{ \cos(2\beta) \, g_1^2 \sum_p \left[ 2\frac{1}{{(m_{\tilde{u}}^p)}^2} - \frac{1}{{(m_{\tilde{d}}^p)}^2} - \frac{1}{{(m_{\tilde{q}}^p)}^2} \right]
    \nonumber \\
    & + 6  \lvert Y_d^{pr}\rvert^2 \left[ \frac{1}{{(m_{\tilde{q}}^p)}^2} + \frac{1}{{(m_{\tilde{d}}^r)}^2} \right]
    - 6 \Big\lvert \cos \beta \, a_d^{pr}  - \tan \beta \, \tilde{\mu} \, Y_d^{pr} \Big\rvert^2 
    \frac{1}{(m_{\tilde{d}}^r)^2} \frac{1}{(m_{\tilde{q}}^p)^2}
    \\
    &+ 6  \lvert Y_u^{pr}\rvert^2 \left[ \frac{1}{{(m_{\tilde{q}}^p)}^2} + \frac{1}{{(m_{\tilde{u}}^r)}^2} \right] 
    - 6 \Big\lvert \sin \beta \, a_u^{pr} - \cot \beta \, \tilde{\mu} \, Y_u^{pr} \Big\rvert^2 
    \frac{1}{(m_{\tilde{u}}^r)^2} \frac{1}{(m_{\tilde{q}}^p)^2} \Bigg\} ,
    \nonumber
\end{align}
where the sum over flavor indices~$p$ is shown explicitly in the first line due to the lack of repeated indices.
In the following, the Einstein summation over the flavor indices~$p$ and~$r$ is left implicit. 
In the limit where stops and sbottoms are much lighter than the other BSM fields 
\begin{align}\label{eq:mass-approx}
m_{\tilde q}^3,\,m_{\tilde u}^3,\,m_{\tilde d}^3 \ll m_{\tilde{q}}^{1,2},\,m_{\tilde{u}}^{1,2},\, m_{\tilde{d}}^{1,2},\,m_{\tilde l,\tilde e}^{r},\,m_\Phi,\,m_{1,2,3} \,,    
\end{align}
such that the masses of the latter can be taken to infinity, the above equation simplifies to:
\begin{align}
    C_{HG} =  
    \frac{g_3^2}{384 \pi^2} \left[
     \frac{Y_b^2 + Y_t^2 - \frac{1}{6}c_{2\beta} \, g_1^2}{{\mQL}} + \frac{Y_t^2 + \frac{1}{3} c_{2\beta} \, g_1^2}{{\mtR}} + \frac{Y_b^2 - \frac{1}{6}c_{2\beta} \, g_1^2}{{\mbR}} - \frac{Y_b^2 |X_b|^2}{\mbR \mQL}
    - \frac{Y_t^2 |X_t|^2}{\mtR \mQL} \right]\! , 
    \label{eq:chg}
\end{align}
where $c_{2\beta}=\cos(2\beta)$ and once again $m_{\tilde t_R} = m_{\tilde u}^3$, $m_{\tilde Q_3} = m_{\tilde q}^3$ and $m_{\tilde b_R} = m_{\tilde d}^3$\,. Furthermore, for convenience, we defined:
\begin{equation}
a_{b} \cos(\beta) \equiv Y_b A_{b},
\quad 
a_{t} \sin(\beta) \equiv Y_t A_{t},
\quad 
X_t \equiv A_t - \tilde{\mu} \cot(\beta),
\quad 
X_b \equiv A_b - \tilde\mu \tan(\beta)
\,.
\end{equation}
Equation~\eqref{eq:chg} reproduces the result from Ref.~\cite{Drozd:2015kva}.

Following the same steps as above, the other Wilson coefficients read:
\begin{align}
    C_{HW} 
    = {}& \frac{g_2^2}{16\pi^2} \frac{1}{16 \mQL} \left( Y_b^2 + Y_t^2 -\frac{c_{2\beta} g_1^2}{6}\right)\\
    & -  \frac{g_2^2}{16\pi^2} X_b^2 Y_b^2 \left[ \left(\mbR + \mQL\right)\frac{\mbRp{4} - 8 \mbR \mQL +  \mQLp{4}}{16 \mQL \left(\mQL-\mbR\right)^4}  + \frac{3}{4} \frac{\mbRp{4} \mQL}{\left(\mQL - \mbR\right)^5} \log\frac{\mQL}{\mbR} \right] \nonumber\\
    & - \frac{g_2^2}{16\pi^2} X_t^2 Y_t^2 \left[ \left(\mtR + \mQL\right)\frac{\mtRp{4} - 8 \mtR \mQL +  \mQLp{4}}{16 \mQL \left(\mQL-\mtR\right)^4}  + \frac{3}{4} \frac{\mtRp{4} \mQL}{\left(\mQL - \mtR\right)^5} \log\frac{\mQL}{\mtR} \right] \! , \nonumber
\end{align}

\begin{align}
    C_{HB} 
    =  \frac{g_1^2}{16\pi^2} \frac{1}{144} & \left[\frac{1}{\mQL}\left( Y_b^2 + Y_t^2 -\frac{g_1^2 c_{2\beta}}{6}\right) +   \frac{4}{\mbR}\left(Y_b^2 -\frac{g_1^2 c_{2\beta}}{6}\right) +  \frac{16}{\mtR}  \left(Y_t^2 + \frac{g_1^2 c_{2\beta}}{3} \right) \right]\nonumber\\
    + \frac{g_2^2}{16\pi^2} X_b^2 Y_b^2  & \left[ \frac{\left(7 \mbRp{6}\mQL -\mbRp{8} + 75 \mbRp{4} \mQLp{4} + 31 \mbR \mQLp{6} - 4 \mQLp{8}\right)}{144 \mbR \mQL \left(\mQL-\mbR\right)^4}  \right] \nonumber\\
    - \frac{g_2^2}{16\pi^2} X_b^2 Y_b^2  & \left[ \frac{\mbR \mQL \left(\mbR + 2 \mQL\right)}{4 \left(\mQL - \mbR\right)^5} \log\left(\frac{\mQL}{\mbR}\right) \right] \nonumber\\
    - \frac{g_2^2}{16\pi^2} X_t^2 Y_t^2  & \left[ \frac{\left(17 \mtRp{6}\mQL +\mtRp{8} - 39 \mtRp{4} \mQLp{4} - 103 \mtR \mQLp{6} + 16 \mQLp{8}\right)}{144 \mtR \mQL \left(\mQL-\mtR\right)^4} \right]\nonumber\\
    + \frac{g_2^2}{16\pi^2} X_t^2 Y_t^2  & \left[ \frac{\mtR \mQL \left(\mtR - 4 \mQL\right)}{4 \left(\mQL - \mtR\right)^5} \log\left(\frac{\mQL}{\mtR}\right) \right]
\end{align}
and
\begin{align}
    C_{HWB} 
    = {}& \frac{g_2 g_1}{16\pi^2} \frac{1}{24 \mQL} \left(Y_b^2- Y_t^2 - \frac{c_{2\beta} g_2^2}{2} \right) \nonumber\\
    & + \frac{g_2 g_1}{16\pi^2} X_b^2 Y_b^2   \left[ \frac{\left(-\mbRp{6} + 9 \mbRp{4} \mQL + 27 \mbR \mQLp{4} + \mQLp{6}\right)}{24 \mQL \left(\mQL-\mbR\right)^4} \right] \nonumber\\
    & - \frac{g_2 g_1}{16\pi^2} X_b^2 Y_b^2   \left[\frac{\mbR \mQL \left(2 \mbR + \mQL \right)}{2 \left(\mQL - \mbR\right)^5} \log\left(\frac{\mQL}{\mbR}\right) \right] \nonumber\\
    & + \frac{g_2 g_1}{16\pi^2} X_t^2 Y_t^2   \left[ \frac{\left(\mtRp{6} -3 \mtRp{4} \mQL + 33 \mtR \mQLp{4} + 5 \mQLp{6}\right)}{24 \mQL \left(\mQL-\mtR\right)^4} \right] \nonumber\\  
    & - \frac{g_2 g_1}{16\pi^2} X_t^2 Y_t^2   \left[ \frac{\mtR \mQL \left(\mtR + 2 \mQL \right)}{2 \left(\mQL - \mtR\right)^5} \log\left(\frac{\mQL}{\mtR}\right) \right]
    \,.
\end{align}
These expressions also agree with Ref.~\cite{Drozd:2015kva}, except for the terms proportional to $Y_b$, which were neglected in that work.

In addition, we have validated our matching conditions for the Warsaw basis Wilson coefficients $C_H$, $C_{HD}$, $C_{H\Box}$, $C_{uH}$, $C_{Hu}$, and $C_{Hq}^{(1,3)}$ against the results presented in \cite{Drozd:2015rsp}, which are limited to the case where only stop contributions are considered.\footnote{Note that Ref.~\cite{Drozd:2015rsp} uses a redundant and incomplete operator set. The Warsaw basis Wilson coefficients~($C_i$) are given in terms of the coefficients~($c_i$) of~\cite{Drozd:2015rsp} by:
\begin{align*}
    C_H = c_6 + \lambda \left( c_R + g_2^2 c_W \right) + \lambda^2 c_D \,,
    \quad
    C_{HD} = g_1^2 c_B - 2c_T \,,
    \quad
    C_{H\Box} = \frac{1}{2}\left(c_R - c_H - c_T\right) + \frac{3}{4} g_2^2 c_W + \frac{1}{4} g_1^2 c_B \,,
    \\
    {C_{uH}}^{pr} = Y_u^{pr} \left( \frac{1}{2} c_R + \lambda c_D + \frac{1}{2} g_2^2 c_W \right) \,,
    \quad
    {C_{Hu}}^{pr} = \frac{1}{3} g_1^2 c_B \delta^{pr} \,,
    \quad
    {C_{Hq}^{(1)}}^{pr} = \frac{1}{12} g_1^2 c_B \delta^{pr} \,,
    \quad
    {C_{Hq}^{(3)}}^{pr} = \frac{1}{4} g_2^2 c_W \delta^{pr} \,.
\end{align*}
}
We find agreement except for contributions to operators of the type $(D_\mu W^{\mu\nu}) (D^\rho W_{\rho\nu})$ and $(D_\mu B^{\mu\nu}) (D^\rho B_{\rho\nu})$, which appear to have been neglected in the previous literature~\cite{Drozd:2015rsp,Huo:2015nka}.\footnote{Notice also that the coefficient~$c_H$ in~\cite{Drozd:2015rsp} is missing a $\bar{X}_t^4$~contribution, which is however present in~\cite{Huo:2015nka}. Similarly, the logarithmic $\bar{X}_t^6$~contribution to $c_6$ in~\cite{Drozd:2015rsp} must have opposite sign, which is again correct in~\cite{Huo:2015nka}.}
These operators are generated in the matching by one-loop stop contributions to the gauge-boson self energies.
When projecting onto a minimal basis (such as the Warsaw basis) these operators are removed in favor of additional contributions to several other operators.
This highlights the importance of using complete and non-redundant operator bases when conducting EFT analyses, especially when performing an off-shell matching, which has become feasible in recent years due to the advancement of automated tools.
The full details on the comparison can be found in three \texttt{Mathematica} notebooks that are released together with our results on GitHub~\cite{mssm-to-smeft:github}.

\subsection{A minimal phenomenological example: the stop-bino case}
\label{subsec:stop-bino}
To give a minimal phenomenological example, we consider a scenario where the right-handed stop~($\tilde{t}_R$) and bino~($\tilde{B}$) fields have masses much smaller than those of the other MSSM particles. In this case, the leading contributions to the Wilson coefficients come from stop and bino loops.
Under these assumptions and {\it keeping only the operators relevant for top-pair production at the LHC}, we obtain:
\begin{align}
\mathcal{L}_{\rm SMEFT} \supset {}& 
\left[C_{uG}^{33} \, \frac{1}{g_3} \left(\bar{t}_L \sigma^{\mu \nu} T^{A} t_R \right)\left( H^c G_{\mu \nu}^{A} \right) + \hc \right]
+  C_{G} \, \frac{1}{g_3^3} \, f^{ABC}G_{\mu}^{A\nu}\, G_{\nu}^{B\rho}\, G_{\rho}^{C\mu} \nonumber \\
& + C_{qu}^{(8)3311} (\bar{ t}_L\gamma^\mu T^A t_L)    \left( \bar{Q} \gamma_\mu T^A Q\right) 
+ C_{qu}^{(8)1133} (\bar{  t}_R\gamma_\mu T^A t_R)  \left(\bar{Q} \gamma_\mu T^A Q + \bar{  t}_L\gamma_\mu T^A t_L\right)   \nonumber \\
& +  4 C_{qq}^{(1)3333}  (\bar{  t}_L \gamma_\mu  t_L) (\bar{  t}_L \gamma_\mu  t_L) 
  +  C_{uu}^{3333}(\bar{  t}_R\gamma_\mu  t_R) (\bar{  t}_R\gamma_\mu  t_R) \nonumber \\
& + \frac{1}{4} C_{qd}^{(1)3311} 
 (\bar{t}_L\gamma^\mu  t_L)   \left( 4 \bar{  d}_R\gamma_\mu  d_R  -2\bar{  Q}_L\gamma_\mu  Q _L + 3 \bar{  t}_L \gamma_\mu  t_L \right)  \nonumber \\
& + C_{qu}^{(1)1133} (\bar{ t}_R\gamma^\mu  t_R)  \left(4 \bar{  u}_R\gamma_\mu  u_R - 2\bar{  d}_R\gamma_\mu  d_R + \bar{  Q}_L\gamma_\mu  Q_L \right) \,, \label{eq:stopbinoA}
\end{align}
where $Q = \left( u, d \right)$ is the first generation quark doublet.
This expression makes use of the following relations valid under the stop-bino assumption:
\begin{subequations}\label{eq:coeffMatch}
\begin{eqnarray}
C_{qq}^{(3)1331} + C_{qq}^{(3)3113} & = & C_{qq}^{(1)1331} + C_{qq}^{(1)3113} = \frac{1}{4} C_{qu}^{(8)3311} = \frac{1}{4} C_{qd}^{(8)3311} \,, 
\\
C_{qq}^{(3)3333} & = &  3 C_{qq}^{(1)3333} + \frac{3}{4} C_{qd}^{(1)3311} \,, 
\\
C_{qq}^{(1)1133} + C_{qq}^{(1)3311} & = & -\frac{2}{3} \left(C_{qq}^{(3)1331} + C_{qq}^{(3)3113}\right) -\frac{1}{2} C_{qd}^{(1)3311} \,, 
\\
C_{ud}^{(1)3311} & = & -2 C_{qu}^{(1)1133} \,, 
\\
C_{qu}^{(8)1133}  & = & C_{qu}^{(8)3333}  = C_{ud}^{(8)3311} = 2 \left(  C_{uu}^{1331} +  C_{uu}^{3113}\right) \,, 
\\
C_{uu}^{1133} + C_{uu}^{3311} & = & -\frac{1}{3} \left( C_{uu}^{1331} + C_{uu}^{3113} \right) + 4 C_{qu}^{(1)1133} \,,
\end{eqnarray}
\end{subequations}
where we symmetrize in the Wilson coefficients that belong to identical operator; for example, we have $Q_{uu}^{1133}=Q_{uu}^{3311}$ by interchanging the two identical fermion currents, and similarly for $Q_{qq}^{(1,3)}$.
The relevant Wilson coefficients are given by:
\begin{subequations}\label{eq:coeffExprGen}
\begin{eqnarray}
    C_{uG} ^{33}& = &  - \frac{g_3 Y_{t}}{384 \pi^2} \frac{8 g_{1}^2}{9} \frac{1}{\mtRp{2}} \frac{1}{\left( 1-x \right)^4}\left[ 1 - 6 x + 3 x^2 + 2 x^3 - 6 x^2 \log(x) \right], \\
    C_{G} & = &  \frac{g_3^3}{5760 \pi^2} \frac{1}{\mtRp{2}} \,, \\    
    C_{qu}^{(8)3311} & = &   -\frac{g_3^4}{960 \pi^2} \frac{1}{\mtRp{2}} \,, \\
    C_{qu}^{(8)1133} & = &  -\frac{g_3^4}{960 \pi^2} \frac{1}{\mtRp{2}} \nonumber \\ 
    & & + \frac{g_3^2}{576 \pi^2} \frac{8 g_{1}^2}{9} 
 \frac{1}{\mtRp{2}} \frac{1}{\left( 1-x \right)^4}\left[ 2 - 9 x + 18 x^2  - 11 x^3 + 6 x^3 \log(x) \right] \,, \\
    C_{qq}^{(1)3333} & = &  -\frac{1}{4}\left[\frac{g_3^4}{5760 \pi^2} \frac{1}{\mtRp{2}}  + \frac{1}{5120 \pi^2} \left(\frac{8 g_{1}^2}{9} \right)^2 \frac{1}{\mtRp{2}}\right] , \\
    C_{uu}^{3333} & = &  -\frac{g_3^4}{5760 \pi^2} \frac{1}{\mtRp{2}} - \frac{1}{1280 \pi^2} \left(\frac{8 g_{1}^2}{9} \right)^2 \frac{1}{\mtRp{2}} \nonumber \\
    & & + \frac{g_3^2}{1728 \pi^2} \frac{8 g_{1}^2}{9} \frac{1}{\mtRp{2}} \frac{1}{\left( 1-x \right)^4}\left[ 2 - 9 x + 18 x^2  - 11 x^3 + 6 x^3 \log(x) \right] \nonumber \\
    & & + \frac{1}{1152 \pi^2} \left(\frac{8 g_{1}^2}{9} \right)^2 \frac{1}{\mtRp{2}} \frac{1}{\left( 1-x \right)^4} \left[ -7 - 36 x + 99 x^2 - 56x^3 \right. \nonumber \\ 
    & & \hspace{6cm}\left. + 6 x \left( -6 + 3 x + 4 x^2 \right) \log(x)\right] , \\    
    C_{qd}^{(1)3311} & = & \frac{1}{5120 \pi^2} \left(\frac{8 g_{1}^2}{9} \right)^2 \frac{1}{\mtRp{2}} \,, \\
    C_{qu}^{(1)1133} & = & -\frac{1}{2560 \pi^2} \left(\frac{8 g_{1}^2}{9} \right)^2 \frac{1}{\mtRp{2}}  \\
    & & + \frac{1}{4608 \pi^2} \left(\frac{8 g_{1}^2}{9}\right)^2 \frac{1}{\mtRp{2}} \frac{1}{\left( 1-x \right)^4}\left[ 2 - 9 x + 18 x^2  - 11 x^3 + 6 x^3 \log(x) \right] 
    \,, 
    \nonumber 
\end{eqnarray}
\end{subequations}
where $x \equiv m_{1}^2/\mtRp{2}$. 

In Ref.~\cite{Lessa:2023tqc}, a similar simplified top partner plus dark matter ($\tilde{t}$--$\chi$) model was considered,  but the $\tilde{t}$--$\chi$ coupling~($y_{DM}$) was taken to be a free parameter and $\chi$~was a Dirac fermion.
Setting $g_1 = \frac{3}{2\sqrt{2}} y_{DM}$ in Eqs.~\eqref{eq:coeffExprGen} and keeping only the terms induced by the stop-bino coupling, i.e.,\ neglecting pure QCD terms or terms induced by electroweak gauge bosons, the above coefficients simplify to:
\begin{subequations}
\begin{eqnarray}
    C_{uG} ^{33}& = &  - \frac{g_3 Y_{t} y_{DM}^2}{384 \pi^2}  \frac{1}{\mtRp{2}} \frac{1}{\left( 1-x \right)^4}\left[ 1 - 6 x + 3 x^2 + 2 x^3 - 6 x^2 \log(x) \right], \\ 
    C_{qu}^{(8)1133} & = & \frac{g_3^2 y_{DM}^2}{576 \pi^2} 
 \frac{1}{\mtRp{2}} \frac{1}{\left( 1-x \right)^4}\left[ 2 - 9 x + 18 x^2  - 11 x^3 + 6 x^3 \log(x) \right] , \\
    C_{uu}^{3333} & = & \frac{g_3^2 y_{DM}^2}{1728 \pi^2}  \frac{1}{\mtRp{2}} \frac{1}{\left( 1-x \right)^4}\left[ 2 - 9 x + 18 x^2  - 11 x^3 + 6 x^3 \log(x) \right] \nonumber \\
    & & - \frac{y_{DM}^4}{128 \pi^2} \frac{1}{\mtRp{2}} \frac{1}{\left( 1-x \right)^3} \left[ 1 + 4 x - 5 x^2 + 2 x(2 +x) \log(x)\right] , \\    
    C_{G} & = & C_{qu}^{(8)3311} = C_{qq}^{(1)3333} = C_{qd}^{(1)3311} = C_{qu}^{(1)1133} =  0 \,,
\end{eqnarray}%
\label{eq:coeffExprSimp} %
\end{subequations}%

\noindent
which agrees with the expressions obtained in Ref.~\cite{Lessa:2023tqc} except for $C_{uu}^{3333}$. The difference in the result for $C_{uu}^{3333}$ is due to the Majorana nature of $\tilde{B}$ in the MSSM and the fact that in Ref.~\cite{Lessa:2023tqc} only the $y_{DM}^4$ term was kept.

\begin{figure}[t]
    \centering
    \includegraphics[width=\linewidth]{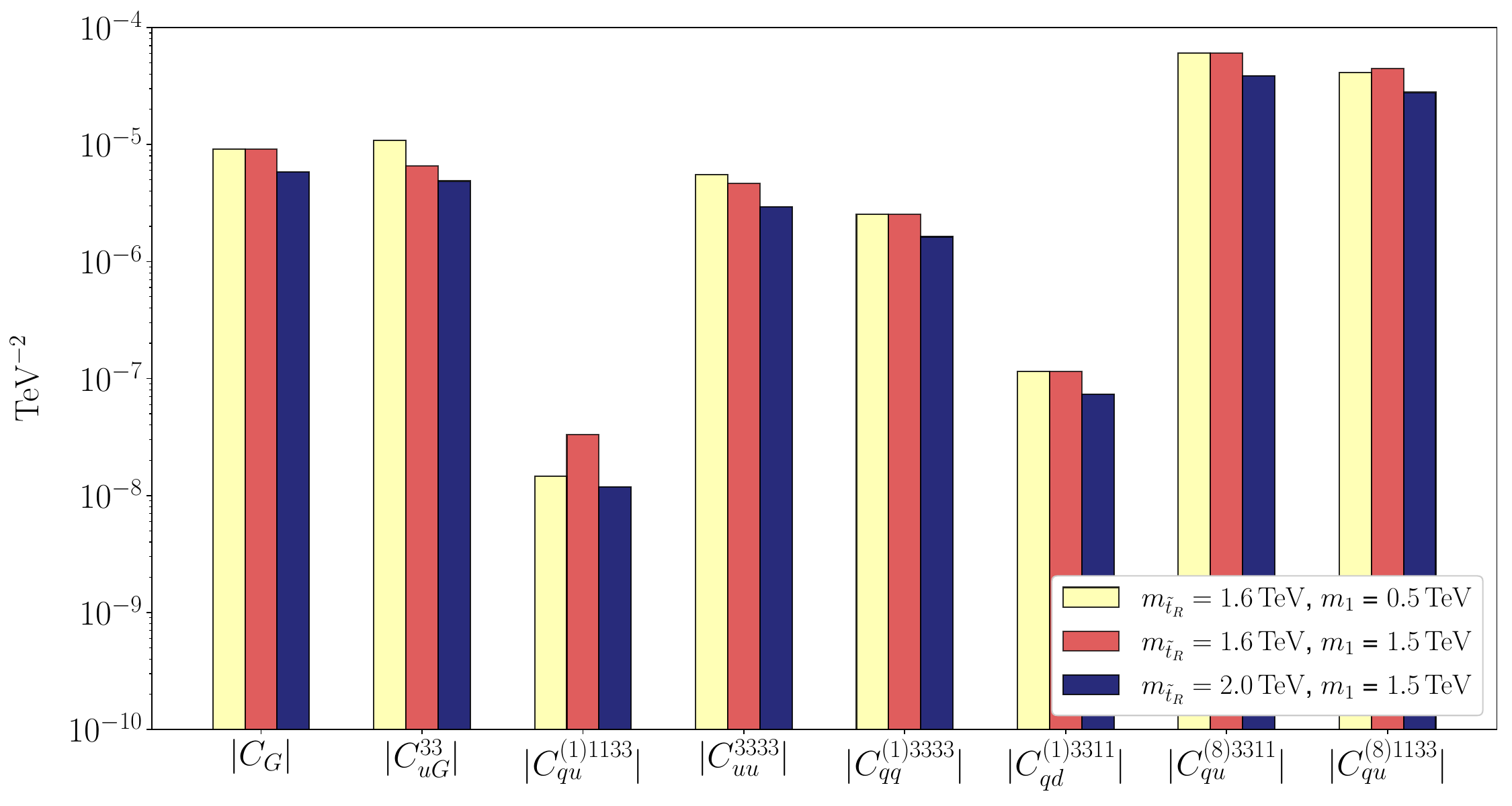}
    \caption{Bar-plot highlighting the order of magnitudes for the various Wilson coefficients relevant for top-pair production at the LHC in the stop-bino model. Considered are three sets of  ($m_{\tilde{t}_R}$, $m_1$) values:   (1.6, 0.5), (1.6, 1.5) and (2.0, 1.5)~TeV. The relevant gauge couplings are taken at 1~TeV, $g_1 \sim 0.37$, $g_3 \sim 1.1$.}
    \label{fig:bar-plot}
\end{figure}

\begin{figure}[t]
    \centering
    \includegraphics[width=0.5\linewidth]{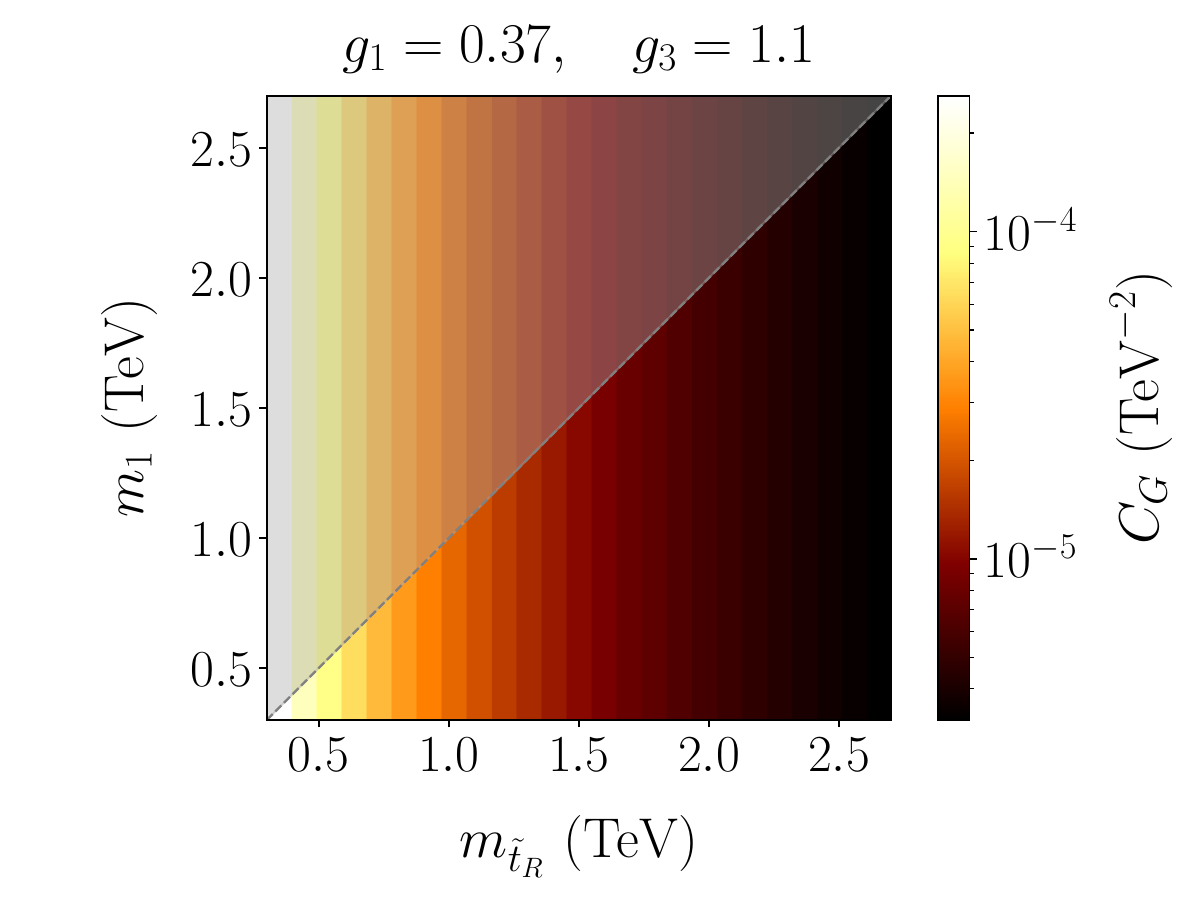}\includegraphics[width=0.5\linewidth]{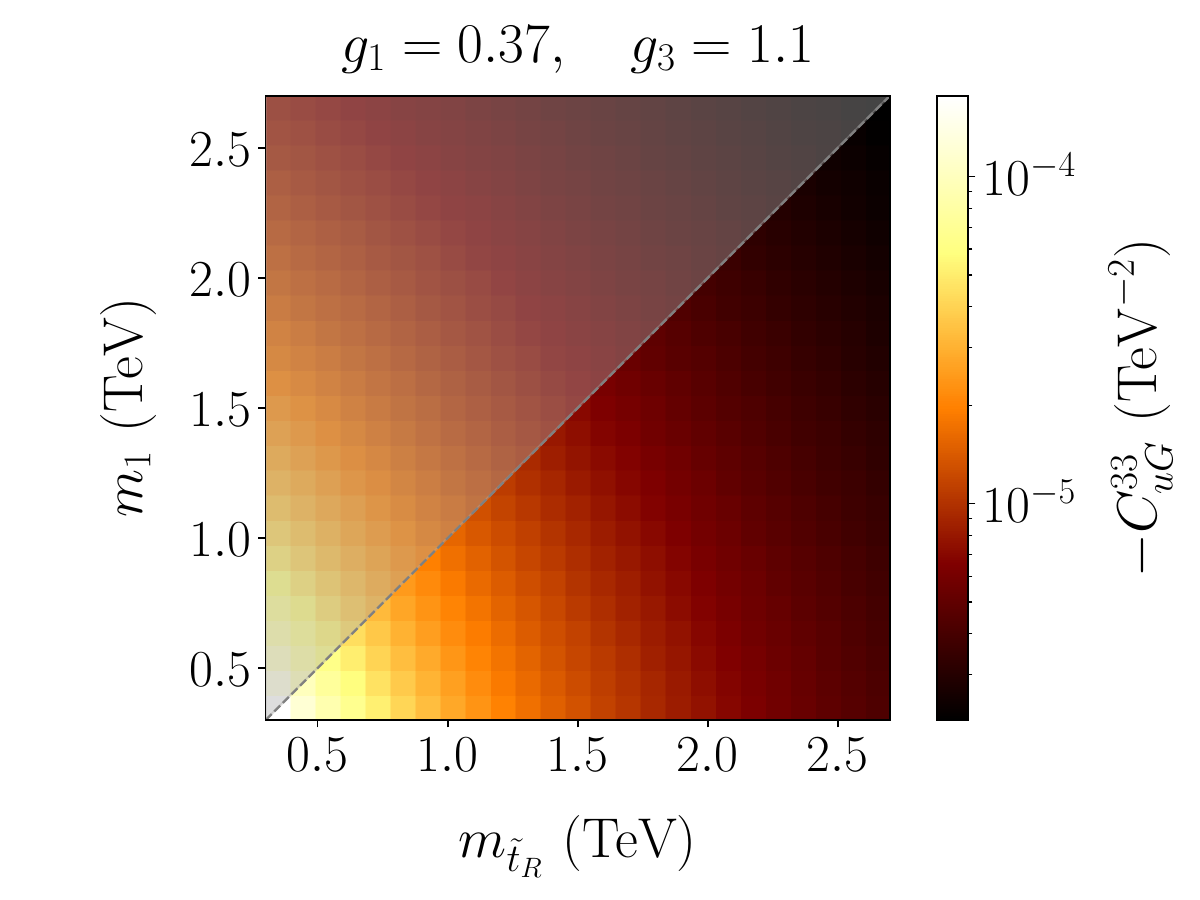}\\
    \includegraphics[width=0.5\linewidth]{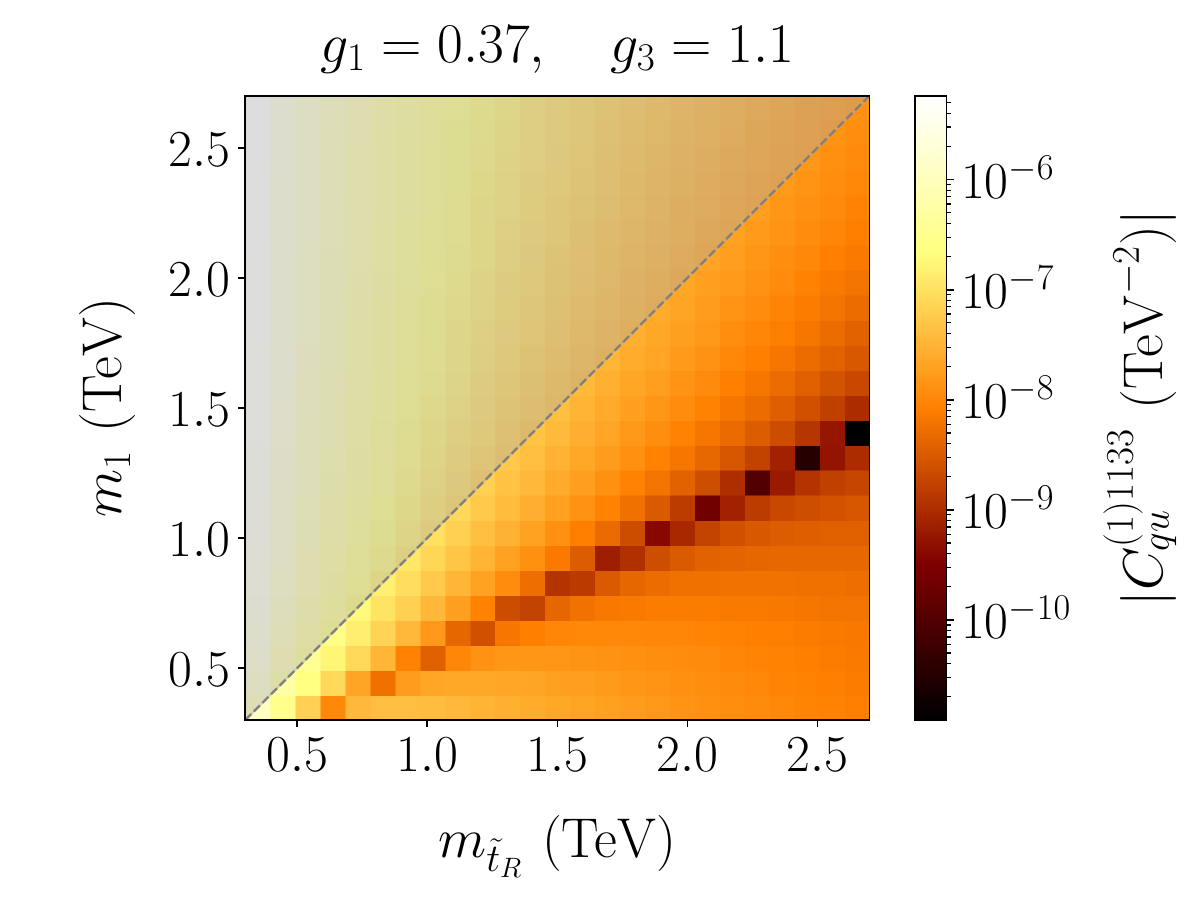}\includegraphics[width=0.5\linewidth]{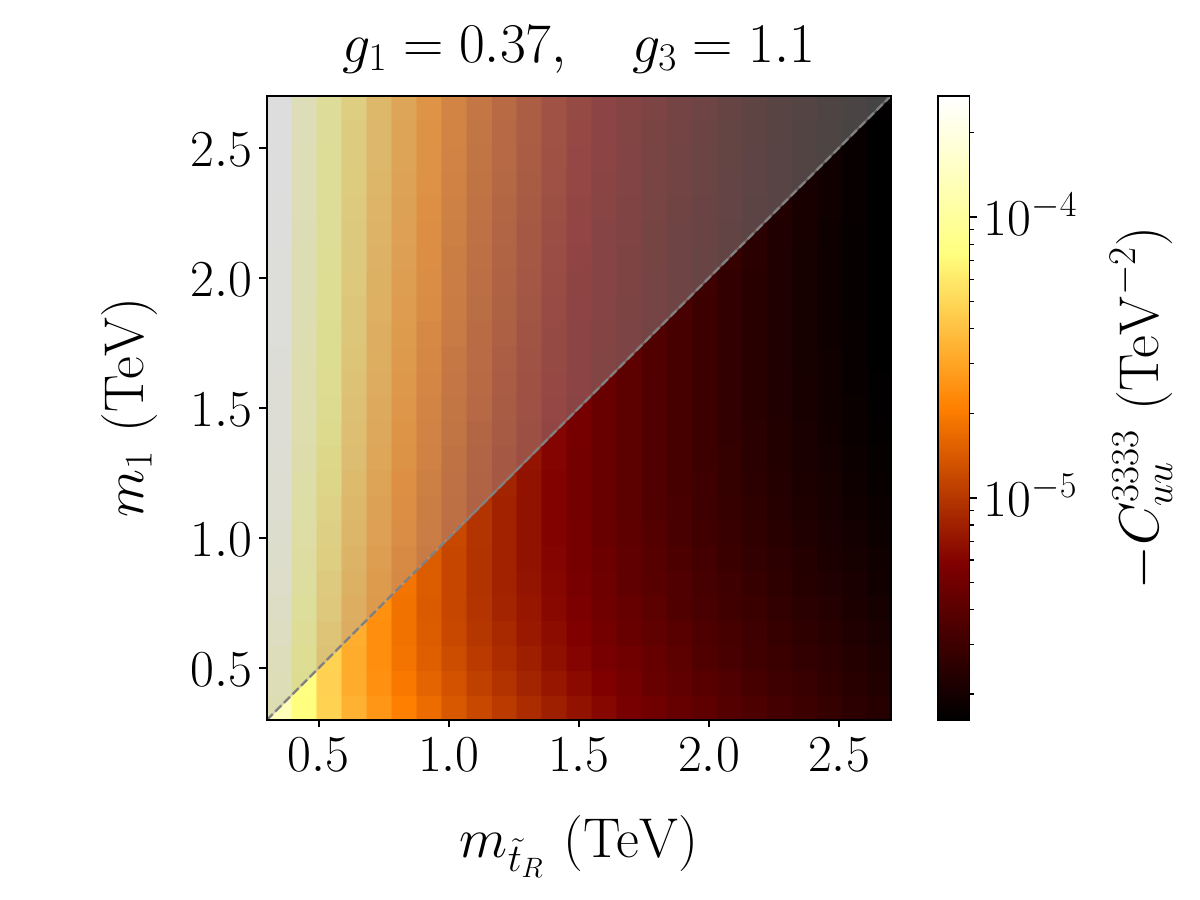}
    \caption{Two-dimensional plots highlighting the patterns of variation of the Wilson coefficients $C_G$, $C_{uG}^{33}$, $C_{qu}^{(1)1133}$, and $C_{uu}^{3333}$ in the ($m_{\tilde{t}_R}$, $m_1$) plane. The shaded regions correspond to $m_1 > m_{\tilde{t}_R}$.}
    \label{fig:2d-plots}
\end{figure}

The magnitude of the Wilson coefficients in Eqs.~\eqref{eq:coeffExprGen} is illustrated in Fig.~\ref{fig:bar-plot} for three choices of $m_{\tilde{t}_R}$ and $m_1$. 
For each case, $g_1$~and $g_3$ are fixed to their values at 1~TeV. 
As expected, the coefficients $C_G$, $C_{qq}^{(1)3333}$, $C_{qd}^{(1)3311}$ and $C_{qu}^{(8)3311}$ are insensitive to variations in~$m_1$. 
Moreover, for fixed $m_1/m_{\tilde t_R}$, all coefficients decrease with $1/m_{\tilde t_R}^2$. 
For the first four coefficients in the bar chart, Fig.~\ref{fig:2d-plots} further illustrates their dependence on~$(m_{\tilde{t}_R}, m_1)$ by means of two-dimensional plots. 
The insensitivity to $m_1$ is confirmed for $C_G$, and also $C_{uu}^{3333}$ depends only little on $m_1$. 
$C_{uG}^{33}$ and $C_{qu}^{(1)1133}$, on the other hand, show interesting patterns. In particular, for $C_{qu}^{(1)1133}$, the \emph{sign} depends on the relative size of $m_{\tilde t_R}$ and $m_1$, and the two mass parameters can conspire such that this Wilson coefficient basically vanishes.

Overall, we see that, as typical for Wilson coefficients that are generated at the loop level only, their size is small and clearly below the current LHC sensitivity~\cite{ATL-PHYS-PUB-2022-037, VandenBossche:2025nvi}. 
Indeed, even the largest values in Figs.~\ref{fig:bar-plot} and \ref{fig:2d-plots} are around two orders of magnitude smaller than those considered in Ref.~\cite{Lessa:2023tqc}. The reason is that Ref.~\cite{Lessa:2023tqc} treated $y_{DM}$ as a free parameter and explored the parameter space corresponding to large $y_{DM}$ values; such large values are not possible in the stop-bino model, where $y_{DM}\propto g_1$.

A comparison between the largest absolute values of the Wilson coefficients in Fig.~\ref{fig:2d-plots}  and the $95\%$ confidence level (CL) bounds from model-independent global fits reveals the following: 
\begin{itemize}
    \item The bounds deduced for $|C_G|$ based on global analyses on flavor universal SMEFT [i.e., with $\mathrm{U}(3)^5$ symmetry] are of ${\cal O}(10^{-1})-{\cal O}(1)$ \cite{Bartocci:2024fmm}. This is around three orders of magnitude weaker than the largest values for $|C_G|$ computed within our setup.

    \item The strongest constraints on (the imaginary part of) $C_{uG}^{33}$ are of the order ${\cal O}(10^{-3})-{\cal O}(10^{-2})$, based on Electric-Dipole-Moment measurements \cite{Aguilar-Saavedra:2018ksv}. These are one order of magnitude weaker than our largest results for $|C_{uG}^{33}|$.

    \item The coefficient $|C_{qu}^{(1)1133}|$ has been studied in top-specific \cite{terHoeve:2025gey,Elmer:2023wtr} as well as $\mathrm{U}(3)^5$ symmetric SMEFT setups \cite{Bartocci:2024fmm} and the constraints are of ${\cal O}(10^{-1})$.  With the scenario considered here, however, the leading contribution to this coefficient is quartic in $g_1$, leading to very tiny values even for small stop masses.

    \item Similarly, the largest values we obtain for $|C_{uu}^{3333}|$, of ${\cal O}(10^{-4})$, are notably smaller than the strongest constraints on this coefficient from model-independent fits, which are of~${\cal O}(1)$ \cite{Bartocci:2024fmm,terHoeve:2025gey}.
\end{itemize}

\begin{figure}[t]
    \centering
    \includegraphics[width=\linewidth]{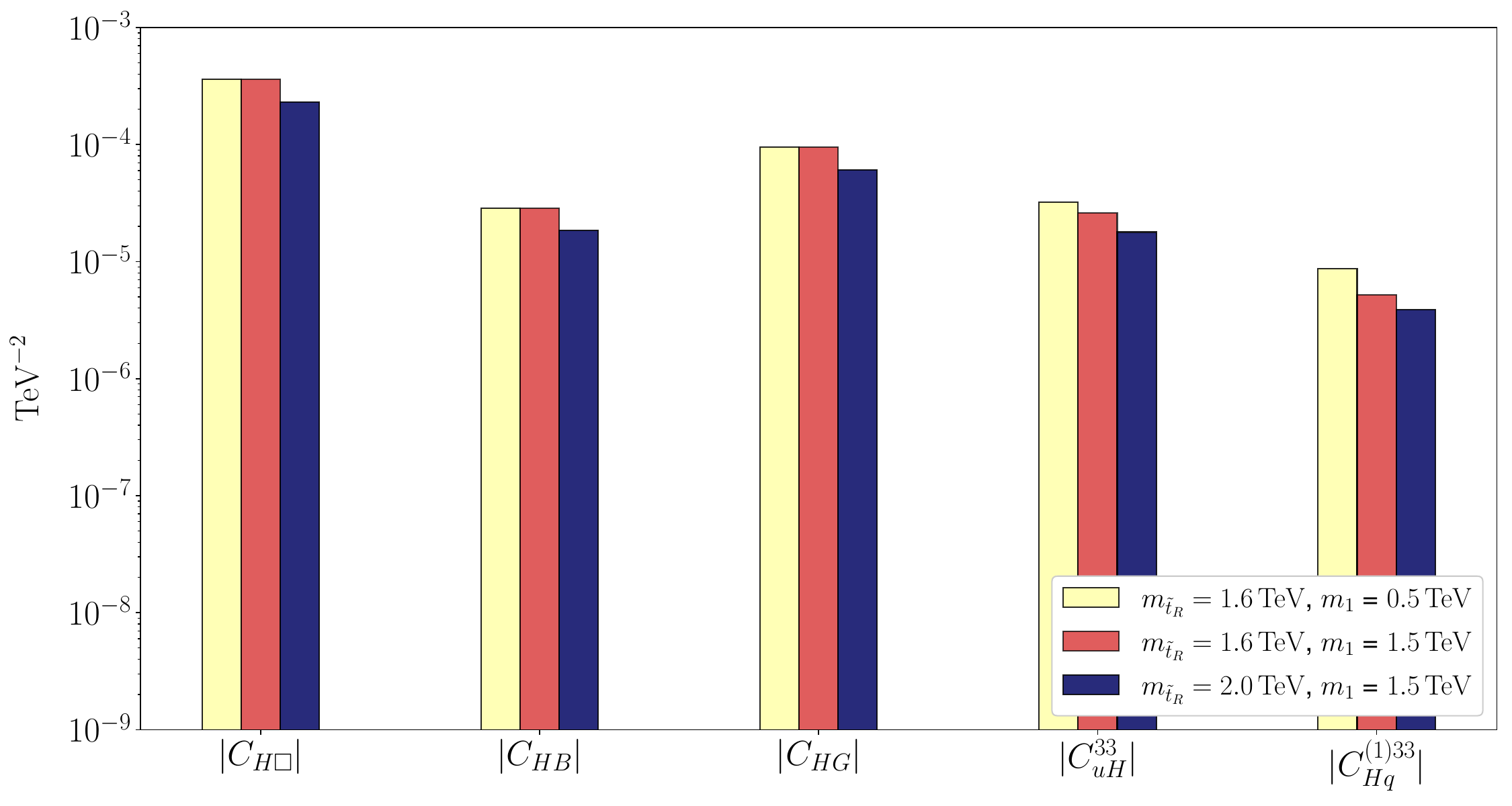}
    \caption{Bar-plot highlighting the order of magnitudes for the Wilson coefficients relevant for Higgs physics in the stop-bino model, for the same sets of  ($m_{\tilde{t}_R}$, $m_1$) as in Fig.~\ref{fig:bar-plot}. The relevant gauge couplings are again taken at 1~TeV, roughly $g_1 = 0.37$, $g_3 = 1.1$.}
    \label{fig:bar-plot-2}
\end{figure}

Looking beyond top physics, integrating out $\tilde t_R$ and $\tilde B$ will also affect the Higgs sector. The magnitudes of the Wilson coefficients relevant in the context of Higgs physics, that have non-zero matching expressions for our stop-bino model, are shown in Fig.~\ref{fig:bar-plot-2}. Of the five  coefficients, three ($C_{H\square}$, $C_{HB}$, $C_{HG}$) depend only on $m_{\tilde t_R}$, while two ($C_{uH}^{33}$, $C_{Hq}^{(1)33}$) depend on both $m_{\tilde t_R}$ and $m_1$. The largest values are attained for $C_{H\square}$, but also these do not exceed a few $\times 10^{-4}$. Overall, the stop-bino model leads to exceedingly small Wilson coefficients, likely below the sensitivity even of a future $e^+e^-$ Higgs factory~\cite{deBlas:2019rxi,deBlas:2022ofj,Celada:2024mcf}. Additional effects, including additional non-zero Wilson coefficients such such as $C_W$ and $C_{HW}$, could be generated through loop contributions of the $\mathrm{SU}(2)$ doublet $(\tilde t_L, \tilde b_L)$, but we do not expect this to dramatically change the picture.

% Conclusions
\section{Conclusions}
\label{sec:conclusion}

We presented the complete one-loop matching of the general $R$-parity conserving MSSM onto the $B$- and $L$-conserving dimension-six SMEFT Lagrangian, done in a fully automated manner with the \matchete package.
Our results take into account all subleading contributions by all superpartners and include all correlations among the different SMEFT Wilson coefficients that are governed by supersymmetry, thus extending previous results for MSSM matching in the literature.

While the one-loop matching of UV theories onto the SMEFT has become standard in the literature, this is the first time a model of such complexity has been treated with an automated tool. 
Indeed, our work necessitated several performance improvements in \matchete, from an optimization of operator reduction and simplification to an improved memory consumption. Moreover, a consistent treatment of heavy flavored particles was incorporated.
Finally, going beyond the SMEFT, we also provided the complete matching onto the 2HDM-EFT.

Our implementation of the MSSM in \matchete is available on \href{https://github.com/BSM-EFT/MSSM-to-SMEFT}{GitHub}~\cite{mssm-to-smeft:github} together with the full matching results in the form of Mathematica and PDF files.
We also provide example Mathematica notebooks showing how to use these results. 
In addition to the validation of our results against known analytical results from the literature, we also provided  numerical results for a minimal phenomenological example, concretely a stop-bino scenario. To enable such numerical analyses, the matched analytical expressions for the full MSSM, produced by \matchete, were translated to C++ classes and methods by means of the \texttt{OperatorToC++} tool, also available on \href{https://github.com/BSM-EFT/OperatorToCpp}{GitHub}~\cite{operator-to-cpp:github}, which will be documented in detail in a separate publication.
  
The work presented here can be extended in several interesting ways. 
First of all, at the moment all SUSY particles are integrated out at the same scale. In general scenarios, however, where the masses of the superpartners span several orders of magnitude (as, for instance, in Split SUSY), a multi-scale matching approach is in order. 
Concretely, one should first integrate out the heaviest superpartners, before using the RGEs of the resulting EFT to resum large logarithms while evolving all couplings to the mass scale of the next lightest superpartners.
The latter should then be removed from the spectrum in a second matching step, and so on, until all BSM fields are integrated out and the remaining low-energy EFT corresponds to the SMEFT. 
(This was partially done in~\cite{Bahl:2019wzx} for the case of Higgs-mass calculation.)
Such a multi-scale analysis is, however, beyond the scope of the current work and thus left for future investigation.

We also note that it would be interesting to consider scenarios where some of the superpartners are retained in the EFT spectrum. This is relevant, for instance, in the case of light Higgsinos (with or without light stops) in the context of natural SUSY as discussed in~\cite{Baer:2015rja}. (See also~\cite{Aebischer:2017aqa}.)
For such scenarios, the corresponding operator bases are unknown, but functional methods provide the ideal tool, as the relevant operators are directly obtained through the matching with no extra cost.
Again, we leave this for future work.

\acknowledgments
We are grateful to Lohan Sartore for collaboration in initial stages of this work.
FW is grateful to J.~Fuentes-Mart\'in, M.~K\"onig, J.~Pag\`es, and A.~E.~Thomsen for the successful collaboration on the \matchete package and fruitful discussions regarding this project.
We thank Admir Greljo for valuable discussions regarding the issue of the sfermion and fermion mass-basis misalignment.

This work was supported in part by the French ANR under project number ANR-22-CE31-0022-02 (EFTatLHC). 
SP acknowledges support by the Agencia Estatal de Investigacion (MCIU/AEI/10.13039/501100011033), the ERDF/EU and the European Union NextGenerationEU/PRTR, through grants PID2020-114473GB-I00 and PID2023-146220NB-I00, CEX2023-001292-S and CNS2022-135595. 
FW acknowledges support by the Deutsche Forschungs\-ge\-mein\-schaft (DFG, German Research Foundation) under grant 396021762 – TRR~257: \textit{Particle Physics Phenomenology after the Higgs Discovery}. 
AL is supported by FAPESP grants
no.\ 2018/25225-9 and 2021/01089-1; 
he also acknowledges financial supprt by the Universit\'e Grenoble Alpes for a one-month visiting professorship in the context of their ``Campagne d’accueil de scientifiques \'etrangers 2025''. Finally, 
AL, SP and FW thank the LPSC Grenoble for hospitality during research visits related to this project. 

\appendix

\section{Conventions}
\label{app:conventions}

This appendix collects all conventions employed in this work.
For that, we adapt the conventions used by the \matchete package~\cite{Fuentes-Martin:2022jrf} to ensure the compatibility of our discussion here with the one-loop matching conditions derived by the code.

We take the Minkowski metric with the \textit{mostly minus-sign} convention given by $g^{\mu\nu}=\mathrm{diag}(+1,-1,-1,-1)^{\mu\nu}$ and the totally anti-symmetric Levi-Civita symbol $\epsilon^{\mu\nu\rho\sigma}$ is defined by $\epsilon^{0123}=-\epsilon_{0123}=+1$.
For the Dirac algebra we choose the chiral Weyl basis with
\begin{align}
	\gamma^\mu &= 
    \begin{pmatrix}
		0 & \sigma^\mu 
        \\[0.2cm]
		\overline{\sigma}^\mu & 0
	\end{pmatrix} \,,
	&
	\gamma^5 &= 
    \begin{pmatrix}
		-\mathbb{1} & 0 
        \\[0.2cm]
		0 & +\mathbb{1}
	\end{pmatrix}\,,
    \label{eq:gamma-matrices}
\end{align}
where $\sigma^{1,2,3}$ are the Pauli matrices with $\sigma^0=\mathbb{1}_{2 \times 2}$ and we have $\overline{\sigma}^0=\sigma^0$ while $\overline{\sigma}^k=-\sigma^k$ for $k\in\{1,2,3\}$. 
In this convention, the charge conjugation matrix is given by $C=i\gamma^2\gamma^0$, and the left- and right-chirality projectors assume the form $P_{L,R}=\frac{1}{2}(\mathbb{1}\mp\gamma^5)$.
Loop integrals are computed in dimensional regularization with $D=4-2\epsilon$ space-time dimensions.
For the treatment of~$\gamma^5$ in $D$~dimensions, we employ the Naive Dimensional Regularization~(NDR) scheme~\cite{Korner:1991sx,Kreimer:1989ke,Nicolai:1980km}, where $\gamma^5$ is taken as fully anti-commuting $\{\gamma^\mu,\gamma^5\}=0$, and $\mathrm{tr}(\gamma^\mu \gamma^\nu \gamma^\rho \gamma^\sigma \gamma^5)=-4i\epsilon^{\mu\nu\rho\sigma}$ is enforced by hand to ensure the correct four-dimensional limit.
This breaks the cyclicity of traces with a $\gamma^5$ and six or more $\gamma^\mu$~ matrices.
This requires the introduction of a reading point, which must be consistently chosen between the matching and subsequent calculations within the EFT~\cite{Fuentes-Martin:2022vvu}. 
For the present analysis, the reading point prescription of \matchete~\cite{Fuentes-Martin:2022jrf} is used.
Spinor indices are suppressed throughout this work.
Furthermore, we introduce the notation $\psi_1 \cdot \psi_2 \equiv \psi_1^\intercal \varepsilon \psi_2$ for the anti-symmetric $\mathrm{SU}(2)$-contraction of two Weyl spinors, where $\varepsilon$~is the totally anti-symmetric tensor of the $\mathrm{SU}(2)$~group. 
Notice that this notation is only used for anti-symmetric spinor contractions. 
For anti-symmetric $\mathrm{SU}(2)_L$ contractions the $\varepsilon$~tensor is displayed explicitly.
For any $\mathrm{SU}(2)$~group, we define the totally anti-symmetric tensor by $\varepsilon = \left(\begin{smallmatrix} 0 & 1 \\ -1 & 0 \end{smallmatrix}\right) = i\sigma^2$ or $\varepsilon^{ij}=-\varepsilon^{ji}$ with $\varepsilon^{12}=+1$.
As usual, $x^\intercal$,~$x^\ast$, and~$x^\dagger$ denote the transposed, complex conjugate, and Hermitian conjugate of the quantity~$x$, respectively.

The covariant derivative is defined as $D_\mu = \partial_\mu - i A_\mu^a T^a$, where the gauge coupling has been absorbed into the gauge fields~$A_\mu^a$ in our convention, while $T^a$~is the generator in the appropriate representation. 
We normalize the generators of the fundamental representation by $\mathrm{tr}\big(T^a_F\,T^b_F\big)=\frac{1}{2}\delta^{ab}$.
Thus, the field-strength tensor of the corresponding gauge group reads $F_{\mu\nu} = \partial_\mu A_\nu^a - \partial_\nu A_\mu^a + f^{abc} A_\mu^b A_\nu^c$ with the respective structure constants~$f^{abc}$.

For completeness, we list the definitions of all SMEFT operators in the Warsaw basis~\cite{Grzadkowski:2010es}, which we adopt for this analysis, in Table~\ref{tab:Warsaw-basis}.
The ones shown in gray are not generated at one loop in the MSSM. 
The full SMEFT Lagrangian in the Warsaw basis then reads
\begin{align}
    \cL_{\mathrm{Warsaw}}
    =
    &-\frac{1}{4 g_3^2} G^A_{\mu\nu} G^{A\,\mu\nu}
    -\frac{1}{4 g_2^2} W^I_{\mu\nu} W^{I\,\mu\nu}
    -\frac{1}{4 g_1^2} B_{\mu\nu} B^{\mu\nu}
    \\
    &+ i \, \bar{q} \slashed{D} q
    + i \, \bar{u} \slashed{D} u
    + i \, \bar{d} \slashed{D} d
    + i \, \bar{\ell} \slashed{D} \ell
    + i \, \bar{e} \slashed{D} e
    \nonumber\\
    &+ (D_\mu H)^\dagger (D^\mu H) 
    - M_H^2 \, H^\dagger H 
    - \frac{\lambda}{2} (H^\dagger H)^2
    \nonumber\\
    &- \Big[ 
    (\bar{q} \, Y_u \, u) H^c 
    + (\bar{q} \, Y_d \, d) H
    + (\bar{\ell} \, Y_e \, e) H
    + \hc \Big]
    \nonumber\\
    &+ \sum_{k \, \in \, \text{Tab.}\,\ref{tab:Warsaw-basis}} C_k Q_k
    \,,
    \nonumber
\end{align}
where the sum in the last line runs over all operators~$Q_k$ listed in Tab.~\ref{tab:Warsaw-basis} with the corresponding Wilson coefficients denoted by~$C_k$. 
Operators violating~$B$ or~$L$ cannot be generated in the $R$-parity-conserving MSSM and are therefore neglected in this work.
In addition, we also have $H^c = \varepsilon H^\ast$.
Notice again that in our conventions the gauge coupling is absorbed into the corresponding gauge field.
Therefore, our definition of the Warsaw basis operators containing field-strength tensors differs by additional inverse powers of the corresponding gauge couplings compared to the definition in~\cite{Grzadkowski:2010es}, in order to ensure that the Wilson coefficients in our work and in~\cite{Grzadkowski:2010es} agree.

\begin{table}[!p]
	\centering
	\newcommand{\OpScale}{.735} %scaling of the table
	\renewcommand{\arraystretch}{1.5}
	\scalebox{\OpScale}{%
	\centering
	\begin{tabular}{| lc || lc || lc | lc |}
		\multicolumn{8}{c}{1--4: Bosonic Operators} \\[.1cm] \hline
		\multicolumn{2}{|c||}{1: $X^3$ } & 
		\multicolumn{2}{c||}{2: $H^6$ } &
		\multicolumn{4}{c|}{4: $X^2 H^2$ } \\ \hline
		$Q_{G}$ & $\frac{1}{g_3^3} f^{ABC} G_\mu^{A\nu} G_\nu^{B\rho} G_\rho^{C\mu}$ &	
		$Q_{H}$ & $(H^\dagger H)^3$ &
		$Q_{HG}$ & $\frac{1}{g_3^2} (H^\dagger H) G_{\mu\nu}^A G^{A\mu\nu}$ &
		$Q_{H B}$ & $\frac{1}{g_1^2} (H^\dagger H) B_{\mu\nu} B^{\mu\nu}$ \\ \cline{3-4}
		{\color{gray} $Q_{\tilde G}$} & {\color{gray} $\frac{1}{g_3^3} f^{ABC} \tilde G_\mu^{A\nu} G_\nu^{B\rho} G_\rho^{C\mu}$} &
		\multicolumn{2}{c||}{3: $H^4 D^2$ } &
		{\color{gray} $Q_{H \tilde G}$} & {\color{gray} $\frac{1}{g_3^2} (H^\dagger H) \tilde G_{\mu\nu}^A G^{A\mu\nu}$} &
		{\color{gray} $Q_{H \tilde B}$} & {\color{gray} $\frac{1}{g_1^2} (H^\dagger H) \tilde B_{\mu\nu} B^{\mu\nu}$} \\ \cline{3-4}
		$Q_{W}$ & $\frac{1}{g_2^3} \varepsilon^{IJK} W_\mu^{I\nu} W_\nu^{J\rho} W_\rho^{K\mu}$ &	
		$Q_{H\Box}$ & $(H^\dagger H) \Box (H^\dagger H)$ & 
		$Q_{HW}$ & $\frac{1}{g_2^2} (H^\dagger H) W_{\mu\nu}^I W^{I\mu\nu}$ &
		$Q_{H W B}$ & $\frac{1}{g_2 g_1} (H^\dagger \sigma^I H) W_{\mu\nu}^I B^{\mu\nu}$ \\ 
		{\color{gray} $Q_{\tilde W}$} & {\color{gray} $\frac{1}{g_2^3} \varepsilon^{IJK} \tilde W_\mu^{I\nu} W_\nu^{J\rho} W_\rho^{K\mu}$} & 
		$Q_{HD}$ & $(H^\dagger D_\mu H)^\ast (H^\dagger D^\mu H)$ & 
		{\color{gray} $Q_{H \tilde{W}}$} & {\color{gray} $\frac{1}{g_2^2} (H^\dagger H) \tilde W_{\mu\nu}^I W^{I\mu\nu}$} & 
		{\color{gray} $Q_{H \tilde W B}$} & {\color{gray} $\frac{1}{g_2 g_1} (H^\dagger \sigma^I H) \tilde W_{\mu\nu}^I B^{\mu\nu}$} \\[0.1cm]
		\hline
	\end{tabular}
	}
	\vspace{0.3cm}
	\\ \centering
	\scalebox{\OpScale}{%
	\centering
	\begin{tabular}{| lc || lc | lc | lc |}
		\multicolumn{8}{c}{5--7: Fermion Bilinears $(\psi^2)$} \\[.1cm] \hline
		\multicolumn{8}{|c|}{non-Hermitian $(\bar L R)$} \\ \hline
		\multicolumn{2}{|c||}{5: $\psi^2 H^3$ + H.c. } & \multicolumn{6}{c|}{6: $\psi^2 X H$ + H.c. }  \\ \hline
		$Q_{eH}$ & $(H^\dagger H)(\bar\ell_p e_r H)$ &	
		$Q_{eW}$ & $\frac{1}{g_2} (\bar\ell_p \sigma^{\mu\nu}e_r)\sigma^I H W_{\mu\nu}^I$ &
		$Q_{uG}$ & $\frac{1}{g_3} (\bar q_p \sigma^{\mu\nu}T^A u_r) H^c G_{\mu\nu}^A$ &
		$Q_{dG}$ & $\frac{1}{g_3} (\bar q_p \sigma^{\mu\nu}T^A d_r)H G_{\mu\nu}^A$ \\ 
		$Q_{uH}$ & $(H^\dagger H)(\bar q_p u_r H^c)$ &
		$Q_{eB}$ & $\frac{1}{g_1} (\bar\ell_p \sigma^{\mu\nu}e_r) H B_{\mu\nu}$ &
		$Q_{uW}$ & $\frac{1}{g_2} (\bar q_p \sigma^{\mu\nu}u_r)\sigma^I H^c W_{\mu\nu}^I$ &
		$Q_{dW}$ & $\frac{1}{g_2} (\bar q_p \sigma^{\mu\nu}d_r)\sigma^I H W_{\mu\nu}^I$ \\ 
		$Q_{dH}$ & $(H^\dagger H)(\bar q_p d_r H)$ &		
		& & 
		$Q_{uB}$ & $\frac{1}{g_1} (\bar q_p \sigma^{\mu\nu}u_r) H^c B_{\mu\nu}$ &
		$Q_{dB}$ & $\frac{1}{g_1} (\bar q_p \sigma^{\mu\nu}d_r) H B_{\mu\nu}$ \\ \hline
		\end{tabular}
	}%
	\vspace{0.1cm} \newline \centering
	\scalebox{\OpScale}{%
	\begin{tabular}{| lc | lc | lc |}
		\hline 
		\multicolumn{6}{|c|}{7: $\psi^2 H^2 D$ ~~~--~~~ Hermitian + $Q_{Hud}$ } \\ \hline
		\multicolumn{2}{|c|}{$(\bar L L)$} & 
		\multicolumn{2}{c|}{$(\bar R R)$} & 
		\multicolumn{2}{c|}{$(\bar R R^\prime)$ + H.c.} \\ \hline
		$Q_{Hl}^{(1)}$ & $(H^\dagger i \overleftrightarrow{D}_\mu H)(\bar\ell_p \gamma^\mu \ell_r)$ & 
		$Q_{H e}$ & $(H^\dagger i \overleftrightarrow{D}_\mu H)(\bar e_p \gamma^\mu e_r)$ & 
		$Q_{Hud}$ & $i(H^{c\,\dagger} D_\mu H)(\bar u_p \gamma^\mu d_r)$ \\
		$Q_{Hl}^{(3)}$ & $(H^\dagger i \overleftrightarrow{D}_\mu^I H)(\bar\ell_p \sigma^I\gamma^\mu \ell_r)$ & 
		$Q_{H u}$ & $(H^\dagger i \overleftrightarrow{D}_\mu H)(\bar u_p \gamma^\mu u_r)$ & 
		& \\
		$Q_{Hq}^{(1)}$ & $(H^\dagger i \overleftrightarrow{D}_\mu H)(\bar q_p \gamma^\mu q_r)$ & 
		$Q_{H d}$ & $(H^\dagger i \overleftrightarrow{D}_\mu H)(\bar d_p \gamma^\mu d_r)$ & 
		& \\
		$Q_{Hq}^{(3)}$ & $(H^\dagger i \overleftrightarrow{D}_\mu^I H)(\bar q_p \sigma^I\gamma^\mu q_r)$ & 
		& & 
		& \\ \hline
	\end{tabular}
	}%
	\vspace{0.3cm}
	\\ \centering
	\scalebox{\OpScale}{%
	\setlength{\tabcolsep}{1.8mm}
	\begin{tabular}{| lc | lc | lc || lc |}
		\multicolumn{8}{c}{8: Fermion Quadrilinears $(\psi^4)$ } \\[.1cm] \hline
		\multicolumn{6}{|c||}{Hermitian} & \multicolumn{2}{c|}{non-Hermitian} \\ \hline
		\multicolumn{2}{|c|}{$(\bar L L)(\bar L L)$} & 
		\multicolumn{2}{c|}{$(\bar R R)(\bar R R)$} & 
		\multicolumn{2}{c||}{$(\bar L L)(\bar R R)$} &
		\multicolumn{2}{c|}{$(\bar L R)(\bar L R)$ + H.c.}
		\\ \hline
		$Q_{ll}$ & $(\bar\ell_p \gamma_\mu \ell_r)(\bar\ell_s \gamma^\mu \ell_t)$ & 
		$Q_{ee}$ & $(\bar e_p \gamma_\mu e_r)(\bar e_s \gamma^\mu e_t)$ & 
		$Q_{l e}$ & $(\bar\ell_p \gamma_\mu \ell_r)(\bar e_s \gamma^\mu e_t)$ & 
		$Q_{quqd}^{(1)}$ & $(\bar q_p^i u_r)\varepsilon_{ij}(\bar q_s^j d_t)$ \\
		$Q_{qq}^{(1)}$ & $(\bar q_p \gamma_\mu q_r)(\bar q_s \gamma^\mu q_t)$ & 
		$Q_{uu}$ & $(\bar u_p \gamma_\mu u_r)(\bar u_s \gamma^\mu u_t)$ & 
		$Q_{l u}$ & $(\bar\ell_p \gamma_\mu \ell_r)(\bar u_s \gamma^\mu u_t)$ &
		$Q_{quqd}^{(8)}$ & $(\bar q_p^i T^A u_r)\varepsilon_{ij}(\bar q_s^j T^A d_t)$ \\
		$Q_{qq}^{(3)}$ & $(\bar q_p \gamma_\mu \sigma^I q_r)(\bar q_s \gamma^\mu \sigma^I q_t)$ & 
		$Q_{dd}$ & $(\bar d_p \gamma_\mu d_r)(\bar d_s \gamma^\mu d_t)$ & 
		$Q_{l d}$ & $(\bar\ell_p \gamma_\mu \ell_r)(\bar d_s \gamma^\mu d_t)$ &
		$Q_{l equ}^{(1)}$ & $(\bar \ell_p^i e_r)\varepsilon_{ij}(\bar q_s^j u_t)$ \\
		$Q_{l q}^{(1)}$ & $(\bar\ell_p \gamma_\mu \ell_r)(\bar q_s \gamma^\mu q_t)$ & 
		$Q_{eu}$ & $(\bar e_p \gamma_\mu e_r)(\bar u_s \gamma^\mu u_t)$ & 
		$Q_{q e}$ & $(\bar q_p \gamma_\mu q_r)(\bar e_s \gamma^\mu e_t)$ &
		$Q_{l equ}^{(3)}$ & $(\bar \ell_p^i \sigma_{\mu\nu} e_r)\varepsilon_{ij}(\bar q_s^j \sigma^{\mu\nu} u_t)$ \\
		$Q_{l q}^{(3)}$ & $(\bar\ell_p \gamma_\mu \sigma^I \ell_r)(\bar q_s \gamma^\mu \sigma^I q_t)$ & 
		$Q_{ed}$ & $(\bar e_p \gamma_\mu e_r)(\bar d_s \gamma^\mu d_t)$ & 
		$Q_{qu}^{(1)}$ & $(\bar q_p \gamma_\mu q_r)(\bar u_s \gamma^\mu u_t)$ & 
		& \\ 
		& & 
		$Q_{ud}^{(1)}$ & $(\bar u_p \gamma_\mu u_r)(\bar d_s \gamma^\mu d_t)$ & 
		$Q_{qu}^{(8)}$ & $(\bar q_p \gamma_\mu T^A q_r)(\bar u_s \gamma^\mu T^A u_t)$ & & \\ \cline{7-8}
		& & 
		$Q_{ud}^{(8)}$ & $(\bar u_p \gamma_\mu T^A u_r)(\bar d_s \gamma^\mu T^A d_t)$ & 
		$Q_{qd}^{(1)}$ & $(\bar q_p \gamma_\mu q_r)(\bar d_s \gamma^\mu d_t)$ &
		\multicolumn{2}{c|}{$(\bar L R)(\bar R L)$ + H.c.} \\ \cline{7-8}
		& & 
		& & 
		$Q_{qd}^{(8)}$ & $(\bar q_p \gamma_\mu T^A q_r)(\bar d_s \gamma^\mu T^A d_t)$ &
		$Q_{l e d q }$ & $(\bar \ell _p^i e_r)(\bar d_s q_{ti})$ \\ \hline
	\end{tabular}
	}%
	\caption{%
    List of all baryon- and lepton-number conserving SMEFT operators at mass-dimension six in the Warsaw basis~\cite{Grzadkowski:2010es}. 
    The division into {classes~1--8} is adopted from~\cite{Alonso:2013hga}.
    The operators shown in gray are not generated at one loop in the MSSM.
    Notice that operators containing field-strength tensors are normalized by appropriate inverse powers of the corresponding gauge coupling due to our convention, which absorbs the gauge coupling into the gauge field.
    Including this normalization in the operators ensures that the Wilson coefficients agree between the different conventions.
    Further note that $\sigma^I$ denotes the Pauli matrices such that the generators of~$\mathrm{SU}(2)_L$ take the form $T^I=\frac{1}{2}\sigma^I$, which should not be confused with $\sigma^{\mu\nu} = \frac{i}{2} [\gamma^\mu,\gamma^\nu]$ in our notation.
    Table adapted from~\cite{Isidori:2023pyp}.
	\label{tab:Warsaw-basis}
	}
\end{table}

\section{From Weyl to Dirac spinors}
\label{app:Weyl-Dirac}
In this appendix, we collect all formulae required to translate the MSSM Lagrangian between the two- and four-component spinor notation, as presented in Sec.~\ref{sec:MSSM-Lagrangians}.
We start with a generic Dirac spinor~$\psi$ and decompose it into a left-handed~$\psi_L$ and a right-handed~$\psi_R$ Weyl spinor, leading to
\begin{subequations}
\begin{align}
    \psi &= \begin{pmatrix}
        \psi_L \\ \psi_R
    \end{pmatrix}\,, 
    &
    \psi^c &= C {\overline{\psi}}^{\,\intercal} = \begin{pmatrix}
        i \sigma^2 \psi_R^\ast \\ -i \sigma^2 \psi_L^\ast
    \end{pmatrix} \equiv \begin{pmatrix}
        \psi_L^c \\ \psi_R^c
    \end{pmatrix}\,, 
    \\[0.2cm]
    \overline{\psi} &= \begin{pmatrix}
        \psi_R^\dagger & \psi_L^\dagger
    \end{pmatrix}\,,
    &
    \overline{\psi^c} &= \psi^\intercal C = \begin{pmatrix}
        i \psi_L^\intercal \sigma^2 & -i \psi_R^\intercal \sigma^2
    \end{pmatrix}\,.
\end{align}  
\end{subequations}
Using the expressions for the Dirac matrices in the Weyl representation from Eq.~\eqref{eq:gamma-matrices}, it is then straight forward to express the kinetic terms of chiral Dirac fermions in terms of Weyl fermions 
\begin{subequations}
\begin{align}
    i (\overline{\psi} \slashed{D} P_L \psi) 
    &=
    i \psi_L^\dagger \overline{\sigma}^\mu D_\mu \psi_L \,,
    \\
    i (\overline{\psi} \slashed{D} P_R \psi) 
    &=
    i \psi_R^\dagger \sigma^\mu D_\mu \psi_R
    = 
    i {(\psi_L^c)}^\dagger \overline{\sigma}^\mu D_\mu \psi_L^c \,,
\end{align}
\end{subequations}
where in the last equality we expressed the right-handed Weyl spinor $\psi_R$ in terms of its charge conjugate $\psi_L^c = i \sigma^2 \psi_R^\ast$, which is left handed.

Similarly, we can decompose a Majorana fermion~$\Psi$ in terms of the left-handed Weyl spinor~$\chi$ and its charge conjugate~$\chi^c=-i\sigma^2\chi^\ast$, which is right-handed, finding
\begin{align}
    \Psi &= \begin{pmatrix}
        \chi \\ \chi^c
    \end{pmatrix} \,,
    &
    \overline{\Psi} &= \begin{pmatrix}
       {\chi^c}^\dagger & \chi^\dagger
    \end{pmatrix} \,,
    &
    \Psi^c &= \Psi \,.
\end{align}
For the kinetic term of a Majorana fermion we thus find
\begin{align}
    \frac{i}{2} \overline{\Psi} \slashed{D} \Psi
    &= i \chi^\dagger \overline{\sigma}^\mu D_\mu \chi \,.
\end{align}

Next, we consider the fermion bilinears without derivatives present in the interaction terms. 
For the Majorana fermions we get the fermion current
\begin{align}
    \overline{\Psi}\Psi &= \chi \cdot \chi + \chi^\dagger \cdot \chi^\dagger \,,
\end{align}
whereas for the Dirac fermions we find
\begin{subequations}
\begin{align}
    \overline{\psi_1} P_L \psi_2 &= \psi_{1,R}^c \cdot \psi_{2,L} \,,
    &
    \overline{\psi_1} P_R \psi_2 &= {(\psi_{2,R}^c)}^\dagger \cdot {(\psi_{1,L})}^\dagger \,,
    \\
    \overline{\psi_1^c} P_L \psi_2 &= \psi_{1,L} \cdot \psi_{2,L} \,,
    &
    \overline{\psi_1^c} P_R \psi_2 &= \psi_{1,R} \cdot \psi_{2,R} \,,
    \\
    \overline{\psi_1} P_L \psi_2^c &= \psi_{1,R}^c \cdot \psi_{2,R}^c \,,
    &
    \overline{\psi_1} P_R \psi_2^c &= \psi_{1,L}^c \cdot \psi_{2,L}^c \,.
\end{align}
\end{subequations}
For the mixed fermion bilinears we obtain
\begin{subequations}
\begin{align}
    \overline{\psi} P_L \Psi &= \psi_L^c \cdot \chi \,,
    &
    \overline{\psi} P_R \Psi &= \psi_L^\dagger \cdot \chi^\dagger \,,
    \\
    \overline{\Psi} P_L \psi &= \chi \cdot \psi_L \,,
    &
    \overline{\Psi} P_R \psi &= \chi^\dagger \cdot {\psi_L^c}^\dagger \,.
\end{align}
\end{subequations}

\section{Full MSSM Lagrangian in the soft-SUSY mass basis}
\label{app:MSSM-Lagrangian}
For completeness, we now provide the full MSSM Lagrangian, expressed in terms of four-component Dirac and Majorana spinors, and written in the soft-SUSY mass basis, i.e., the mass eigenstate basis after soft SUSY breaking but before EWSB,
\begin{align}
    \cL_\mathrm{MSSM}
    &=
    \cL_\mathrm{MSSM}^\mathrm{free} 
    + \cL_\mathrm{MSSM}^\mathrm{Herm} 
    + \big[ \cL_\mathrm{MSSM}^\mathrm{non-Herm} + \hc \big]
    + \cL_\mathrm{MSSM}^\mathrm{reg}
    \,,
    \label{eq:full-MSSM-Lag}
\end{align}
where the Lagrangian is split for convenience into a piece~$\cL_\mathrm{MSSM}^\mathrm{free}$ containing kinetic and mass terms, a Hermitian part~$\cL_\mathrm{MSSM}^\mathrm{Herm}$, a non-Hermitian one~$\cL_\mathrm{MSSM}^\mathrm{non-Herm}$, and a part containing the $\overline{\mathrm{DR}}$--$\overline{\mathrm{MS}}$ scheme change counterterms.
For simplicity we write $g_i^{\scriptscriptstyle \overline{\rm MS}} = g_i$ in the equations below.
The free part reads
\begin{align}
    \cL_\mathrm{MSSM}^\mathrm{free} 
    =
    &- \frac{1}{4 \, g_3^2} G_{\mu\nu}^A G^{\mu\nu,A}
    - \frac{1}{4 \, g_2^2} W_{\mu\nu}^I W^{\mu\nu,I}
    - \frac{1}{4 \, g_1^2} B_{\mu\nu} B^{\mu\nu}
    \\
    &+\frac{i}{2} \overline{\tilde{G}}^A \slashed{D} \tilde{G}^A - \frac{m_3}{2} \overline{\tilde{G}}^A \tilde{G}^A
    +\frac{i}{2} \overline{\tilde{W}}^I \slashed{D} \tilde{W}^I - \frac{m_2}{2} \overline{\tilde{W}}^I \tilde{W}^I
    +\frac{i}{2} \overline{\tilde{B}} \slashed{D} \tilde{B} - \frac{m_1}{2} \overline{\tilde{B}} \tilde{B}
    \nonumber\\
    &+ i\overline{q} \slashed{D} P_L q
    + i\overline{u} \slashed{D} P_R u
    + i\overline{d} \slashed{D} P_R d
    + i\overline{\ell} \slashed{D} P_L \ell
    + i\overline{e} \slashed{D} P_R e
    \nonumber\\
    &+ (D_\mu \tilde{q})^\dagger (D^\mu \tilde{q}) - \tilde{q}^\dagger m_{\tilde{q}}^2 \tilde{q}
    + (D_\mu \tilde{u})^\dagger (D^\mu \tilde{u}) - \tilde{u}^\dagger m_{\tilde{u}}^2 \tilde{u}
    + (D_\mu \tilde{d})^\dagger (D^\mu \tilde{d}) 
    \nonumber\\
    &- \tilde{d}^\dagger m_{\tilde{d}}^2 \tilde{d} + (D_\mu \tilde{\ell})^\dagger (D^\mu \tilde{\ell}) - \tilde{\ell}^\dagger m_{\tilde{\ell}}^2 \tilde{\ell}
    + (D_\mu \tilde{e})^\dagger (D^\mu \tilde{e}) - \tilde{e}^\dagger m_{\tilde{e}}^2 \tilde{e}
    \nonumber\\
    &+ (D_\mu H)^\dagger (D^\mu H) - m_H^2 H^\dagger H
    + (D_\mu \Phi)^\dagger (D^\mu \Phi) - m_\Phi^2 \Phi^\dagger \Phi
    + i\overline{\Sigma} \slashed{D} \Sigma - \tilde{\mu} \overline{\Sigma} \Sigma
    \,.
    \nonumber
\end{align}
The Hermitian terms of the full MSSM Lagrangian are given by
\begin{align}
    \cL_\mathrm{MSSM}^\mathrm{Herm}
    =
    &- c_{2\gamma}^2 \frac{g_1^2 + g_2^2}{8} (H^\dagger H)^2
    \\
    &- \tilde{q}^\dagger \! \left( s_\gamma^2 y_u y_u^\dagger + c_{2\gamma} \frac{3 g_2^2 - g_1^2}{12} \right) \! \tilde{q} \, (H^\dagger H)
    + (\tilde{q}^\dagger H) \! \left( s_\gamma^2 y_u y_u^\dagger - c_\gamma^2 y_d y_d^\dagger + c_{2\gamma} \frac{g_2^2}{2} \right) \! (H^\dagger \tilde{q})
    \nonumber\\
    &- \tilde{u}^\dagger \left( s_\gamma^2 y_u^\dagger y_u + c_{2\gamma} \frac{g_1^2}{3} \right) \tilde{u} \, (H^\dagger H)
    - \tilde{d}^\dagger \left( c_\gamma^2 y_d^\dagger y_d - c_{2\gamma} \frac{g_1^2}{6} \right) \tilde{d} \, (H^\dagger H)
    \nonumber\\
    &- c_{2\gamma} \frac{g_1^2 + g_2^2}{4} (\tilde{\ell}^\dagger \tilde{\ell}) \, (H^\dagger H)
    - (\tilde{\ell}^\dagger H) \left( c_\gamma^2 y_e y_e^\dagger - c_{2\gamma} \frac{g_2^2}{2} \right) (H^\dagger \tilde{\ell})
    \nonumber\\
    &- \tilde{e}^\dagger \left( c_\gamma^2 y_e^\dagger y_e - c_{2\gamma} \frac{g_1^2}{2} \right) \tilde{e} \, (H^\dagger H)
    \nonumber\\
    &+ \frac{g_2^2 (c_{4\gamma}-3) + g_1^2 (c_{4\gamma}+1)}{8} (\Phi^\dagger \Phi) (H^\dagger H)
    + \frac{g_2^2 (c_{4\gamma}+3) + g_1^2 (c_{4\gamma}-1)}{8} (\Phi^\dagger H) (H^\dagger \Phi)
    \nonumber\\
    &- \tilde{q}^\dagger \left( c_\gamma^2 y_u y_u^\dagger - c_{2\gamma} \frac{3 g_2^2 - g_1^2}{12} \right) \tilde{q} \, (\Phi^\dagger \Phi)
    + (\tilde{q}^\dagger \Phi) \! \left( c_\gamma^2 y_u y_u^\dagger - s_\gamma^2 y_d y_d^\dagger - c_{2\gamma} \frac{g_2^2}{2} \right) \! (\Phi^\dagger \tilde{q})
    \nonumber\\
    &- \tilde{u}^\dagger \left( c_\gamma^2 y_u^\dagger y_u - c_{2\gamma} \frac{g_1^2}{3} \right) \tilde{u} \, (\Phi^\dagger \Phi)
    - \tilde{d}^\dagger \left( s_\gamma^2 y_d^\dagger y_d + c_{2\gamma} \frac{g_1^2}{6} \right) \tilde{d} \, (\Phi^\dagger \Phi)
    \nonumber\\
    &+ c_{2\gamma} \frac{g_2^2 + g_1^2}{4} (\tilde{\ell}^\dagger \tilde{\ell}) (\Phi^\dagger \Phi)
    - (\tilde{\ell}^\dagger \Phi) \left( s_\gamma^2 y_e y_e^\dagger + c_{2\gamma} \frac{g_2^2}{2} \right) (\Phi^\dagger \tilde{\ell}) 
    \nonumber\\
    &- \tilde{e}^\dagger \left( s_\gamma^2 y_e^\dagger y_e + c_{2\gamma} \frac{g_1^2}{2} \right) \tilde{e} \, (\Phi^\dagger \Phi)
    - c_{2\gamma}^2 \frac{g_2^2 + g_1^2}{8} (\Phi^\dagger \Phi)^2
    \nonumber\\
    %%%
    &{\color{gray} + \frac{6 g_3^2 + 9 g_2^2 - g_1^2}{72} (\tilde{q}^{\dagger} \tilde{q}) (\tilde{q}^{\dagger} \tilde{q})
    - \frac{1}{4} \left[ g_3^2 (\delta^{ps} \delta^{rt}) + g_2^2 (\delta^{pt} \delta^{rs}) \right] \tilde{q}_{ai}^{s\dagger} \tilde{q}^{ajr} \tilde{q}_{bj}^{t\dagger} \tilde{q}^{bip}}
    \nonumber\\
    &{\color{gray} + \frac{1}{36} \left[ 3 g_3^2 (\delta^{pt}\delta^{rs} - 3\delta^{ps}\delta^{rt}) - 8 g_1^2 \delta^{pt}\delta^{rs} \right] (\tilde{u}^{s\dagger} \tilde{u}^{r}) (\tilde{u}^{t\dagger} \tilde{u}^{p})}
    \nonumber\\
    &{\color{gray} + \frac{1}{36} \left[ 3 g_3^2 (\delta^{pt}\delta^{rs}-3\delta^{ps}\delta^{rt}) - 2 g_1^2 \delta^{pt}\delta^{rs}\right] (\tilde{d}^{s\dagger} \tilde{d}^{r}) (\tilde{d}^{t\dagger} \tilde{d}^{p})}
    \nonumber\\
    &{\color{gray} + \frac{1}{8} \left[ g_2^2 (\delta^{pt}\delta^{rs}-2\delta^{ps}\delta^{rt}) - g_1^2 \delta^{pt}\delta^{rs}\right] (\tilde{\ell}^{s\dagger} \tilde{\ell}^{r}) (\tilde{\ell}^{t\dagger} \tilde{\ell}^{p})
    - \frac{g_1^2}{2} (\tilde{e}^\dagger \tilde{e}) (\tilde{e}^\dagger \tilde{e})}
    \nonumber\\
    &{\color{gray} - \frac{3 g_3^2 - 2 g_1^2}{18} (\tilde{q}^\dagger \tilde{q}) (\tilde{u}^\dagger \tilde{u})
    - \left( y_u^{rs\ast} y_u^{tp} - \frac{g_3^2}{2} \delta^{ps} \delta^{rt} \right) (\tilde{q}^{t\dagger} \tilde{u}^{p}) (\tilde{u}^{s\dagger} \tilde{q}^{r})}
    \nonumber\\
    &{\color{gray} - \frac{3 g_3^2 + g_1^2}{18} (\tilde{q}^\dagger \tilde{q}) (\tilde{d}^\dagger \tilde{d})
    - \left( y_d^{rs\ast} y_d^{tp} - \frac{g_3^2}{2} \delta^{ps} \delta^{rt} \right) (\tilde{q}^{t\dagger} \tilde{d}^{p}) (\tilde{d}^{s\dagger} \tilde{q}^{r})}
    \nonumber\\
    &{\color{gray} 
    + \frac{3 g_2^2 + g_1^2}{12} (\tilde{q}^\dagger \tilde{q}) (\tilde{\ell}^\dagger \tilde{\ell})
    - \frac{g_2^2}{2} (\tilde{q}^\dagger \tilde{\ell}) (\tilde{\ell}^\dagger \tilde{q})
    - \frac{g_1^2}{6} (\tilde{q}^\dagger \tilde{q}) (\tilde{e}^\dagger \tilde{e})
    }
    \nonumber\\
    &{\color{gray}
    + \frac{3 g_3^2 + 4 g_1^2}{18} (\tilde{u}^\dagger \tilde{u}) (\tilde{d}^\dagger \tilde{d})
    - \frac{g_3^2}{2} (\tilde{d}^\dagger \tilde{u}) (\tilde{u}^\dagger \tilde{d})
    - \frac{g_1^2}{3} (\tilde{u}^\dagger \tilde{u}) (\tilde{\ell}^\dagger \tilde{\ell})
    + \frac{2 g_1^2}{3} (\tilde{u}^\dagger \tilde{u}) (\tilde{e}^\dagger \tilde{e})
    }
    \nonumber\\
    &{\color{gray}
    + \frac{g_1^2}{6} (\tilde{d}^\dagger \tilde{d}) (\tilde{\ell}^\dagger \tilde{\ell})
    - \frac{g_1^2}{3} (\tilde{d}^\dagger \tilde{d}) (\tilde{e}^\dagger \tilde{e})
    - \left( y_e^{rs\ast} y_e^{tp} - \frac{g_1^2}{2} \delta^{ps} \delta^{rt} \right) (\tilde{\ell}^{t\dagger} \tilde{\ell}^r) (\tilde{e}^{s\dagger} \tilde{e}^p)
    }
    \,,\nonumber
\end{align}
where the terms in the last nine rows (shown in gray) cannot contribute to the one-loop matching at mass-dimension six, but only at higher orders in perturbation theory.\footnote{%
Terms containing a single heavy field contribute to the tree-level matching. 
Such terms are only present for the heavy Higgs~$\Phi$ in the non-Hermitian part of the Lagrangian.
Terms containing two heavy states, except for~$\Phi$, start contributing at one loop, whereas terms with three or four heavy fields, except for~$\Phi$, start contributing at two-loop level only.
Terms with a higher number of $\Phi$~fields can contribute at lower loop-orders, since $\Phi$ does not necessarily have to run in the loop, but can also be replaced by its equation of motion.}
The non-Hermitian terms are given by
\begin{align}
\label{eq:full_non-Hermitian_MSSM_Lagrangian}
    \cL_\mathrm{MSSM}^\mathrm{non-Herm}
    =
    &- s_\gamma (\overline{q} \, y_u \, P_R \, u) \varepsilon H^\ast
    - c_\gamma (\overline{q} \, y_d \, P_R \, d) H
    - c_\gamma (\overline{\ell} \, y_e \, P_R \, e) H
    \\
    &- c_\gamma (\overline{q} \, y_u \, P_R \, u) \, \varepsilon \Phi^\ast
    + s_\gamma (\overline{q} \, y_d \, P_R \, d) \Phi
    + s_\gamma (\overline{\ell} \, y_e \, P_R \, e) \Phi
    \nonumber\\
    &+ \tilde{q}^{\dagger} (c_\gamma \tilde{\mu} y_u - s_\gamma a_u) \tilde{u} \, \varepsilon H^\ast
    + \tilde{q}^{\dagger} (s_\gamma \tilde{\mu} y_d - c_\gamma a_d) \tilde{d} H
    + \tilde{\ell}^{\dagger} (s_\gamma \tilde{\mu} y_e - c_\gamma a_e) \tilde{e} H
    \nonumber\\
    &- \tilde{q}^\dagger (s_\gamma \tilde{\mu} y_u + c_\gamma a_u) \tilde{u} \, \varepsilon \Phi^\ast
    + \tilde{q}^\dagger (c_\gamma \tilde{\mu} y_d + s_\gamma a_d) \tilde{d} \Phi
    + \tilde{\ell}^\dagger (c_\gamma \tilde{\mu} y_e + s_\gamma a_e) \tilde{e} \Phi
    \nonumber\\
    &- 
    (\overline{q} \varepsilon P_R \Sigma^c) y_u \tilde{u}
    + (\overline{q} P_R \Sigma) y_d \tilde{d}
    + (\overline{\ell} P_R \Sigma) y_e \tilde{e}
    \nonumber\\
    &- 
    \tilde{q}^\dagger \varepsilon y_u (\overline{\Sigma} P_R u)
    + \tilde{q}^\dagger y_d (\overline{d^c} P_R \Sigma)
    + \tilde{\ell}^\dagger y_e (\overline{e^c} P_R \Sigma)
    \nonumber\\
    &- \sqrt{2} g_2 \big[\,\overline{\Sigma} (c_\gamma P_L + s_\gamma P_R) \tilde{W} \big] H
    - \frac{g_1}{\sqrt{2}} \big[\,\overline{\Sigma} (c_\gamma P_L + s_\gamma P_R) \tilde{B}\big] H
    \nonumber\\
    &- \sqrt{2} \, g_2 \Phi^\dagger \big[ \overline{\tilde{W}^c} (c_\gamma P_L - s_\gamma P_R) \Sigma \big]
    - \frac{g_1}{\sqrt{2}} \Phi^\dagger \big[ \overline{\tilde{B}^c} (c_\gamma P_L - s_\gamma P_R) \Sigma \big]
    \nonumber\\
    &- \sqrt{2} g_3 \tilde{q}^\dagger (\overline{{\tilde{G}}^{c}} P_L q)
    + \sqrt{2} g_3 \tilde{u}^\dagger (\overline{{\tilde{G}}^{c}} P_R u)
    + \sqrt{2} g_3 \tilde{d}^\dagger (\overline{{\tilde{G}}^{c}} P_R d)
    \nonumber\\
    &- \sqrt{2} g_2 \tilde{q}^\dagger (\overline{{\tilde{W}}^{c}} P_L q)
    - \sqrt{2} g_2 \tilde{\ell}^\dagger (\overline{{\tilde{W}}^{c}} P_L \ell)
    - \frac{g_1}{3\sqrt{2}} \tilde{q}^\dagger (\overline{{\tilde{B}}^c} P_L q)
    \nonumber\\
    &+ \frac{2\sqrt{2} \, g_1}{3} \overline{\tilde{u}} (\overline{{\tilde{B}}^c} P_R u)
    - \frac{\sqrt{2} \, g_1}{3} \tilde{d}^\dagger (\overline{{\tilde{B}}^c} P_R d)
    + \frac{g_1}{\sqrt{2}} \tilde{\ell}^\dagger (\overline{{\tilde{B}}^c} P_L \ell)
    - \sqrt{2} g_1 \tilde{e}^\dagger (\overline{{\tilde{B}}^c} P_R e)
    \nonumber\\
    &- \frac{s_{2\gamma}}{12} \tilde{q}^\dagger (6 y_u y_u^\dagger - 3 g_2^2 + g_1^2) \tilde{q} (\Phi^\dagger H)
    + \frac{s_{2\gamma}}{2} (\tilde{q}^\dagger H) (y_d y_d^\dagger + y_u y_u^\dagger - g_2^2) (\Phi^\dagger \tilde{q})
    \nonumber\\
    &+ \frac{s_{2\gamma}}{6} \tilde{u}^\dagger (2 g_1^2 - 3 y_u^\dagger y_u) \tilde{u} (\Phi^\dagger H)
    + \frac{s_{2\gamma}}{6} \tilde{d}^\dagger (3 y_d^\dagger y_d - g_1^2) \tilde{d} (\Phi^\dagger H)
    - (\tilde{u}^\dagger y_u^\dagger y_d \tilde{d}) (\Phi^\intercal \varepsilon H)
    \nonumber\\
    &+ s_{2\gamma} \frac{g_1^2 + g_2^2}{4} (\tilde{\ell}^\dagger \tilde{\ell}) (\Phi^\dagger H)
    + \frac{s_{2\gamma}}{2} (\tilde{\ell}^\dagger H) (y_e y_e^\dagger - g_2^2) (\Phi^\dagger \tilde{\ell}) 
    + \frac{s_{2\gamma}}{2} \tilde{e}^\dagger (y_e^\dagger y_e - g_1^2) \tilde{e} (\Phi^\dagger H)
    \nonumber\\
    &+ s_{4\gamma} \frac{g_1^2 + g_2^2}{8} (\Phi^\dagger H)(H^\dagger H)
    - s_{2\gamma}^2 \frac{g_1^2 + g_2^2}{8} (\Phi^\dagger H)^2
    - s_{4\gamma} \frac{g_1^2 + g_2^2}{8} (\Phi^\dagger \Phi) (\Phi^\dagger H)
    \nonumber\\
    &{\color{gray}- (\tilde{q}^\dagger y_d \tilde{d}) (\tilde{e}^\dagger y_e^\dagger \tilde{\ell})}
    \nonumber
\end{align}
and their Hermitian conjugates are included in Eq.~\eqref{eq:full-MSSM-Lag}. 
Notice also, that we use the notation $\tilde{G}=\tilde{G}^A T^A$ (and similar for the other gauginos), where the generators are not explicitly shown.
Again, the term in gray in the last line can only contribute at two loops.

The $\overline{\mathrm{DR}}$--$\overline{\mathrm{MS}}$ scheme change counterterm Lagrangian reads
\begin{align}
    \label{eq:MSSM_scheme-ct_Lagrangian}
    16\pi^2 \, \cL_\mathrm{MSSM}^\mathrm{reg}
    &= \frac{1}{8}  \left[ g_1^4 + 2 g_1^2 g_2^2 + g_2^4 \left( 3 - \frac{2}{3}c_{2\gamma}^2 \right) \right] (H^\dagger H)^2 
    \\
    &\quad
    + \bigg[ 
    s_{4\gamma}\frac{g_2^4}{12}(\Phi^\dagger H) (H^\dagger H)
    \nonumber\\
    &\qquad\quad 
    - s_\gamma \left( \frac{4}{3} g_3^2 - \frac{3}{8} g_2^2 - \frac{1}{72} g_1^2 \right) (\overline{q} \, y_u \, P_R \, u) \varepsilon H^\ast
    \nonumber\\
    &\qquad\quad
    - c_\gamma \left( \frac{4}{3} g_3^2 - \frac{3}{8} g_2^2 - \frac{13}{72} g_1^2 \right) (\overline{q} \, y_d \, P_R \, d) H
    \nonumber\\
    &\qquad\quad
    - c_\gamma \frac{3}{8} \left( g_1^2 - g_2^2 \right) (\overline{\ell} \, y_e \, P_R \, e) H
    \nonumber\\
    &\qquad\quad
    - s_\gamma \left( \frac{4}{3} g_3^2 - \frac{3}{8} g_2^2 - \frac{1}{72} g_1^2 \right) (\overline{q} \, y_u \, P_R \, u) \varepsilon \Phi^\ast
    \nonumber\\
    &\qquad\quad
    + c_\gamma \left( \frac{4}{3} g_3^2 - \frac{3}{8} g_2^2 - \frac{13}{72} g_1^2 \right) (\overline{q} \, y_d \, P_R \, d) \Phi
    \nonumber\\
    &\qquad\quad
    + c_\gamma \frac{3}{8} \left( g_1^2 - g_2^2 \right) (\overline{\ell} \, y_e \, P_R \, e) \Phi + \mathrm{H.c.} \bigg] + \ldots
    \,, \nonumber
\end{align}
where the term in the second line does not originate from a direct contribution of the $\overline{\mathrm{DR}}$--$\overline{\mathrm{MS}}$ scheme change but from the redefinition of the gauge coupling in Eq.~\eqref{eq:D-term_gaugino_shift}.
The ellipsis denote additional scheme-change counterterms that do not contribute to the one-loop matching at mass-dimension six, since they multiply operators that contribute only at higher orders.

In the above, all indices are suppressed when the contraction is unambiguous, and, if possible, implicit indices are contracted within brackets.
In particular, $\mathrm{SU}(2)_L$, $\mathrm{SU}(3)_c$, and flavor indices are only shown if necessary, and for flavor indices we employ a matrix notation.
Since the Lagrangian is defined in the soft-SUSY mass basis, the $3 \times 3$~matrices $m_{\tilde q,\tilde u,\tilde d,\tilde \ell,\tilde e}$ are taken as diagonal, while $y_{u,d,e}$ and $a_{u,d,e}$ can also have off-diagonal entries.
If shown explicitly, $p,r,s,t$ denote flavor indices, $i,j$ are fundamental $\mathrm{SU}(2)_L$ indices, and $a,b$ represent fundamental $\mathrm{SU}(3)_c$ indices.

\section{Higgs basis and alignment limit}
\label{app:Higgs-basis_alignment-limit}

In Sec.~\ref{sec:Higgs}, we have discussed our treatment of the Higgs sector in the soft-SUSY mass basis.
In addition, we have performed the SMEFT and HEFT matching, finding full agreement for the scalar potential of the physical Higgs boson in both methods, validating our use of the SMEFT approach.
This appendix provides additional details on the Higgs sector in the matching.
First, we review the alignment limit in 2HDM theories, before discussing the matching before EWSB in the so-called \emph{Higgs basis}.
Finally, we compare the results obtained following this approach with those obtained in the soft-SUSY mass basis in Sec.~\ref{sec:Higgs}.

\paragraph{The alignment limit.}
As already discussed at the end of Sec.~\ref{sec:EWSB-MSSM}, we find two generally independent rotation angles to translate between the gauge and the mass eigenstates in the Higgs sector. 
The angles~$\alpha$ (of the neutral CP-even sector) and $\beta$~(for the neutral CP-odd and charged Higgs sector) have been defined in Eqs.~\eqref{eq:rotation-EWSB}--\eqref{eq:rotation-matrices_Higgs-basis}.
In the aforementioned \emph{alignment limit}, which is defined by $\alpha = \beta - \frac{\pi}{2}$, all the rotation matrices in Eq.~\eqref{eq:rotation-matrices_Higgs-basis} agree ({$R_\alpha=R_\beta$}) and hence we are left with one SM-like and one pure BSM Higgs doublet.
If this condition is satisfied, all rotations in Eqs.~\eqref{eq:rotation-EWSB}--\eqref{eq:rotation-matrices_Higgs-basis} can be expressed simply as the single linear transformation shown in Eq.~\eqref{eq:rotation-to-matching-basis} by replacing $\vartheta \to \beta = \alpha + \frac{\pi}{2}$.
Only if this limit holds to good approximation can we match directly onto the SMEFT.
Otherwise, the matching has to be performed onto the more general Higgs Effective Field Theory~(HEFT)~\cite{Feruglio:1992wf,Alonso:2012px,Buchalla:2013rka}, where the physical Higgs~$h^0$ and the Goldstone bosons~$G^{0,\pm}$ are treated separately (reminiscent of the two independent rotation angles) and are not both embedded into a single $\mathrm{SU}(2)_L$-doublet~$H$, such as in the SMEFT. 

As is well known, the \textit{decoupling limit}~\cite{Haber:1989xc,Gunion:2002zf,Haber:2006ue}, i.e., the absence of BSM states at the electroweak scale, which is at the core of our EFT assumption, naturally leads to the \textit{alignment limit}~\cite{Gunion:2002zf,Bernon:2015qea}.
This can be seen by introducing a small parameter~$\delta$, as was done in Eq.~\eqref{eq:misalignment}, as a measure of the deviation from alignment.
Based on the experimental lower bound of $\textrm{m}_{A^0}\gtrsim 500\,\text{GeV}$ for the mass of the CP-odd neutral scalar~$A^0$, obtained by the CMS collaboration~\cite{CMS:2022goy}, we find
\begin{align}
    \delta \equiv \alpha - \beta + \frac{\pi}{2} = \frac{\textrm{m}_Z^2}{\textrm{m}_{A^0}^2} \frac{\sin (4\beta)}{2} + \cO\!\left( \frac{\textrm{m}_Z^4}{\textrm{m}_{A^0}^4} \right) \lesssim 0.015 \,.
\end{align}
This once again demonstrates the power suppression of the misalignment~$\delta = \cO\!\left( \textrm{m}_Z^2 \middle/ \textrm{m}_{A^0}^2 \right)$.
Although a misalignment of this magnitude cannot be probed in current experiments, future experimental facilities, such as a dedicated Higgs factory, could allow testing such a small misalignment. 
A~study of these effects is beyond the scope of the present work.

Notice that when directly integrating out the Higgs doublet~$\Phi$ in the soft-SUSY mass basis, as done in Sec.~\ref{sec:soft-SUSY_Higgs-mass-basis}, we implicitly work close to the alignment limit from the onset, therefore bypassing the complications of rotating the neutral CP-even and CP-odd/charged scalar sectors by different rotation angles, as done in the HEFT scenarios.
Moreover, the EFT approach employed in this work  assumes by definition that the BSM states are heavy and decouple, which, as we have shown, naturally leads to alignment.
The alignment limit also significantly simplifies the EFT analysis, since it allows us to use standard tools for the matching onto SMEFT that are not available for the matching onto HEFT in the misaligned scenario.

\paragraph{Matching in the Higgs basis before EWSB.}
In Sec.~\ref{sec:soft-SUSY_Higgs-mass-basis}, we have rotated the two MSSM Higgs doublets by the angle~$\gamma$, defined in Eq.~\eqref{eq:rotation-angle_mass-basis}, to what we have called the \emph{soft-SUSY mass basis}, where the mass mixing of the doublets vanishes ($\Delta=0$).
A~notable alternative is to rotate the doublets instead by the angle~$\beta$ defined by $\tan\beta=v_u/v_d$, leading to Eq.~\eqref{eq:beta-definition}, which was also done in~\cite{Dawson:2023ebe}.
By replacing~$\vartheta$ by~$\beta$ in Eqs.~\eqref{eq:rotation-to-matching-basis} and~\eqref{eq:MSSM_Lagrangian-matching-basis}, we rotate to what we dub the \emph{unbroken-phase Higgs basis}, defined before EWSB where only a single rotation angle~($\beta$ in our case) can be present.
Notice that the Higgs basis employed in Sec.~\ref{sec:EWSB-MSSM} for the full MSSM has been defined after EWSB instead, where two independent rotation angles~($\alpha,\beta$) are present.
Hence, the two Higgs bases differ by a rotation with the angle~$\delta$ in the neutral CP-even sector.
After rotating the two Higgs doublets by the angle~$\beta$ before EWSB, the mass terms of the SMEFT Higgs potential in Eq.~\eqref{eq:MSSM_Lagrangian-matching-basis} take the form
\begin{subequations}
\begin{align}
    m^{\prime \, 2}_H &= - \frac{\cos^2 (2\beta)}{2} \textrm{m}_Z^2 
    \,,
    \label{eq:Higgs-basis_mass-parameter}
    \\
    m^{\prime \, 2}_\Phi &= \textrm{m}_{A^0}^2 + \frac{\cos^2(2\beta)}{2} \textrm{m}_Z^2
    \,,
    \\
    \Delta^\prime 
    &= \frac{m_{H_u}^2 - m_{H_d}^2}{2} \sin(2\beta) - b \cos(2\beta) 
    = - \tan(2\beta) \, m_H^{\prime\,2}
    = \textrm{m}_Z^2 \, \frac{\sin (4\beta)}{4} \,,
\end{align}
\label{eq:Higgs-basis_masses}%
\end{subequations}
where the primes indicate that these parameters are defined in the unbroken-phase Higgs basis, and should not be confused with the corresponding Lagrangian parameters in the soft-SUSY mass basis, where we also have $\Delta = 0$, for example. 
Notice that in Eq.~\eqref{eq:Higgs-basis_masses} we have expressed the Lagrangian parameters before EWSB in terms of the physical masses after EWSB as computed in the full MSSM [cf. Eqs.~\eqref{eq:Z-mass_Higgs-basis} and~\eqref{eq:physical-masses}] to make the power counting in terms of UV and IR scales manifest ($\textrm{m}_{A^0} \gg \textrm{m}_Z$).
In fact, we find a Higgs mass parameter~$m_H^{\prime\,2}$ of the order of the electroweak scale, which is negative and can therefore trigger EWSB. 
Similarly, we find that the mass mixing of the doublets is suppressed in the EFT power counting, as it is of the order of the electroweak scale as well~$|\Delta'| \leq \frac{1}{4} \textrm{m}_Z^2$.
On the other hand, we obtain a positive and heavy mass squared~$m_\Phi^{\prime\,2}$ for the second doublet. 
This motivates that by identifying and decoupling the latter doublet as the UV~degree of freedom, while treating $m_H^\prime$ and~$\Delta^\prime$ as IR~scales, reproduces the desired low-energy limit.\footnote{Technically, we can treat~$m_H^{(\prime)}$ and~$\Delta^{(\prime)}$ as two-point interactions for the matching (since they are power suppressed and thus expanded out after applying the method of regions), rather than including them in the propagators.}

Next, we can invert Eq.~\eqref{eq:rotation-to-matching-basis} with $\vartheta$ replaced by~$\beta$, which is exactly the definition of the unbroken-phase Higgs basis, and then compare this to the decomposition of the $\mathrm{SU}(2)$~doublets in terms of the physical degrees of freedom in the full MSSM, given in Eq.~\eqref{eq:rotation-EWSB} with $\beta=\beta_\pm=\beta_0$.
This yields
\begin{subequations}
\begin{align}
    H^\prime &= \begin{pmatrix}
        G^+ \\ \frac{1}{\sqrt{2}} \big[v + \sin(\beta\!-\!\alpha) h^0 + \cos(\beta\!-\!\alpha) H^0 + i G^0\big]
    \end{pmatrix}\,,
    \\[0.1cm]
    \Phi^\prime &= \begin{pmatrix}
        H^+ \\ \frac{1}{\sqrt{2}} \big[ \cos(\beta\!-\!\alpha) h^0 - \sin(\beta\!-\!\alpha) H^0 + i A^0 \big]
    \end{pmatrix}\,,
\end{align}
\label{eq:Higgs-embedding}%
\end{subequations}
where the primes are used again to distinguish the doublets in the unbroken-phase Higgs basis ($H^\prime,\Phi^\prime$) from the ones in the soft-SUSY mass basis ($H,\Phi$), discussed previously in Sec.~\ref{sec:soft-SUSY_Higgs-mass-basis}.
Thus, when integrating out~$\Phi^\prime$, we also integrate out parts of the physical Higgs~$h^0$, while retaining parts of the UV state~$H^0$ in the spectrum. 
However, in the alignment limit, both contributions are small, since we have $\smash{\lim_{\alpha\to\beta+\pi/2} \cos(\beta-\alpha) = 0}$.
Therefore, the correct low-energy limit can be obtained by integrating out~$\Phi^\prime$ close to alignment.

When matching the MSSM in the unbroken-phase Higgs basis onto the SMEFT, we obtain the following tree-level matching conditions:
\begin{subequations}
\begin{align}
    M_H^2 &= m_H^{\prime \, 2} \left[ 1 - \tan^2(2\beta) \frac{m_H^{\prime \, 2}}{m_\Phi^{\prime \, 2}} \right]
    \,,
    \\
    \lambda 
    &=
    \cos^2(2\beta) \frac{ g_1^2 + g_2^2 }{4} \left[ 1 - 4 \tan^2(2\beta) \frac{m_H^{\prime \, 2}}{m_\Phi^{\prime \, 2}} \right] \,,
    \\
    C_H
    &=
    \frac{\sin^2(4\beta)}{m_\Phi^{\prime \, 2}} \frac{\left( g_1^2 + g_2^2 \right)^2}{64} 
    \,,
\end{align}
\end{subequations}
which agree with the results in Eqs.~\eqref{eq:threshold-corrections} and~\eqref{eq:tree-level_matching-conditions} after expressing $m_H^{\prime\,2}$ in terms of~$v^2$ and $m_\Phi^{\prime\,2}$ in terms of~$m_\Phi^{2}$.
This shows that the two possibilities we have considered for decoupling~$\Phi^{(\prime)}$, ($i$)~the soft-SUSY mass basis, where the mass mixing between the two doublets vanishes~$\Delta=0$; and ($ii$)~the aligned Higgs basis, where we have a non-vanishing mixing~$\Delta \neq 0$, yield the same result.
Since the two bases are perturbatively aligned, that is, their difference is suppressed in the EFT power counting by terms of order~$\mathcal{O}\!\left({\textrm{m}_Z^2}\middle/{\textrm{m}_{A^0}^2}\right)$ as we have seen in Eq.~\eqref{eq:tan2gamma-tan2beta}, the matching can be performed equally well in either basis. 

Matching in the soft-SUSY mass basis, as we do in this work (see Sec.~\ref{sec:soft-SUSY_Higgs-mass-basis}), leads to some technical simplifications and more compact results. It is also more in line with the assumptions of the EFT approach, where we integrate out mass eigenstates above the electroweak scale, which are thus decoupled by assumption, while also not significantly affecting EWSB.
Moreover, we avoid the issues of the alignment limit mentioned before.

Our discussion of the unbroken-phase Higgs basis here is thus intended only to touch base with the usual MSSM perspective, where the Higgs basis is commonly employed.
Notice, however, that the supplementary material we provide on GitHub also allows us to perform the matching in the unbroken-phase Higgs basis.
To do so, the \matchete model file for the MSSM has to be modified as described within this file.
Otherwise, the same code for the matching can be used.

\section{Matching onto the 2HDM-EFT}
\label{app:2HDM-EFT}

So far, we have only considered the scenario where all BSM states are integrated out of the MSSM resulting in the SMEFT as the effective low-energy description.
An interesting alternative is provided by the case where all $R$-parity even states are kept in the spectrum of the EFT and only their superpartners are integrated out.
That is, compared to the SMEFT case, we also include the second Higgs doublet~$\Phi$ in the spectrum of the EFT.
The extension of the 2HDM model to an EFT (including dimension-six operators) was discussed in \cite{Crivellin:2016ihg,Anisha:2019nzx,Dermisek:2024ohe}.
Here, we adopt the (complete) basis provided in Ref.~\cite{Dermisek:2024ohe}.

A couple of comments are in order regarding the actual implementation. 
First we notice that, when keeping both Higgs light, it is not necessary to rotate to the mass basis of the two doublets, as was done in the SMEFT case in Sec.~\ref{sec:Higgs}.
Instead, we can work in the original basis of the MSSM Lagrangian, see e.g.~Eq.~\eqref{eq:Higgs-potential}.
Since most 2HDM formulations (such as the one in~\cite{Dermisek:2024ohe}) are expressed in terms of two doublets with identical Hypercharge, it is convenient to redefine the MSSM Higgs fields $H_u\sim(\mathbf{1},\mathbf{2})_{1/2}$ and $H_d\sim(\mathbf{1},\mathbf{2})_{-1/2}$ by introducing
\begin{align}
    \phi_1 \equiv H_u &\sim (\mathbf{1},\mathbf{2})_{1/2}\,,
    &
    &\text{and}
    &
    \phi_2 \equiv H_d^c = \varepsilon H_d^\ast &\sim (\mathbf{1},\mathbf{2})_{1/2}\,,
\end{align}
where we use the new field labels~$\phi_{1,2}$ to distinguish this case from the SMEFT scenario.
Furthermore, we realize that this scenario requires a fine-tuning in order to obtain two light Higgs doublets while keeping the Higgsinos heavy, i.e., $\tilde\mu=\cO(m_\mathrm{soft})$, where $m_\mathrm{soft}$ is the soft SUSY-breaking scale in the~UV.
Comparing to Eq.~\eqref{eq:Higgs-potential}, we find the requirements $\hat{m}_{H_{u,d}}^2 \equiv |\tilde\mu|^2+m_{H_{u,d}}^2 = \cO(\textrm{m}_Z^2)$ and $b = \cO(\textrm{m}_Z^2)$ in order to obtain two light Higgs doublets, where $\textrm{m}_Z$ indicates the weak scale in the~IR.

A~minor simplification arises compared to the SMEFT case, since we are now integrating out only $R$-parity odd states, while only $R$-parity even states remain in the EFT. Hence, no tree-level contributions are generated and the matching starts at one loop. Nonetheless, due to the greater complexity of the 2HDM-EFT basis compared to the Warsaw basis, some technical modifications to \matchete (included with \texttt{v0.4.0}) were required in order to make this mapping feasible.

The matching results for the 2HDM case, together with the adjusted MSSM and 2HDM-EFT model files used in \matchete, as well as example notebooks for performing this matching are provided on \href{https://github.com/BSM-EFT/MSSM-to-SMEFT}{GitHub} in the directory named \texttt{2HDM-EFT}.

\section{Reading the one-loop matching results}
\label{app:reading-results}

\subsection{Loop functions}
The one-loop matching conditions provided in the supplementary material are expressed in terms of loop functions for brevity. 
These loop functions are defined as the finite piece of the loop integrals
\begin{align}
    \mathrm{LF}_{i_1,i_2,\ldots,i_{n+1}}[m_1,m_2,\ldots,m_n]
    &\equiv \left. \int \frac{\dd^D k}{(2\pi)^D} \left( \prod_{k=1}^n \frac{1}{(k^2-m_k^2)^{i_k}} \right) \frac{1}{(k^2)^{i_{n+1}}} \right\rvert_\text{finite} \,.
    \label{eq:LF}
\end{align}
When matching (the MSSM onto SMEFT) at dimension six, at most six different masses can appear in a single loop function, for example, for contributions proportional to six powers of the couplings~$a_{u,d,e}$.
For convenience, the matching results can be expressed as a function of the masses and the matching scale directly in \matchete by applying the routine \texttt{EvaluateLoopFunctions}.
In addition, their explicit form in terms of the masses is also provided on \href{https://github.com/BSM-EFT/MSSM-to-SMEFT/matching-results/PDFs}{GitHub} in the file named \texttt{LoopFunction-definition.pdf}.

To decrease the number of loop functions appearing in the output, \matchete simplifies the loop functions by removing all duplicate masses and sorting the masses so that $i_1,\ldots,i_n$ appear in decreasing order in Eq.~\eqref{eq:LF}.
Notice that this does not comprise a minimal set of loop functions. 
Further simplifications could be achieved using, e.g., partial fraction identities. 
However, this can significantly complicate the expressions for the matching conditions, which is why we refrain from doing so, and keep the non-minimal set of loop functions.

\subsection{Redefining renormalizable SMEFT couplings}\label{app:threshold-corrections-details}

\begin{table}[tbp]
    \centering
    {\renewcommand{\arraystretch}{1.1}
    \begin{tabular}{c|c|cccc}
         & Parameter & This work & \matchete & Theory & Scheme
         \\\hline\hline
         \multirow{6}{*}{Gauge couplings} & \multirow{2}{*}{$\mathrm{U}(1)_Y$} & $g_1^{\scriptscriptstyle \overline{\mathrm{MS}}}$ & $g_1$ & MSSM & $\overline{\mathrm{MS}}$
         \\
         & & $g_1$ & $c_{B^2}$ & SMEFT & $\overline{\mathrm{MS}}$
         \\\cline{2-6}
         & \multirow{2}{*}{$\mathrm{SU}(2)_L$} & $g_2^{\scriptscriptstyle \overline{\mathrm{MS}}}$ & $g_2$ & MSSM & $\overline{\mathrm{MS}}$
         \\
         & & $g_2$ & $c_{W^2}$ & SMEFT & $\overline{\mathrm{MS}}$
         \\\cline{2-6}
         & \multirow{2}{*}{$\mathrm{SU}(3)_c$} & $g_3^{\scriptscriptstyle \overline{\mathrm{MS}}}$ & $g_3$ & MSSM & $\overline{\mathrm{MS}}$ 
         \\
         & & $g_3$ & $c_{G^2}$ & SMEFT & $\overline{\mathrm{MS}}$
         \\\hline
         \multirow{6}{*}{Yukawa couplings} & \multirow{2}{*}{up-type${}^\ast$} & $y_u$ & $y_u$ & MSSM & $\overline{\mathrm{DR}}$
         \\
         & & $Y_u$ & $c_{Hqu}$ & SMEFT & $\overline{\mathrm{MS}}$
         \\\cline{2-6}
         & \multirow{2}{*}{down-type${}^\ast$} & $y_d$ & $y_d$ & MSSM & $\overline{\mathrm{DR}}$
         \\
         & & $Y_d$ & $c_{Hqd}$ & SMEFT & $\overline{\mathrm{MS}}$
         \\\cline{2-6}
         & \multirow{2}{*}{charged-lepton${}^\ast$} & $y_e$ & $y_e$ & MSSM & $\overline{\mathrm{DR}}$ 
         \\
         & & $Y_e$ & $c_{Hle}$ & SMEFT & $\overline{\mathrm{MS}}$
         \\\hline
         \multirow{4}{*}{Higgs} & \multirow{2}{*}{mass} & $m_H$ & $m_H$ & MSSM & $\overline{\mathrm{DR}}$ 
         \\
         & & $M_H$ & $c_{H^2}$ & SMEFT & $\overline{\mathrm{MS}}$
         \\\cline{2-6}
         & quartic coupling & $\lambda$ & $\lambda$ & SMEFT & $\overline{\mathrm{MS}}$
         \\\hline
         \multirow{13}{*}{BSM} & Mixing angle${}^\dagger$ & $\gamma,\beta$ & $\gamma$ & MSSM & $\overline{\mathrm{DR}}$
         \\\cline{2-6}
         & Heavy Higgs mass & $m_\Phi$ & $m_\Phi$ & MSSM & $\overline{\mathrm{DR}}$
         \\\cline{2-6}
         & \multirow{5}{*}{sfermion masses${}^\ast$} & $m_{\tilde{q}}$ & $m_{\tilde{q}}$ & MSSM & $\overline{\mathrm{DR}}$
         \\
         & & $m_{\tilde{u}}$ & $m_{\tilde{u}}$ & MSSM & $\overline{\mathrm{DR}}$
         \\
         & & $m_{\tilde{d}}$ & $m_{\tilde{d}}$ & MSSM & $\overline{\mathrm{DR}}$
         \\
         & & $m_{\tilde{l}}$ & $m_{\tilde{l}}$ & MSSM & $\overline{\mathrm{DR}}$
         \\
         & & $m_{\tilde{e}}$ & $m_{\tilde{e}}$ & MSSM & $\overline{\mathrm{DR}}$
         \\\cline{2-6}
         & \multirow{3}{*}{gaugino masses} & $m_1$ & $m_1$ & MSSM & $\overline{\mathrm{DR}}$
         \\
         & & $m_2$ & $m_2$ & MSSM & $\overline{\mathrm{DR}}$
         \\
         & & $m_3$ & $m_3$ & MSSM & $\overline{\mathrm{DR}}$
         \\\cline{2-6}
         & \multirow{3}{*}{scalar trilinears${}^\ast$} & $a_u$ & $a_u$ & MSSM & $\overline{\mathrm{DR}}$
         \\
         & & $a_d$ & $a_d$ & MSSM & $\overline{\mathrm{DR}}$
         \\
         & & $a_e$ & $a_e$ & MSSM & $\overline{\mathrm{DR}}$
         \\\hline
    \end{tabular}
    }
    \caption{List of all renormalizable parameters and the labels used to denote them in this work and in the \matchete output when interpreted as MSSM or SMEFT parameters. 
    In addition, the employed renormalization scheme is shown.
    The SMEFT parameters are defined to absorb all the threshold corrections they receive in the one-loop matching, with the exception of the quartic Higgs coupling~$\lambda$, 
    which is not a free parameter in the MSSM that can be replaced in favor of the SMEFT coupling.
    \\
    ${}^\dagger$: The angle~$\gamma$ is defined in Eq.~\eqref{eq:rotation-angle_mass-basis} to diagonalize the Higgs-doublet mass matrix at tree level only.
    \\
    ${}^\ast$: All flavorful parameters are in the sfermion mass basis as discussed in Sec.~\ref{sec:MSSM-Lagrangian-Dirac}.
    \label{tab:parameters}}
\end{table}

Separating the tree-level and one-loop threshold corrections within the matching conditions relating the SMEFT renormalizable couplings to the MSSM parameters, as shown in Eq.~\eqref{eq:threshold-corrections}, can be useful from a top-down point of view.
However, for a phenomenological study, it is more convenient to adopt a bottom-up perspective, because the numerical values for the renormalizable SMEFT (or rather SM) couplings are known at low energies from experiments.
This approach involves shifting the renormalizable SMEFT couplings in order to absorb the threshold corrections from the matching. 
We illustrate this with the up-type Yukawa matrix as an example, whose matching condition reads
\begin{align}
    Y_u^{pr} &= s_\beta\, y_u^{pr} + \frac{1}{16\pi^2} \Delta_{Y_u} \equiv c_{Hqu}
    \,,
\end{align}
where we have defined an effective coupling~$c_{Hqu}$ to absorb the threshold corrections. 
Subsequently, we can invert the above relation to substitute every occurrence of the MSSM Yukawa matrix~$y_u$ in the matching conditions by the newly defined effective coupling~$c_{Hqu}$. 
The advantage of expressing the matching conditions in terms of~$c_{Hqu}$ is that this coupling is equivalent to the full SMEFT Yukawa matrix in the Warsaw basis. 
Therefore, its value is determined by experimental data.

A~similar approach is adopted for the down-type and charged-lepton Yukawa matrices, as well as the Higgs mass, for which the effective couplings are labeled as $c_{Hqd}$, $c_{Hle}$, and~$c_{H^2}$, respectively. No effective coupling is introduced for the quartic Higgs coupling~$\lambda$, since it does not have an equivalent parameter in the full MSSM.

The matching conditions expressed in terms of these effective couplings can be obtained directly from \matchete with the \texttt{MapEffectiveCouplings} routine with the option \texttt{ShiftRenCouplings -> True}. 
Doing so also introduces additional effective couplings, labeled as $c_{B^2}$, $c_{W^2}$, and~$c_{G^2}$, for the gauge couplings $g_1$, $g_2$, and~$g_3$, respectively, in order to absorb the corresponding threshold corrections, shown on the right-hand sides of Eq.~\eqref{eq:gauge-coupling-corrections}.
To include the definitions of the effective couplings in terms of the MSSM parameters in the output, the option \texttt{AppendEffectiveCouplingsDefs -> True} should also be specified.
The complete list of parameters and the symbols used for them in the SMEFT and the MSSM is collected for convenience in Table~\ref{tab:parameters}, together with the one-loop definitions of all parameters.

We emphasize that the one-loop matching conditions presented in the PDF~files on \href{https://github.com/BSM-EFT/MSSM-to-SMEFT/matching-results/PDFs}{GitHub} are written without introducing the effective couplings.
That is, they are expressed in terms of the MSSM couplings.
However, the \texttt{Mathematica} results are provided in both ways, i.e., expressed in terms of MSSM or SMEFT couplings.

% - - - - Bibliography - - - - %
{
\addcontentsline{toc}{section}{References}
\bibliographystyle{JHEP}
\footnotesize
\bibliography{references}
}

\end{document}